\newif\iffigs\figstrue

\documentclass[12pt]{article}
\iffigs
  \input epsf
\else
\fi

\batchmode
\newfont{\footscrfont}{rsfs10}
  \newfont{\footbbbfont}{msbm10}
  \newfont{\manfont}{manfnt}
\errorstopmode

\newif\ifscrf\scrftrue
\ifx\footscrfont\nullfont
  \scrffalse
\fi

\newif\ifamsf\amsftrue
\ifx\footbbbfont\nullfont
  \amsffalse
\fi

\def\ppnumber{\vbox{\baselineskip14pt\hbox{CU-TP 924}
\hbox{DUKE-CGTP-99-05}
\hbox{hep-th/9907186}}}
\def\ppdate{July 1999}
\def\pplogo{\vbox{\kern-\headheight\kern -15pt
\halign{##&##\hfil\cr&{
\ppnumber}\cr\rule{0pt}{2.5ex}&\ppdate\cr}
}}

\makeatletter
\date{}
\def\dedicatory#1{\def\@date{\normalsize\it#1}}
\def\subjclass#1{\def\@thefnmark{}\@footnotetext{1991
    {\it Mathematics Subject Classification.} #1}}
\def\keywords#1{\def\@thefnmark{}\@footnotetext{
    {\it Key words and phrases.} #1}}

\def\ps@firstpage{\ps@empty \def\@oddhead{\hss\pplogo}%
  \let\@evenhead\@oddhead 
}
\def\maketitle{\par
 \begingroup
 \def\thefootnote{\fnsymbol{footnote}}
 \def\@makefnmark{\hbox
 to 0pt{$^{\@thefnmark}$\hss}}
 \if@twocolumn
 \twocolumn[\@maketitle]
 \else \newpage
 \global\@topnum\z@ \@maketitle \fi\thispagestyle{firstpage}\@thanks
 \endgroup
 \setcounter{footnote}{0}
 \let\maketitle\relax
 \let\@maketitle\relax
 \gdef\@thanks{}\gdef\@author{}\gdef\@title{}\let\thanks\relax}

\def\abstract{\if@twocolumn
\section*{Abstract}
\else \small
\begin{center}
{\bf ABSTRACT}
\end{center}
\quotation
\fi}

\def\thebibliography#1{\section*{References\@mkboth
 {REFERENCES}{REFERENCES}}\small\list
 {[\arabic{enumi}]}{\settowidth\labelwidth{[#1]}\leftmargin\labelwidth
 \advance\leftmargin\labelsep
 \usecounter{enumi}}
 \def\newblock{\hskip .11em plus .33em minus .07em}
 \sloppy\clubpenalty4000\widowpenalty4000
 \sfcode`\.=1000\relax}

\newif\iffn\fnfalse

\@ifundefined{reset@font}{\let\reset@font\empty}{} 
\long\def\@footnotetext#1{\insert\footins{\reset@font\footnotesize
    \interlinepenalty\interfootnotelinepenalty
    \splittopskip\footnotesep
    \splitmaxdepth \dp\strutbox \floatingpenalty \@MM
    \hsize\columnwidth \@parboxrestore
   \edef\@currentlabel{\csname p@footnote\endcsname\@thefnmark}\@makefntext
    {\rule{\z@}{\footnotesep}\ignorespaces
      \fntrue#1\fnfalse\strut}}}

\makeatother

\ifamsf
  \newfont{\bigbbbfont}{msbm10 scaled\magstep2}
  \newfont{\bbbfont}{msbm10 scaled\magstep1}  
  \newfont{\smallbbbfont}{msbm8}
  \newfont{\tinybbbfont}{msbm6}
  \newfont{\smallfootbbbfont}{msbm7}
  \newfont{\tinyfootbbbfont}{msbm5}
  \newfont{\biggthfont}{eufm10 scaled\magstep2}
  \newfont{\gthfont}{eufm10 scaled\magstep1}  
  \newfont{\smallgthfont}{eufm8}
  \newfont{\tinygthfont}{eufm6}
  \newfont{\footgthfont}{eufm10}
  \newfont{\smallfootgthfont}{eufm7}
  \newfont{\tinyfootgthfont}{eufm5}
\fi

\ifscrf
  \newfont{\scrfont}{rsfs10 scaled\magstep1}  
  \newfont{\smallscrfont}{rsfs7}
  \newfont{\tinyscrfont}{rsfs7}
  \newfont{\smallfootscrfont}{rsfs7}
  \newfont{\tinyfootscrfont}{rsfs7}
\fi

\ifamsf
  \newcommand{\Bbb}[1]{\iffn
      \mathchoice{\mbox{\footbbbfont #1}}{\mbox{\footbbbfont #1}}
      {\mbox{\smallfootbbbfont #1}}{\mbox{\tinyfootbbbfont #1}}\else
      \mathchoice{\mbox{\bbbfont #1}}{\mbox{\bbbfont #1}}
      {\mbox{\smallbbbfont #1}}{\mbox{\tinybbbfont #1}}\fi}
  
\else
  \def\bigbbbfont{\bf}
  \def\Bbb{\bf}
  
\fi

\ifscrf
  \newcommand{\Scr}[1]{\iffn
    \mathchoice{\mbox{\footscrfont #1}}{\mbox{\footscrfont #1}}
    {\mbox{\smallfootscrfont #1}}{\mbox{\tinyfootscrfont #1}}\else
    \mathchoice{\mbox{\scrfont #1}}{\mbox{\scrfont #1}}
    {\mbox{\smallscrfont #1}}{\mbox{\tinyscrfont #1}}\fi}
\else
  \def\Scr{\cal}
\fi

\def\C{{\Bbb C}}

\def\P{{\Bbb P}}

\def\R{{\Bbb R}}
\def\Z{{\Bbb Z}}

\def\bearray{\begin{eqnarray}}
\def\eearray{\end{eqnarray}}
\def\bearraynn{\begin{eqnarray*}}
\def\eearraynn{\end{eqnarray*}}
\def\bfig{\begin{figure}}
\def\efig{\end{figure}}

\def\opeq#1{\advance\lineskip#1 \advance\baselineskip#1
        \advance\lineskiplimit#1}

\def\eqalign#1{\null\,\vcenter{\opeq{2.5\jot}\mathsurround=0pt
        \everycr={}\tabskip=0pt
        \halign{\strut\hfil$\displaystyle{##}$&$\displaystyle{{}##}$\hfil
        \crcr#1\crcr}}\,\null}

\def\cM{{\Scr M}}

\def\cD{{\Scr D}}

\def\cMc{{\hfuzz=100cm\hbox to 0pt{$\;\overline{\phantom{X}}$}\cM}}
\def\barcD{{\hfuzz=100cm\hbox to 0pt{$\;\overline{\phantom{X}}$}\cD}}

\ifamsf

\else

\fi

\def\boldone{\relax{\rm 1\kern-.35em 1}}

\newtheorem{Proposition}{Proposition}[section]

\newtheorem{Theorem}{Theorem}[section]
\newtheorem{Lemma}{Lemma}[section]
\newtheorem{Corrolary}{Corrolary}[section]

\newcommand{\be}{\begin{equation}}
\newcommand{\ee}{\end{equation}}
\newcommand{\bea}{\begin{eqnarray}}
\newcommand{\eea}{\end{eqnarray}}

\newcommand{\bp}{\begin{Proposition}}
\newcommand{\ep}{\end{Proposition}}
\newcommand{\bt}{\begin{Theorem}}
\newcommand{\et}{\end{Theorem}}
\newcommand{\bl}{\begin{Lemma}}
\newcommand{\el}{\end{Lemma}}
\newcommand{\bc}{\begin{Corrolary}}
\newcommand{\ec}{\end{Corrolary}}
\newcommand{\nn}{\nonumber}

\newread\epsffilein    
\newif\ifepsffileok    
\newif\ifepsfbbfound   
\newif\ifepsfverbose   
\newdimen\epsfxsize    
\newdimen\epsfysize    
\newdimen\epsftsize    
\newdimen\epsfrsize    
\newdimen\epsftmp      
\newdimen\pspoints     
\def\epsfbox#1{\global\def\epsfllx{72}\global\def\epsflly{72}%
   \global\def\epsfurx{540}\global\def\epsfury{720}%
   \def\lbracket{[}\def\testit{#1}\ifx\testit\lbracket
   \let\next=\epsfgetlitbb\else\let\next=\epsfnormal\fi\next{#1}}%
\def\epsfgetlitbb#1#2 #3 #4 #5]#6{\epsfgrab #2 #3 #4 #5 .\\%
   \epsfsetgraph{#6}}%
\def\epsfnormal#1{\epsfgetbb{#1}\epsfsetgraph{#1}}%
\def\epsfgetbb#1{%
\openin\epsffilein=#1
\ifeof\epsffilein\errmessage{I couldn't open #1, will ignore it}\else
   {\epsffileoktrue \chardef\other=12
    \def\do##1{\catcode`##1=\other}\dospecials \catcode`\ =10
    \loop
       \read\epsffilein to \epsffileline
       \ifeof\epsffilein\epsffileokfalse\else
          \expandafter\epsfaux\epsffileline:. \\%
       \fi
   \ifepsffileok\repeat
   \ifepsfbbfound\else
    \ifepsfverbose\message{No bounding box comment in #1; using defaults}\fi\fi
   }\closein\epsffilein\fi}%
\def\epsfclipstring{}
\def\epsfsetgraph#1{%
   \epsfrsize=\epsfury\pspoints
   \advance\epsfrsize by-\epsflly\pspoints
   \epsftsize=\epsfurx\pspoints
   \advance\epsftsize by-\epsfllx\pspoints
   \epsfxsize\epsfsize\epsftsize\epsfrsize
   \ifnum\epsfxsize=0 \ifnum\epsfysize=0
      \epsfxsize=\epsftsize \epsfysize=\epsfrsize
      \epsfrsize=0pt
     \else\epsftmp=\epsftsize \divide\epsftmp\epsfrsize
       \epsfxsize=\epsfysize \multiply\epsfxsize\epsftmp
       \multiply\epsftmp\epsfrsize \advance\epsftsize-\epsftmp
       \epsftmp=\epsfysize
       \loop \advance\epsftsize\epsftsize \divide\epsftmp 2
       \ifnum\epsftmp>0
          \ifnum\epsftsize<\epsfrsize\else
             \advance\epsftsize-\epsfrsize \advance\epsfxsize\epsftmp \fi
       \repeat
       \epsfrsize=0pt
     \fi
   \else \ifnum\epsfysize=0
     \epsftmp=\epsfrsize \divide\epsftmp\epsftsize
     \epsfysize=\epsfxsize \multiply\epsfysize\epsftmp   
     \multiply\epsftmp\epsftsize \advance\epsfrsize-\epsftmp
     \epsftmp=\epsfxsize
     \loop \advance\epsfrsize\epsfrsize \divide\epsftmp 2
     \ifnum\epsftmp>0
        \ifnum\epsfrsize<\epsftsize\else
           \advance\epsfrsize-\epsftsize \advance\epsfysize\epsftmp \fi
     \repeat
     \epsfrsize=0pt
    \else
     \epsfrsize=\epsfysize
    \fi
   \fi
   \ifepsfverbose\message{#1: width=\the\epsfxsize, height=\the\epsfysize}\fi
   \epsftmp=10\epsfxsize \divide\epsftmp\pspoints
   \vbox to\epsfysize{\vfil\hbox to\epsfxsize{%
      \ifnum\epsfrsize=0\relax
        \includegraphics{#1}%
      \else
        \epsfrsize=10\epsfysize \divide\epsfrsize\pspoints
        \includegraphics{#1}%
      \fi
      \hfil}}%
\global\epsfxsize=0pt\global\epsfysize=0pt}%
{\catcode`\%=12 \global\let\epsfpercent=
\long\def\epsfaux#1#2:#3\\{\ifx#1\epsfpercent
   \def\testit{#2}\ifx\testit\epsfbblit
      \epsfgrab #3 . . . \\%
      \epsffileokfalse
      \global\epsfbbfoundtrue
   \fi\else\ifx#1\par\else\epsffileokfalse\fi\fi}%
\def\epsfempty{}%
\def\epsfgrab #1 #2 #3 #4 #5\\{%
\global\def\epsfllx{#1}\ifx\epsfllx\epsfempty
      \epsfgrab #2 #3 #4 #5 .\\\else
   \global\def\epsflly{#2}%
   \global\def\epsfurx{#3}\global\def\epsfury{#4}\fi}%
\def\epsfsize#1#2{\epsfxsize}
\usepackage{graphics}

\begin{document}

\title{D3-branes on partial resolutions of 
abelian quotient singularities of Calabi-Yau threefolds}

\author{Chris~Beasley$^{1,a}$, Brian~R.~Greene$^{2,b}$,
C.~I.~Lazaroiu$^{3,c}$, M.~R.~Plesser$^{1,d}$}


\maketitle

\vbox{
\centerline{$^{1}$ Center for Geometry and Theoretical Physics}
\centerline{Box 90318, Duke University}
\centerline{Durham, NC 27708-0318}

\medskip

\centerline{$^2$Departments of Physics and Mathematics}
\centerline{$^3$Department of Physics}

\centerline{Columbia University}
\centerline{N.Y., N.Y. 10027}
\medskip
\medskip
\bigskip
}

\abstract{We investigate field theories on the worldvolume of a
D3-brane transverse to partial resolutions 
of a $\Z_3\times\Z_3$ Calabi-Yau threefold quotient singularity. 
We deduce the 
field content and lagrangian of such theories and present a systematic 
method for mapping the moment map levels characterizing the partial 
resolutions of the singularity to the Fayet-Iliopoulos parameters of the 
D-brane worldvolume theory. As opposed to the simpler cases studied before, 
we find a complex web of partial resolutions and associated field-theoretic 
Fayet-Iliopoulos deformations. The analysis is performed by toric methods, 
leading to a structure which can be efficiently described 
in the language of convex geometry. 
For the worldvolume 
theory, the analysis of the moduli space  
has an elegant description in terms of quivers. As a by-product, we present 
a systematic way of extracting the birational geometry of the classical moduli 
spaces, thus generalizing previous work on resolution of singularities 
by D-branes.
}

\vskip .6in

$^a$ ceb5@cgtp.duke.edu

$^b$ greene@phys.columbia.edu

$^c$ lazaroiu@phys.columbia.edu

$^d$ plesser@cgtp.duke.edu

\pagebreak

\section*{Introduction}

In this paper we perform a systematic study of the worldvolume field theory 
of one D3-brane transverse to partial resolutions of a Calabi-Yau quotient 
singularity, by generalizing and improving on the approach pioneered in 
\cite{branes3}.
Particular examples of
such systems have been studied before in \cite{branes3,branes3'}, 
where it was shown that 
they exhibit interesting phenomena such as topology change and 
projection of non-geometric phases. The main purpose of the present paper is to
improve on the previous analysis of such systems, and to give a systematic 
and computationally efficient way to approach more complicated singularities. 
As a simple illustration of our methods, we re-analyze the case of 
$\C^3/\Z_2\times\Z_2$ Gorenstein singularities (a first analysis of which was 
presented in \cite{branes3'}), then we proceed to the analysis of the new and 
considerably more complicated case of a D3-brane transverse to a 
$\C^3/\Z_3\times \Z_3$ singularity.

The motivation of the present work is to prepare the ground for a 
detailed analysis of the conformal field theory on a large number of 
D3-branes  transverse to 
partial resolutions of a $\C^3/\Z_3\times\Z_3$ Gorenstein singularity
\cite{us}. Such an analysis is motivated by investigations of the 
AdS/CFT conjecture \cite{Maldacena} for 
spaces of the form $AdS_5\times X_5$ with $X_5$ an Einstein-Sasaki 
five-manifold
describing the angular part of the tangent cone to a partial resolution of 
such a singularity. As explained in \cite{Figueroa,MP}, the partial 
resolutions of interest for the AdS/CFT conjecture are those%
\footnote{We exclude the cases $X_5=S^5$ and 
$X_5=S^5/\Z_3$, which can be studied by direct methods.} 
for which the tangent cone at 
the singular point can be written as a complex cone over a del Pezzo surface, 
so that $X_5$ is a circle bundle over the del Pezzo. Since one knows how to 
obtain the associated field theory only in the case when the partial 
resolution is toric, one is currently restricted to toric del Pezzo's, 
which amounts to considering only the surfaces 
$F_0=\P^1\times\P^1$ and  $dP_1,dP_2,dP_3$ 
(the blow-ups of $P^2$ at one, two and three points 
respectively). Of these, only $dP_3$ and $F_0$ are known to admit a circle 
bundle $X_5$ over themselves which carries a regular Einstein-Sasaki 
structure, a result
which follows from the work of Tian and Yau on the positive case of the Calabi 
conjecture \cite{Tian_Yau,Tian,Tian_moduli}. Their investigations 
showed that all del Pezzo surfaces $dP_k$ 
with $k\geq 3$ admit a K\"ahler-Einstein metric of positive curvature. This 
allows one to prove the existence of regular Einstein-Sasaki 
five-manifolds $X_5$ which form the total space of a  circle bundle over 
$sP_3$.The case of $F_0$ follows from more elementary results. 

From the work of \cite{Tian_Yau}, it is also known that 
$dP_1$ and $dP_2$ do 
not admit such metrics, which raises the difficult mathematical question of 
whether a (possibly singular) associated `five-manifold', carrying a 
(possibly non-regular) Einstein-Sasaki structure exists, and the physical 
question of what is the status of the AdS/CFT conjecture in such a situation. 
It is known (see, for example \cite{Altmann,MP}) that the complex cone 
over each toric del Pezzo surface can be realized as the tangent cone 
to some partial resolution of $\C^3/\Z_3\times \Z_3$. However, it is in 
principle possible that such partial resolutions are simply 
inaccessible from the point of view of the worldvolume field theory of a 
D3-brane transverse to our space, 
namely that they cannot be realized physically by turning on Fayet-Iliopoulos
parameters in our theory. As discussed in \cite{branes3,branes3'}, the partial 
resolutions of the singularity are realized as moduli spaces of the D-brane 
theory, in the presence of certain Fayet-Iliopoulos terms. Namely, the space 
of all possible Fayet-Iliopoulos parameters admits a partition into cones, 
and the complex structure of the classical moduli space of the worldvolume 
theory does not change as long as the Fayet-Iliopoulos parameters remain 
inside of a given cone. However, there is no apriori reason to expect that 
all partial resolutions can be obtained as moduli spaces in this manner. 
In other words, it is in principle possible that some partial resolutions 
are never realized in the field theory in the way outlined above, no matter 
how one chooses the Fayet-Iliopoulos terms. 
One purpose of the present paper is to test this 
possibility directly, by investigating the full list of field theoretic 
realizations of the partial resolutions of interest. As we will discover, 
these partial resolutions are in fact realized for some choices of 
Fayet-Iliopoulos terms. This shows that the 
solution of the puzzle is necessarily more subtle. The 
complexity of the analysis is quite pronounced, in marked contrast with cases 
considered before. Therefore, it turns out to be necessary to refine 
the techniques available for studying the problem, and to follow a  
systematic approach towards its resolution.

The structure of this paper is as follows. In Section 1, we present 
an overview  of our results for the case of $\Z_3\times \Z_3$, in 
nontechnical language.  In Section 2, we give a systematic 
presentation of our method in a very general context. Since this approach 
leads to cumbersome notation, we illustrate the abstract 
methods by an application to the example of a D3-brane transverse to a  
$\C^3/\Z_2\times\Z_2$ quotient singularity and its partial 
resolutions. In Section 3, we give a systematic presentation of the 
case $\C^3/\Z_3\times \Z_3$. Section 4 presents our conclusions. The 
appendix lists certain data relevant to the model 
$\C^3/\Z_3\times\Z_3$. Throughout the paper, we assume some 
familiarity with toric geometry as well as with some basic concepts of 
algebraic and symplectic geometry.

\section{Overview of the case $\C^3/\Z_3\times \Z_3$}

\subsection{The geometric realization of the complex cones over $F_0,dP_1,
dP_2$ and $dP_3$}

In \cite{branes3'} it was shown how a certain partial resolution of the 
$\C^3/\Z_2\times\Z_2$ singularity (which leads to the conifold singularity)
can be realized in the moduli space of D-branes. This approach was used in 
\cite{MP} to deduce the worldvolume theory of $N$ parallel D3 branes 
transverse to a conifold singularity, as well as the corresponding worldvolume 
theories at certain other conical singularities which can be obtained as 
partial resolutions of $\C^3/\Z_2\times\Z_2$. In this manner, a systematic 
procedure was presented for deducing the CFT side of the AdS/CFT 
correspondence in the case of nonspherical horizons. In particular, the 
results of \cite{Klebanov} (originally 
obtained by making use of arguments entirely dependent on the high symmetry of 
the conifold), were reobtained in a systematic manner and generalized. 
In this section, we will apply methods similar to \cite{MP} to the conical 
singularities discussed in the introduction. 

As it turns out, each of these singularities can be realized as a partial 
resolution of $\C^3/\Z_3\times\Z_3$. To explain this,
consider a set of generators $g_1=({\hat 1},{\hat 0}), 
g_2=({\hat 0},{\hat 1})$ of the group $\Z_3 \times \Z_3$.  We choose
the action of $\Z_3 \times \Z_3$ on $\C^3$ to be given by: 
\be
\label{R}
\eqalign{
R(g_1):(X,Y,Z) &\mapsto (\omega X,\omega^{-1} Y,Z)\cr
R(g_2):(X,Y,Z) &\mapsto (\omega X,Y,\omega^{-1} Z)\ ,\cr }
\ee
where $\omega=e^{\frac{2\pi i}{3}}$. 
As is the case with any abelian quotient singularity, this is an affine 
toric variety. 
Thus, it can be described by a cone $C$ in $\R^3$ which cuts the plane 
$x+y+z=1$ along a convex polygon. 
In our case, this polygon is a triangle with vertices $v_1,v_2,v_3$
which contains $7$ other integral points $\{w_1,\ldots,w_7\}$, 
only one of which lies in its interior. We label these vectors as follows
(see Figure1(a)):
\be
\label{vecs}
\eqalign{ &v_1 = (0,3) \quad v_2= (0,0) \quad v_3 = (3,0) \quad w_1 = (0,2) 
\quad w_2 = (0,1) \cr
&w_3 = (1,0) \quad w_4 = (2,0) \quad w_5 = (1,2) \quad w_6 = (2,1)
\quad w_7 = (1,1) \cr}~~.
\ee

$$\matrix{\centerline {\epsfxsize=1in\epsfbox{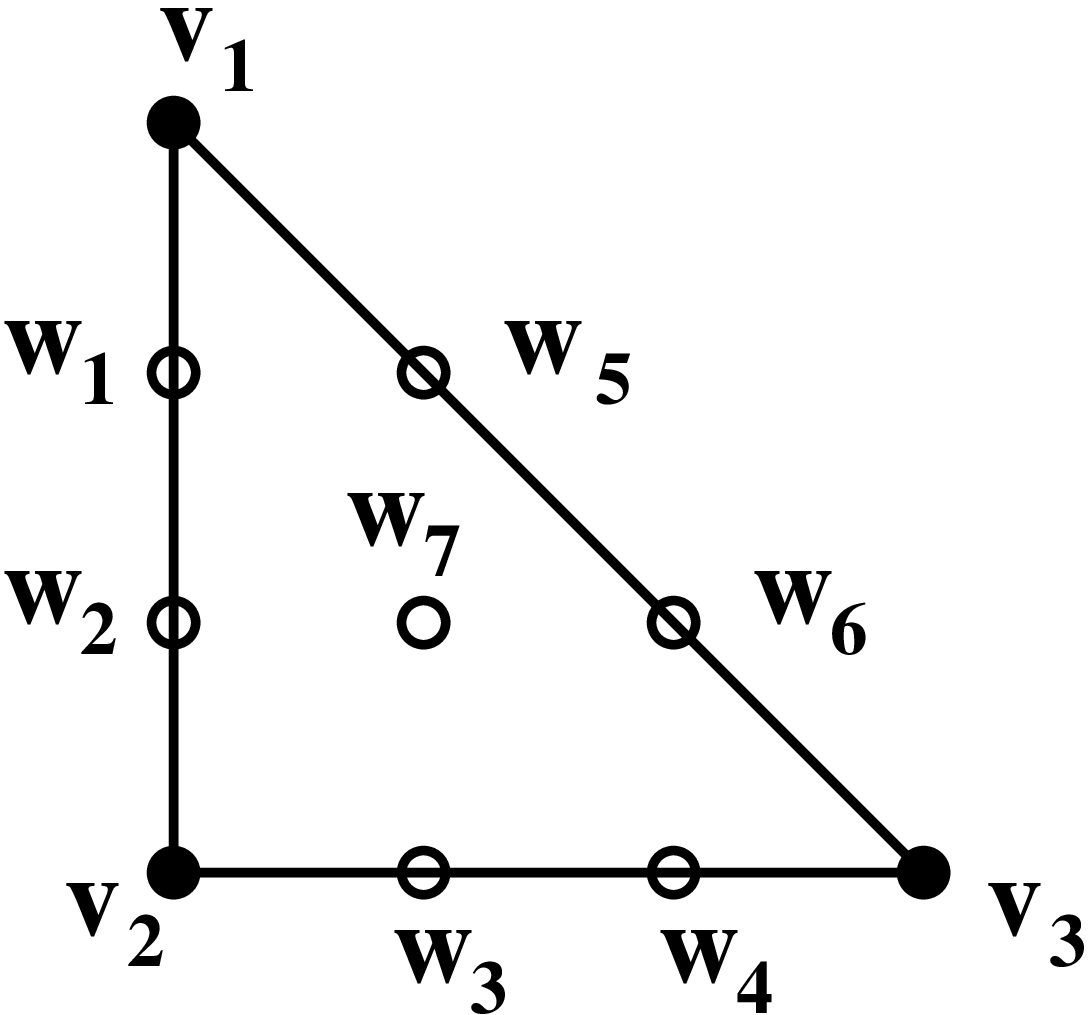}}\cr
\hbox{Figure 1(a). {\footnotesize $\Bbb{Z}_3\times\Bbb{Z}_3$ orbifold.}}
\cr \cr
{\epsfxsize=1in\epsfbox{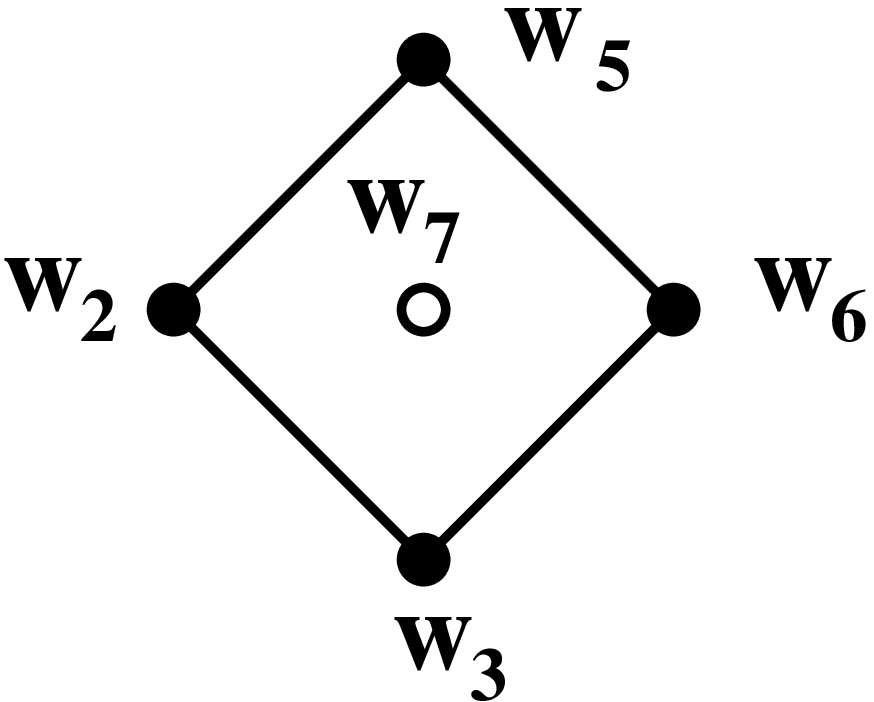}}
\cr 
\hbox{Figure 1(b). 
{\footnotesize Cone over $F_0$.}}
\cr \cr
{\epsfxsize=1in\epsfbox{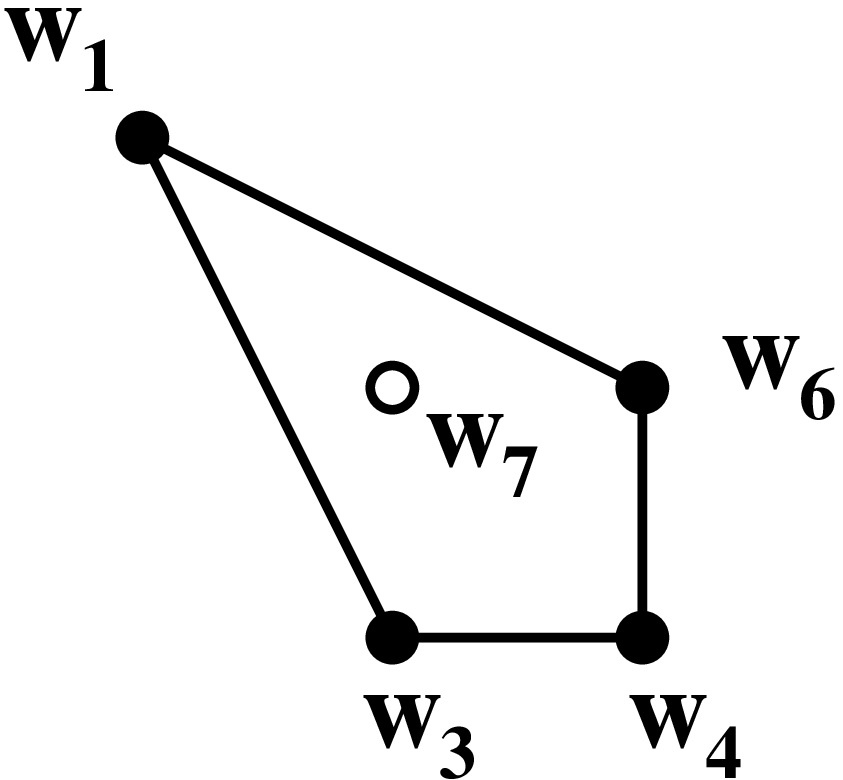}}
\cr 
\hbox{Figure 1(c). {\footnotesize Cone over $dP_1$}}
\cr \cr
{\epsfxsize=1in\epsfbox{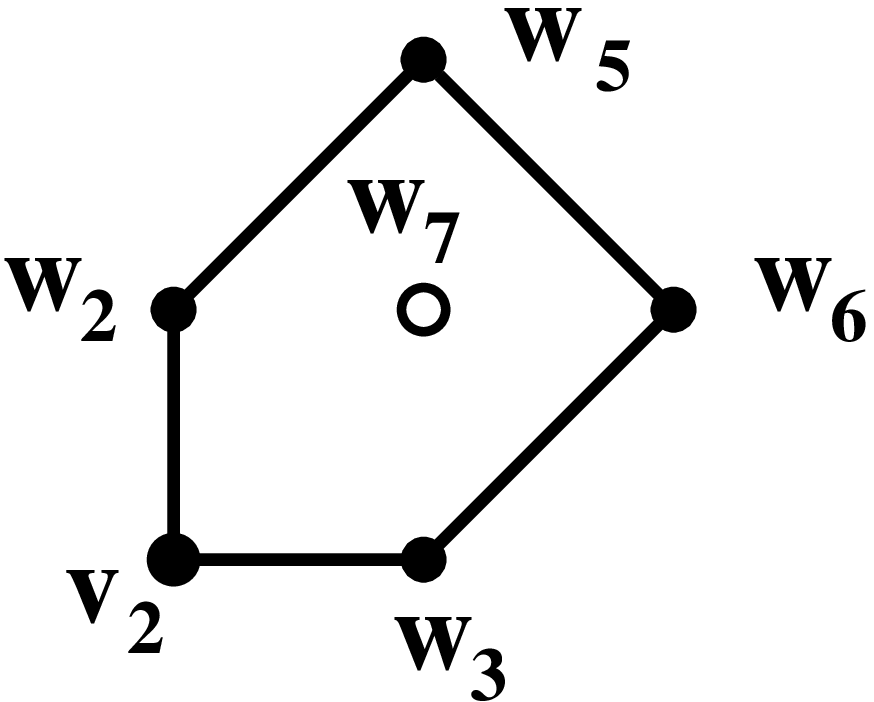}}
\cr 
\hbox{Figure 1(d). {\footnotesize Cone over $dP_2$}}
\cr \cr
{\epsfxsize=1in\epsfbox{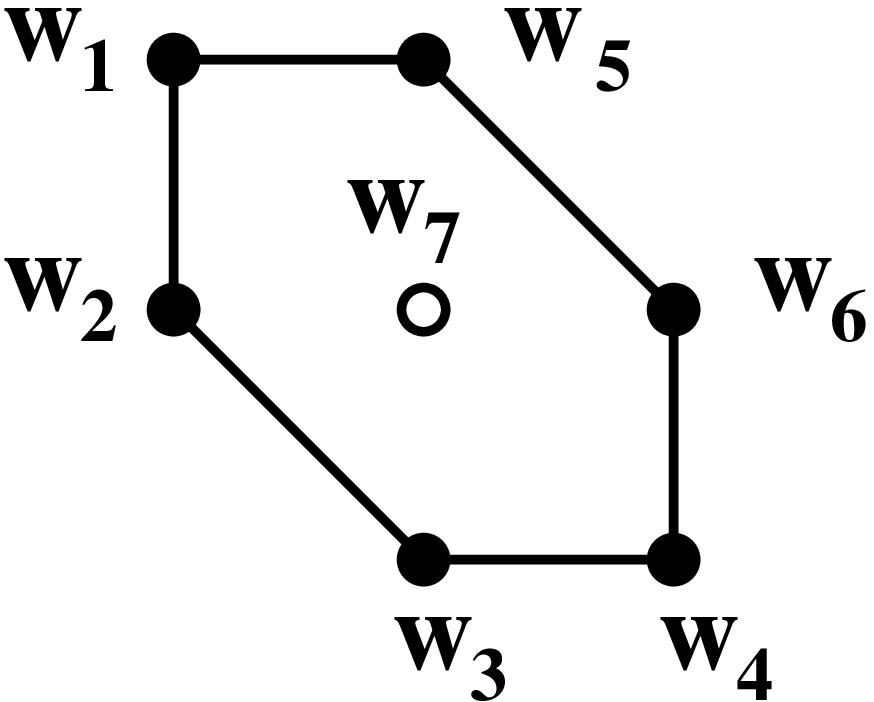}}
\cr  
\hbox{Figure 1(e). {\footnotesize Cone over $dP_3$}} 
\cr \cr}$$
\centerline{Figure 1. {\footnotesize Toric presentations of the partial 
resolutions of interest}}

\

\

To find the associated symplectic quotient description of this toric 
singularity, consider the basis of linear relations between these vectors 
given by\footnote{We consider only linear relations such that the sum of all 
charge vectors equals zero. Equivalently, we consider linear relations 
between the lifts of the vectors in Figure 1(a) to the plane $z=1$ in $\R^3$.}:
\be
\label{basis}
\begin{array}{ccc}v_{{3}}-3\,w_{{3}}+2\,v_{{2}}&=&0
\\\noalign{\medskip}w_{{6}}-2\,w_{{3}}-w_{{2}}+2\,v_{{2}}&=&0
\\\noalign{\medskip}w_{{4}}-2\,w_{{3}}+v_{{2}}&=&0\\\noalign{\medskip}w_
{{5}}-w_{{3}}-2\,w_{{2}}+2\,v_{{2}}&=&0\\\noalign{\medskip}w_{{7}}-w_{{3
}}-w_{{2}}+v_{{2}}&=&0\\\noalign{\medskip}v_{{1}}-3\,w_{{2}}+2\,v_{{2}}&=&0
\\\noalign{\medskip}w_{{1}}-2\,w_{{2}}+v_{{2}}&=&0\end {array}
\ee
Introduce homogeneous coordinates $x_i, y_j$ on $\C^{10}$ 
corresponding to $v_i,w_j$.  Then the 
$U(1)^7$-action is determined by the matrix of linear relations: 
\be
\label{chrgmat}
\left [\begin {array}{cccccccccc}
x_3&y_6&y_4&y_5&y_7&y_3&x_1&y_1&y_2&x_2 
\\\noalign{\medskip}1&0&0&0&0&-3&0&0&0&2
\\\noalign{\medskip}0&1&0&0&0&-2&0&0&-1&2\\\noalign{\medskip}0&0&1&0&0
&-2&0&0&0&1\\\noalign{\medskip}0&0&0&1&0&-1&0&0&-2&2
\\\noalign{\medskip}0&0&0&0&1&-1&0&0&-1&1\\\noalign{\medskip}0&0&0&0&0
&0&1&0&-3&2\\\noalign{\medskip}0&0&0&0&0&0&0&1&-2&1\end {array}\right 
]
\ee
while the associated  moment map equations, for a central level
$\zeta=(\zeta_1...\zeta_7)$ are:  
\be
\label{moment}
\begin {array}{ccc} {|x_{{3}}|}^{2}-3\,{|y_{{3}}|}^{2}+2\,{|x_{{2}}|}^{
2}&=&\zeta_1\\\noalign{\medskip}{|y_{{6}}|}^{2}-2\,{|y_{{3}}|}^{2}-{|y_{{2}}|}^{2}+2
\,{|x_{{2}}|}^{2}&=&\zeta_2\\\noalign{\medskip}{|y_{{4}}|}^{2}-2\,
{|y_{{3}}|}^{2}+
{|x_{{2}}|}^{2}&=&\zeta_3\\\noalign{\medskip}{|y_{{5}}|}^{2}-{|y_{{3}}|}^{2}-
2\,{|y_{{2}}|}^{2}+2\,{|x_{{2}}|}^{2}&=&\zeta_4\\
\noalign{\medskip}{|y_{{7}}|}^{2}-{|y_{{3}}|}^{2}-{|y_{{2}}|}^{2}+
{|x_{{2}}|}^{2}&=&\zeta_5\\\noalign{\medskip}{|x_{{1}}|}^{2}-3\,
{|y_{{2}}|}^{2}+2\,{|x_{{2}}|}^{2}&=&\zeta_6\\
\noalign{\medskip}{|x_{{1}}|}^{2}-2\,{|y_{{2}}|}^{2}+{|x_{{2}}|}^{2}&=&
\zeta_7\end {array}
\ee
If all $\zeta_i = 0$, then we obtain the unresolved
$\C^3/\Z_3 \times \Z_3$ orbifold singularity.  

Conversely, for generic values of $\zeta_i$, the singularity is
completely resolved and the symplectic quotient leads to a smooth
manifold. However, if the $\zeta_i$ lie in particular cones of some
codimension, the symplectic quotient is singular.  The singularity is
determined by the cone, and is the same for all points in its
interior.  

As an extreme example, consider the one-dimensional cone in which all
$\zeta_i=0$ except for $\zeta_5 > 0$.  In this cone, we see from
(\ref{moment}) that there is a vicinity of the origin of the space of 
homogeneous coordinates on which $y_7$ cannot vanish (this coordinate 
corresponds to the
center of the triangle in figure 1(a)).  We can thus perform
explicitly one of the $U(1)$ quotients by fixing its value to be real
and positive.  Since its value is then determined by the remaining
homogeneous variables, we have effectively eliminated one coordinate
and one quotient from the problem.  Note that the higher the dimension
of the cone in which the $\zeta_i$ lie, the larger is the number of
coordinates we can eliminate, and the milder the singularity we obtain.
The extreme case of the $\C^3/\Z_3\times\Z_3$ singularity 
corresponds to $\zeta=0$ (a zero-dimensional cone), while a complete 
resolution corresponds to a `generic' vector $\zeta$, i.e. one which 
belongs to a seven-dimensional cone.

The procedure described above gives a symplectic description 
of the various resolutions.%
\footnote{To be rigorous, this procedure yields not the partial
resolutions but their tangent cones.  In fact, the partial resolutions
are only quasiprojective varieties, while the tangent cones are
affine. As discussed for example in \cite{Oda}, the partial
resolutions are associated to triangulations of the first set of
points in Figure 1, and therefore their fans are not of affine type,
i.e. do not coincide with the fan of all faces of a single
3-dimensional cone. However, the procedure above suffices to correctly
identify the chambers in $\R^7(\zeta)$ parameterizing the resolutions.
Note that the singularities we wish to realize (the complex cones over
the toric del Pezzo surfaces) are affine toric varieties, and
therefore they will only correspond to the tangent cones to the
partial resolutions we will identify below.}  It is not hard to see
that, with appropriate choices of levels $\zeta$, one can realize the
singularity corresponding to any subpolygon of the triangle shown in
Figure 1(a).

For the geometrically-inclined reader, we mention that 
the discussion above is a concrete realization of very general results of 
Symplectic Geometry and Geometric Invariant Theory. Such results assure us that
the space of moment map levels has a canonical partition into conical 
chambers,
which together form a fan $\Psi$ in 
$\R^7(\zeta)$. As $\zeta$ varies inside of a given chamber, the algebraic 
structure of our toric variety does not change 
(even though the K\"ahler metric 
induced by the symplectic reduction changes). When $\zeta$ crosses a wall 
separating two chambers, the toric variety undergoes a birational 
transformation known as a toric flop; the variety is singular precisely when 
$\zeta$ lies on a wall (this is the same mathematical structure which 
underlies topology change in the moduli space of conformal field theories
\cite{top_change}). The procedure we just described gives one practical 
way of identifying these walls and chambers.

The complex cones over the del Pezzo surfaces of interest are well-known to 
be affine toric varieties themselves\cite{Altmann}, 
and are described by the polygons listed in Figure 1. 
The particular presentations shown there allow us 
to realize these cones as tangent cones to 
partial resolutions of the $\C^3/\Z_3\times\Z_3$ 
singularity. This  amounts to viewing the polygons as 
inscribed in the triangle of Figure 1(a) and turning on appropriate 
levels $\zeta$ in order to eliminate precisely the integral points which lie 
on the triangle but not on the del Pezzo polytope.
The subspaces of $\R^7(\zeta)$ leading to the desired partial resolutions, 
as well as the corresponding charge matrices describing these as 
toric varieties are listed in Table 1.

\

$$\vbox{\offinterlineskip\halign{\strut # height 37pt depth 26pt&
\quad#\quad\hfill\vrule&\quad$#$\quad\hfill\vrule&
\quad$#$\quad\hfill\vrule\cr  \noalign{\hrule} \vrule&$
\matrix{&\hbox{cone over}\cr & F_0}$&
\footnotesize{\hskip 0.05in\matrix{\zeta_{{2}}-2\zeta_{{5}},\zeta_{{4}}-2
\zeta_{{5}} &=0&\cr \zeta_{{1}}-2\zeta_{{5}},\zeta_{{3}}-\zeta_{{5}},
-2\zeta_{{5}}+\zeta_{{6}} &>0& \cr -\zeta
_{{5}}+\zeta_{{7}},\zeta_{{5}} &>0&\cr}}
&\footnotesize{\pmatrix{y_{{6}}&y_{{5}}&y_{{7}}&y_{{3}}&y_{{2}}\cr 
1&0&-2&0&1\cr 0&1&-2&1&0\cr}}
\cr \noalign{\hrule} \vrule &$\matrix{&
\hbox{cone over }\cr & dP_1}$&
\footnotesize{\matrix{\zeta_{{2}}-3\,\zeta_{{5}}+\zeta_{{7}},\zeta_{{3}}-2\,
\zeta_{{5}}+\zeta_{{7}} &=0&\cr 
\zeta_{{1}}-4\,\zeta_{{5}}+2\,\zeta_{{7}},\zeta_{
{4}}-2\,\zeta_{{5}} &>0&
\cr -\zeta_{{5}}+\zeta_{{6}}-\zeta_{{7}}&>0&
\cr \zeta_{{5}}-\zeta_{{7}},2\,\zeta_{{5}}-\zeta_{{7}}&>0&\cr }}&
\footnotesize{\pmatrix{y_{{6}}&y_{{4}}&y_{{7}}&y_{{3}}&y_{{1}}\cr
1&0&-3&1&1\cr 0&1&-2&0&1\cr}}
\cr \noalign{\hrule} \vrule &$\matrix{&\hbox{cone over }\cr & dP_2}$ 
&\hskip 0.97in \footnotesize{\matrix{\zeta_{{2}},\zeta_{{4}},\zeta_{{5}}&=0& \cr\cr
\zeta_{{1}},\zeta_{{3}},\zeta_{{6}},\zeta_{{7}} &>0& \cr }}&
\footnotesize{\pmatrix{y_{{6}}&y_{{5}}&y_{{7}}&y_{{3}}&y_{{2}}&
x_{{2}}\cr 1&0&0&-2&-1&2\cr 0&1&0&-1&-
2&2\cr 0&0&1&-1&-1&1\cr}}\cr \noalign{\hrule} \vrule &$\matrix{& \hbox{cone
over }\cr & dP_3}$ &\hskip 0.14in\footnotesize{\matrix{\zeta_{{2}}-2\,\zeta_{{7}},
\zeta_{{3}}-\zeta_{{7}},\zeta_{{4}}-2\,\zeta_{{7}}&=0&\cr
\zeta_{{5}}-\zeta_{{7}}&=0&\cr \zeta_{{1}}-2\,\zeta_{{7}},\zeta_{{6}}-2\zeta_7,
\zeta_7\ &>0&\cr}}&\footnotesize{\pmatrix{y_{{6}}&y_{{4}}&y_{{5}}&y_{{7}}&y_{{3}}
&y_{{1}}&y_{{2}}\cr 1&0&0&0&-2&-2&3
\cr 0&1&0&0&-2&-1&2\cr 0&0&1&0&-1&-2&2
\cr 0&0&0&1&-1&-1&1\cr}}
\cr \noalign{\hrule} }}$$ \bigskip
\hskip 0.8 in Table 1. {\footnotesize Charge matrices for partial 
resolutions of the $\Bbb{Z}_3\times\Bbb{Z}_3$ orbifold.}  
 
\bigskip

Being affine varieties, the complex cones over the del Pezzo surfaces of 
interest can be presented as the affine spectrum $Spec(R)$ of their 
coordinate ring
$R$. This ring admits a presentation $R=\C[z]/I$ with $\C[z]$ a polynomial 
ring and $I$ an ideal of relations. In the context of toric geometry, 
the generators
$z_i$ are the invariant coordinates under the complex torus 
action in the holomorphic quotient description, while $I$ is the ideal of 
monomial relations constraining them. It is well-known \cite{Oda, Fulton} that 
the invariants $z$ and monomial relations can be determined 
as follows. If $C$ is the cone over the polygon $P$ describing our variety 
(where the polygon is embedded in the affine plane $x+y+z=1$ of $\R^3$), 
then one considers the dual cone $C^{\rm v}$, which is the cone 
over the dual (polar) polygon $P^{\rm v}$. While the cone $C$ and the polygon $P$ 
are appropriate for the holomorphic quotient description of the toric variety, 
their duals $C^{\rm v},P^{\rm v}$ are appropriate for the description in terms of 
invariants $z$ and monomial relations $I$. More precisely, there will be 
one invariant $z_i$ associated to each integral point of the dual polygon 
$P^{\rm v}$. Moreover, any integral linear relation between the primitive 
integral vectors lying in the cone $C^{\rm v}$ corresponds to a monomial relation 
among the variables $z$. Explicitly, a linear relation of the form 
$\sum_{i}{a_i~u_i}$ between the vectors 
$u_i$ associated to the invariant coordinates $z_i$ yields the monomial 
relation $\Pi_{i}{z_i^{a_i}}=1$. 
Hence the entire information 
characterizing the ring $R$ is contained in the dual polygons, which
are listed in Figure 2.  

\

$$\matrix{{\epsfxsize=1in\epsfbox{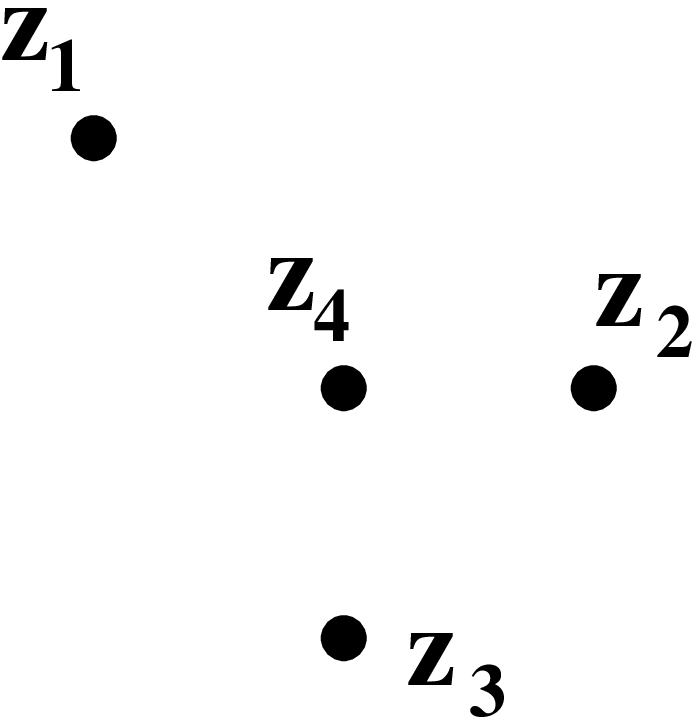}}&
{\epsfxsize=0.9in\epsfbox{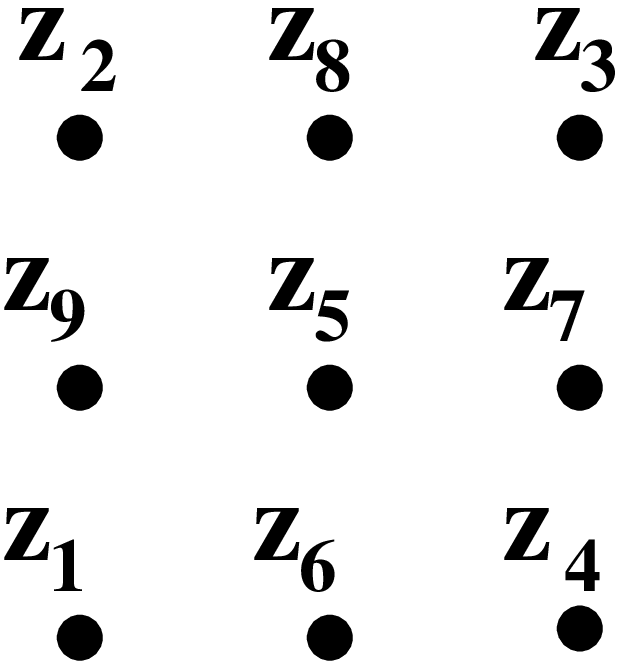}}
\cr\cr 
\hbox{Figure 2(a). {\footnotesize $\Bbb{Z}_3\times\Bbb{Z}_3$ orbifold.}}
&  
\hbox{Figure 2(b). {\footnotesize Cone over $F_0$.}}
\cr\cr\cr
{\epsfxsize=1in\epsfbox{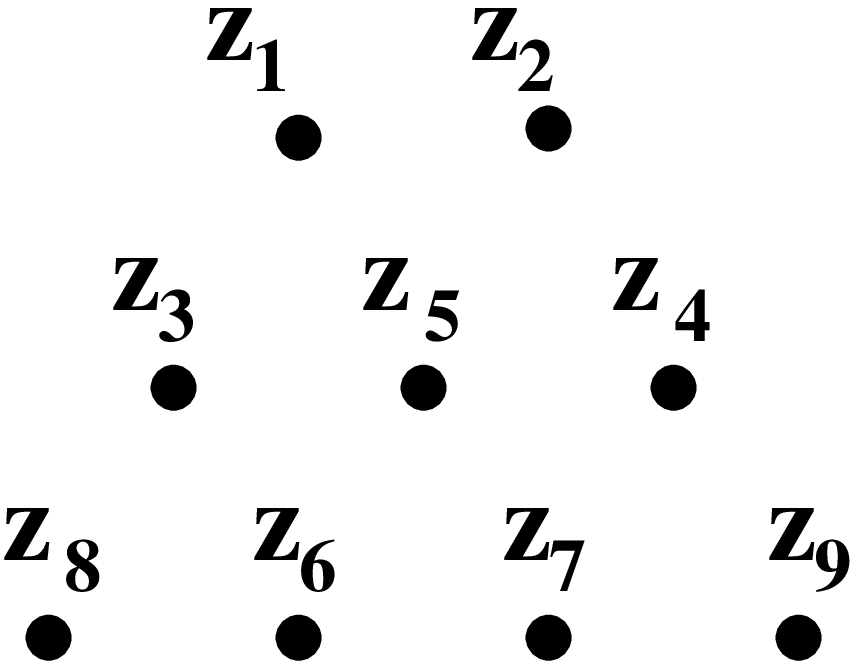}}&
{\epsfxsize=0.7in\epsfbox{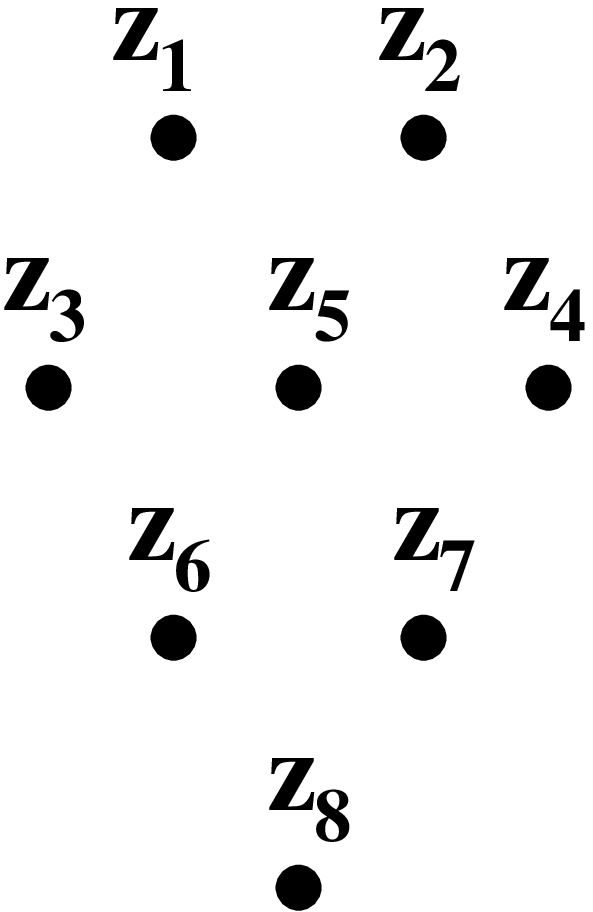}}
\cr\cr 
\hbox{Figure 2(c). {\footnotesize $\matrix{\hbox{Cone over }
\cr dP_1 }$}}&  
\hbox{Figure 2(d). {\footnotesize $\matrix{\hbox{Cone over}\cr dP_2}$}}
\cr\cr\cr
{\epsfxsize=0.9in\epsfbox{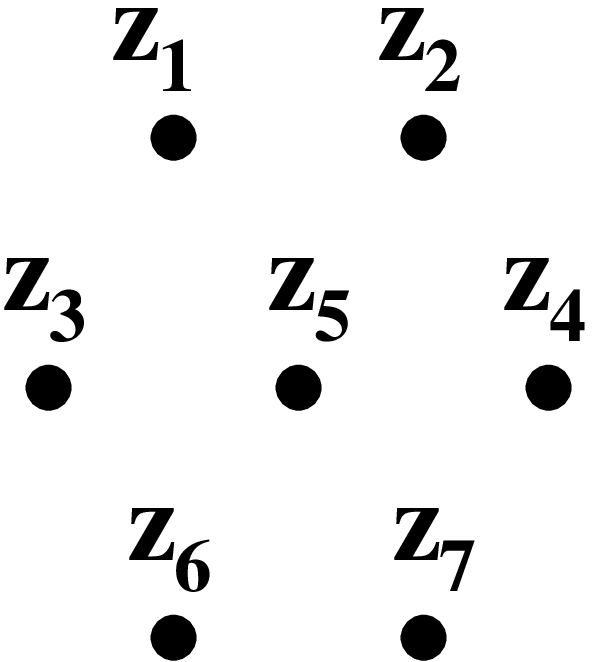}}\cr\cr 
\hbox{Figure 2(e). {\footnotesize $\matrix{\hbox{Cone over}\cr dP_3}$}}
\cr\cr}
$$ 

\bigskip 

\

\centerline{Figure 2. {\footnotesize Geometric picture of the 
generators of $R$ and the relations between them.}}

\subsection{Partial resolutions of the moduli space}

The moduli space for one D-brane in the presence of Fayet-Iliopoulos terms 
can be analyzed by the methods pioneered in \cite{branes3}. 
In the case of $\C^3/\Z_3\times \Z_3$, the main points of the analysis 
are as follows.
The classical moduli space in the presence of
Fayet-Iliopoulos terms is obtained by imposing the D- and F-flatness
constraints and further dividing by the gauge group of the worldvolume 
theory, which in this case turns out to be the torus $U(1)^8$. If we
denote the solution set of the F-flatness constraints by ${\cal Z}$,
then the symplectic quotient which gives the moduli space (where the
D-flatness constraints play the role of moment map equations) is
isomorphic, as a complex variety, with the holomorphic quotient of
${\cal Z}$ by the complexification $(\C^*)^8$ of the gauge group. For
one D-brane on an abelian quotient singularity, the set ${\cal Z}$
(which in the case $\C^3/\Z_3\times \Z_3$ is complex eleven-dimensional, 
see Section 3) has the structure of
an algebraic variety defined by a collection of monomial relations. 
Therefore, ${\cal Z}$ is an affine toric variety, and by virtue of
\cite{Cox} it admits an alternate presentation as a holomorphic
quotient: 
\be 
{\cal Z}=\C^{42}/(\C^*)^{31}~~. 
\ee 
The procedure for precisely identifying the quotient above 
(the mysterious numbers $42, 31$ and the precise $(\C^*)^{31}$ action) is
slightly technical and is explained in detail in section 2. Then
the moduli space can be obtained in the form: 
\be
\label{double_holo} 
{\cal M}_\xi=({\cal Z}-S_\xi)/(\C^*)^{8}=(\C^{42} - Z_\xi)
/(\C^*)^{39}~~, 
\ee where
$S_\xi, Z_\xi$ are certain `exceptional sets' which have complex
codimension at least two in the respective spaces. 
Note that $Z_\xi$ depends on the choice of Fayet-Iliopoulos
parameters $\xi$, which explains the dependence of the complex geometry
of the moduli space on these parameters. In symplectic quotient language, 
the symplectic reduction of ${\cal Z}$ by $U(1)^8$ is performed at a level 
characterized by the Fayet-Iliopoulos parameters $\xi\in \R^8$. 
When re-writing 
${\cal Z}$ itself as a symplectic quotient (with {\em zero} moment map level
$0\in \R^{31}$)
we are rewriting ${\cal M}_\xi$ as a double symplectic quotient. To arrive 
at a presentation equivalent to the last entry in (\ref{double_holo}), we
then rewrite this double quotient as a single symplectic quotient, with 
certain `overall' moment map levels $\eta(\xi) \in \R^{39}$. The crucial 
point of this analysis is that the levels $\eta$ are not arbitrary in 
$\R^{39}$, but are determined by Fayet-Iliopoulos terms $\xi\in \R^8$. 
Therefore, as $\xi$ covers the whole of $\R^8$, the parameters $\eta(\xi)$
will cover only a subspace of $\R^{39}$. As shown in Sections 2 and 3, the 
precise relation between $\eta$ and $\xi$ is given by an injective linear 
map:
\be
\R^8(\xi)\stackrel{w}{\longrightarrow} \R^{39}(\eta)~~,
\ee
such that $\eta(\xi)=w \xi$. Therefore, $\eta(\xi)$ will only vary 
inside of an $8$-dimensional subspace $W$ of $\R^{39}$. This non-genericity 
of the level $\eta$ is responsible for the fact that the symplectic 
quotient description 
${\cal M}_\xi=\{x \in \C^{42}|\mu(x)=\eta(\xi)\}/U(1)^{39}$
is not minimal, in the sense that, depending on the sign of the various 
components of $\eta(\xi)$, it will in general be possible to eliminate 
many of the homogeneous variables of $\C^{42}$ essentially 
by the same procedure as that discussed in subsection 1.1. 
In the holomorphic 
quotient language, this is reflected by the fact that the toric generators 
one obtains for the description $(\C^{42} - Z_\xi)/(\C^*)^{39}$ appear with 
multiplicities, so the toric description is non-minimal as well. 
In order to obtain a
minimal description, one must eliminate all multiplicities in the list 
of toric generators or, equivalently, eliminate the redundant homogeneous 
coordinates in the manner outlined above. 
In our case, this 
reduces the full quotient to the form $(\C^{10}-F_\xi)/(\C^*)^7$, which is 
precisely what we would expect from the geometric description of the partial 
resolutions. This reduced quotient 
has a symplectic quotient description as well, and we let 
$\zeta(\xi)\in \R^7$ denote the associated moment map levels.

As in the previous subsection, 
which homogeneous coordinates can be eliminated depends on the sign of the 
various components of $\eta(\xi)$. A systematic analysis shows that this 
procedure defines a certain fan $\Sigma$ 
in $\R^{39}(\eta)$. Since $w$ is an injective linear function, 
it follows that there exists 
a similar division of $\R^8(\xi)$ into cones $\Xi_I$ obtained by taking the 
inverse image of the cones $\Sigma$ through the map 
$w$ (geometrically, 
this amounts to intersecting the system of cones $\Sigma$ of $\R^{39}$ with 
the subspace $W$). Which homogeneous coordinates of the full quotient can be 
eliminated, and hence the minimal description of ${\cal M}_\xi$ 
as a toric variety will thus depend on where $\xi$ lies  in $\R^8$ with 
respect to the cones $\Xi$. Note that not all of the cones $\Xi_I$ will be 
$8$-dimensional. In fact, it is not hard to see that those cones $\Xi_I$
which have dimension less than $8$ will form faces of the $8$-dimensional ones.

As we discuss in Sections 2 and 3, the reduction  
of the quotient $(\C^{42} - Z_\xi)/(\C^*)^{39}$ to the quotient 
$(\C^{10}-F_\xi)/(\C^*)^7$ induces a piecewise-linear map
\be
\R^{39}(\eta)\stackrel{\pi}{\longrightarrow}\R^7(\zeta)~~,
\ee 
such that $\zeta=\pi(\eta)$. Therefore, the dependence of $\zeta$ on $\xi$ 
is given by the composite map:
\be
\phi:=\pi\circ w:\R^8(\xi)\rightarrow \R^7(\zeta)~~,
\ee 
so that $\zeta=\phi(\xi)$. Clearly $\phi$ is a piecewise-linear map, 
which is linear on each of the cones $\Xi_I$. Note that $\phi$ 
need not have 
different linear expressions on each of the cones $\Xi_I$
\footnote{In technical 
terms, the fan associated to $\phi$ as a piecewise-linear function is 
subordinate to the fan defined by the cones $\Xi_I$, i.e. this later fan 
is a refinement of the fan of $\phi$.}. 

The discussion above shows that understanding how to realize the 
various partial resolutions of 
$\C^3/\Z_3\times \Z_3$ in terms of the worldvolume theory of the D3-brane 
amounts to computing the map $\phi$. Indeed, this map connects 
the Fayet -Iliopoulos parameters $\xi$ of the D-brane worldvolume theory to
the toric levels $\zeta$ discussed in section 2. Therefore, given a choice of 
partial resolution (described in the language of section 2 by a choice of 
$\zeta$), the map $\phi$ allows us to determine if and how this geometric 
choice is physically realized in the D-brane theory.

Clearly $\phi$ must take 
the fan $\Xi$ into a refinement of the natural fan $\Psi$ carried by 
$\R^7(\zeta)$, which 
was discussed in the previous subsection. Hence the fan $\Xi$ is a refinement 
of the preimage of $\Psi$ through $\phi$.  
The full information about the D-brane realization of 
the various partial resolutions is encoded in this preimage. By using our 
methods, one can in principle reconstruct the preimage from the 
smaller pieces provided by the cones $\Xi_I$.

A systematic procedure for determining the map $\phi$ is developed in 
the next section. In the case of $\C^3/\Z_3\times \Z_3$, the complexity of the 
computation turns out to be markedly greater than in cases considered before 
\cite{branes3, branes3'}. As we explain in Section 2, the complexity is 
characterized by an integer $c$\footnote{$c$ is the number of facets of the 
cone of exponents associated to the F-flatness constraints, as discussed in 
subsection 2.7.}, which depends on the 
combinatorial data of the problem, and for which we do not have an analytic 
expression.
In a typical example considered 
in \cite{branes3}, such as a $\Z_2\times \Z_2$ quotient singularity, 
one had $c=9$, while our case case we have $c=42$.

The result of the analysis is that each of the complex cones over 
$F_0,dP_1,dP_2$ and $dP_3$ is indeed realized in the moduli space of the 
worldvolume theory. The details of how this conclusion can be reached are 
discussed in the next sections.

\section{The classical moduli space for one D-brane transverse to 
an abelian Calabi-Yau quotient singularity}

In this section, we spell out in detail our algorithm for the analysis of the 
moduli space of one D-brane on an abelian quotient singularity. 

The algorithm presented here is essentially a systematic version of the 
approach taken in \cite{branes3} and is carried out in the language of 
the quiver formalism \cite{Infirri} 
(see \cite{nonabelian} for a discussion in the context of D-brane moduli 
spaces), which turns out to be a very effective way of implementing the 
projection conditions and analyzing the D- and F- flatness constraints. 
The main novel result is the construction of the map $\phi$ in subsection 
2.8. As a simple illustration, we re-consider the realization of the conifold 
in the moduli space of a D-brane transverse to a $\Z_2\times \Z_2$ 
Calabi-Yau quotient singularity \cite{branes3'}.

\subsection{The quotient group and its representation on $\C^3$}

Consider a finite abelian group $\Gamma$. Choosing a system of generators 
allows us to present the group as:
\be
\Gamma=\Z_{t_1}\times ...\times \Z_{t_d}~~,
\ee
where the integers $t_1..t_d\geq 2$ are the torsion indices of $\Gamma$. 
The elements of the group are then written in the form:
\be
u=(u_1..u_d)~~,
\ee
with $u_a \in Z_{t_a}~(a=1..d)$, while the group operation becomes:
\be
u+v=(~(u_1+v_1) {\rm mod}~t_1 , ... ,(u_d+v_d) {\rm mod}~t_d~)~~.
\ee

The irreducible representations of our group  are given by its 
characters $\chi \in Hom(\Gamma,U(1))$, which are in one to one 
correspondence with the elements of $\Gamma$. Indeed, every character is of 
the form:
\be
\chi_w(u)=\Pi_{a=1..d}{e^{\frac{2\pi i}{t_a}w_a}}~~ (u \in \Gamma),
\ee
where $w=(w_1...w_d)$ is an element of $\Gamma$ called the {\em weight} 
associated to $\chi$. In fact, this correspondence gives an isomorphism 
between $\Gamma$ and the multiplicative group of its characters:
\bea
\chi_{w+w'}=\chi_{w}\chi_{w'} \nn\\
\chi_{w}^{-1}=\chi_{-w}~~.
\eea  

We let $\Gamma$ act on $\C^3$ by a special unitary representation $\theta$. 
Since this representation decomposes into 3 irreducibles, we can always 
find coordinates $(x^1,x^2,x^3)$ on $\C^3$ such that $\theta$ takes the form:
\be
\label{base_action}
(x^1, x^2, x^3)\stackrel{(u \in \Gamma)}{\longrightarrow} \theta(u)(x)=
(\chi_{w_1}(u)x^1, \chi_{w_2}(u)x^2, \chi_{w_3}(u)x^3)~~.
\ee
Special unitarity requires that $w_1+w_2+w_3=0$ (a relation which 
has to be understood as holding in the group $\Gamma$). 

\

{\bf Example}: The $\Z_2\times \Z_2$ quotient singularity

\

The group $\Gamma=\Z_2\times \Z_2$ has $d=2$ torsion indices $t_1=2$, $t_2=2$. 
Consider its action on $\C^3$ given by the weights $w_1=(1,1)$, $w_2=(1,0)$, 
$w_3=(0,1)$ (which do satisfy $w_1+w_2+w_3=(0,0)$ in our group) 
and associated characters:
\bea
\chi_1(u_1,u_2)=e^{i\pi(u_1+u_2)}=(-1)^{(u_1+u_2)}\nn\\
\chi_2(u_1,u_2)=e^{i\pi u_1}=(-1)^{u_1}\nn\\
\chi_3(u_1,u_2)=e^{i\pi u_2}=(-1)^{u_2}~~,
\eea
with $u=(u_1,u_2)$, $u_1,u_2=0,1$. The generators $(1,0)$, $(0,1)$ of the 
group act as:
\bea
(x_1,x_2,x_3)\stackrel{(1,0)}{\longrightarrow}(-x_1,-x_2,x_3)\nn\\
(x_1,x_2,x_3)\stackrel{(0,1)}{\longrightarrow}(-x_1,x_2,-x_3)~~.
\eea

\subsection{Solving the projection constraints}

The low-energy theory on the worldvolume of a D-brane transverse to a
quotient singularity is given by a projection of a supersymmetric
gauge theory with 16 supercharges\cite{douglas-moore}.  The gauge
group is determined by the singularity.  The projection will be
described in more detail below.  It results in an Abelian gauge theory
with a reduced supersymmetry and, when appropriate, a superpotential.
The classical moduli space of supersymmetric vacua of this theory is
found to be the tangent cone to the transverse space; this is
reasonable, since motion of the brane along this space is a
supersymmetry-preserving deformation, while the low-energy limit
restricts us to small motions.  The gauge theory admits
Fayet-Iliopoulos terms; these are expected to parameterize the
deformations of the singularity itself.  How this works is the subject
of this paper, and will be made clear in what follows.  We note here
that this picture is a bit imprecise.  As noted in
\cite{douglas-moore,MP} the Abelian gauge symmetry is in fact broken
by twisted closed-string modes.  As described in detail in \cite{MP},
this fact may safely be ignored in our discussion below.

Let $|\Gamma|$ denote  the number of elements of our group.  The
low-energy theory for a D3-brane near this singularity is found as a
projection of the theory of $|\Gamma|$ branes moving on the covering
space.  This is an ${\cal N}=4$ theory with $U(|\Gamma|)$ gauge
symmetry, containing (in ${\cal N}=1$ language) a vector multiplet and
three chiral multiplets in the adjoint representation of the gauge
group.  As is by now familiar, the three chiral multiplets represent a
nonabelian version of the positions of the $|\Gamma|$ branes in the
three complex dimensional transverse space.  The quotient theory is
obtained by a projection, restricting attention to fields invariant
under the action of $\Gamma$, which we lift to the Chan-Paton indices
via its regular representation.

We label the entries of the adjoint fields by
\be
X^i=(X^i_{v'v})_{v',v\in \Gamma}~~.
\ee
The spacetime indices of the
matrices $X^i$ transform
in the representation (\ref{base_action}) and the chiral fields
surviving the projection are thus those modes which satisfy
\be
\label{proj1}
R(u)X^iR(u)^{-1}=\chi_{-w_i}(u)X^i~~,
\ee
where $R(u)$ is the regular representation of $\Gamma$. Since $\Gamma$ is 
abelian, $R$ decomposes into the sum of all irreducible 
representations, each taken with multiplicity one. Therefore, we can always 
assume (after performing a unitary transformation $X^i\rightarrow UX^iU^{-1}$)
that $R(u)$ are given by the diagonal matrices:
\be
R(u)_{v',v}=\chi_v(u)\delta_{v',v}~~(v',v\in \Gamma)~.
\ee 
Then the projection conditions take the form:
\be
\chi_{v'-v+w_i}(u)X^i_{v',v}=X^i_{v',v}~~,
\ee
which require that all elements of $X^i$ must be zero except for the 
following entries:
\be
\label{surviving}
x^i(v):=X^i_{v-w_i,v}~~.
\ee

The set of surviving fields $x^i(v)$ can be elegantly 
described in the language of graph theory\cite{Infirri}. 
For this, consider a set of points (nodes) which are in one to one 
correspondence with the elements of $\Gamma$. For each node $v\in \Gamma$, 
and for each $i=1..3$, 
draw an edge from $v$ to $v-w_i$. Such an edge is associated with 
a surviving component $x^i(v)$ of the matrix $X^i$ and will be called 
an edge {\em of~type~i}. 
If $w_i= 0$, then the edge connects the node $v$ with 
itself, and is thus a loop, which cannot carry an orientation. 
If $w_i \neq 0$, the edge is given the orientation from $v$ to $v-w_i$ 
(such oriented edges will be called {\em arrows}). The graph ${\cal Q}$ 
thus obtained is called the {\em McKay quiver} of the quotient singularity 
$\C^3/\Gamma$. We denote the set of its nodes by ${\cal Q}_0\approx \Gamma$ 
and the set of its edges by ${\cal Q}_1$. Since we have 3 types of edges 
leaving each vertex, the total number of edges in the graph is equal to 
$3|\Gamma|$. The McKay quiver can be thought of as the superposition of 
3 graphs ${\cal Q}^i$~($i=1..3$), where ${\cal Q}^i$ is obtained from 
${\cal Q}$ by keeping only the edges of type $i$. The edges of ${\cal Q}^i$
represent the surviving components of the matrix $X^i$. In the absence of the 
projection conditions, ${\cal Q}^i$ would coincide with the full graph on the 
set of nodes $\Gamma$, i.e. it would contain an edge connecting any two 
elements of $\Gamma$ (in particular, it would contain a loop at each vertex). 
The projection conditions eliminate some of these edges, in a manner controlled
by the weight $w_i$. The surviving components of $X^i$ can be read from the 
quiver as follows. If $a \in {\cal Q}_1$ is an edge of ${\cal Q}$, then the 
associated field is given by:
\be
\label{edge_var}
x(a):=x^{type(a)}(tail(a))=X^{type(a)}_{head(a),tail(a)}~~,
\ee
where $tail(a), head(a)\in \Gamma$ are the tail, respectively head of the 
edge $a$ (in case $a$ is a loop at a node $v$, we define $tail(a)=head(a)$ to 
be given by $v$; otherwise, $a$ is an arrow going from $tail(a)$ to 
$head(a)$). Throughout this paper, we will represent arrows of type 1 by 
light grey lines, arrows of type 2 by dark grey lines and arrows of type 
3 by black lines. In a color postscript rendering, these arrows appear 
respectively as green, blue and red. 

\

{\bf Example}:$\Z_2\times \Z_2$ singularity

\

In this case, we have $|\Gamma|=4$ so the quiver will have $4$ nodes. 

The edges $v\rightarrow v-w_i$ are:

(1)type $1$ (the surviving components of $X^1$):
\be
\begin{array}{cccc}
(0,0)\rightarrow (1,1),&~
(0,1)\rightarrow (1,0),&~
(1,0)\rightarrow (0,1),&~
(1,1)\rightarrow (0,0)~;
\end{array}
\ee

(2)type $2$ (the surviving components of $X^2$):
\be
\begin{array}{cccc}
(0,0)\rightarrow (1,0),&~
(0,1)\rightarrow (1,1),&~
(1,0)\rightarrow (0,0),&~
(1,1)\rightarrow (0,1)~;
\end{array}
\ee

(3)type $3$ (the surviving components of $X^3$):
\be
\begin{array}{cccc}
(0,0)\rightarrow (0,1),&~
(0,1)\rightarrow (0,0),&~
(1,0)\rightarrow (1,1),&~
(1,1)\rightarrow (1,0)~~.
\end{array}
\ee

Indexing the group elements as follows:
\bea
(0,0)\leftrightarrow 1\nn\\ 
(1,0)\leftrightarrow 2\nn\\ 
(0,1)\leftrightarrow 3\nn\\ 
(1,1)\leftrightarrow 4\nn 
\eea
we can rewrite the edges as:

(1) type 1:
\be
\begin{array}{cccc}
1\rightarrow 4,&~
3\rightarrow 2,&~
2\rightarrow 3,&~
4\rightarrow 1 ~;
\end{array}
\ee

(2)type 2:
\be
\begin{array}{cccc}
1\rightarrow 2,&~
3\rightarrow 4,&~
2\rightarrow 1,&~
4\rightarrow 3~;
\end{array}
\ee

(3)type 3:
\be
\begin{array}{cccc}
1\rightarrow 3,&~
3\rightarrow 1,&~
2\rightarrow 4,&~
4\rightarrow 2  
\end{array}~.
\ee

The quiver is drawn below:

\iffigs
\vskip 1 in
\hskip 1.3in\resizebox{5cm}{!}{\includegraphics[0in,0in][2in,3in]{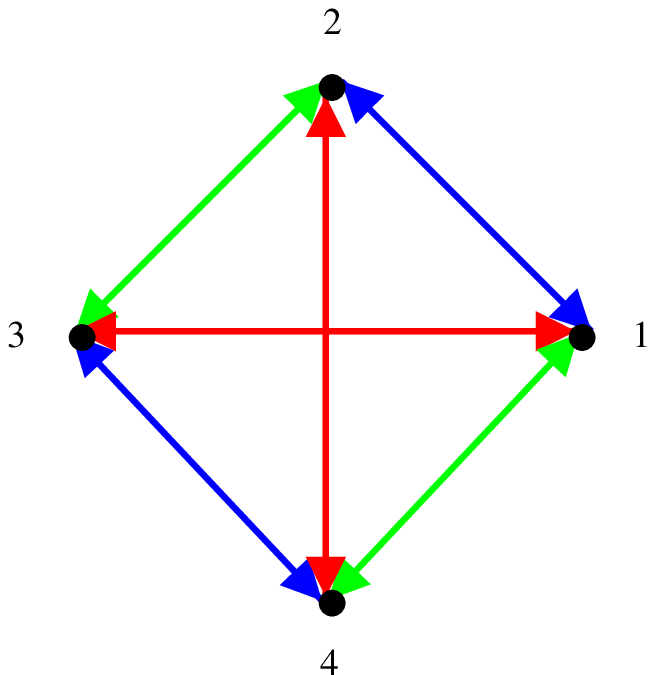}}\begin{center}
Figure 3. {\footnotesize The quiver for the isolated $\Z_2\times \Z_2$
quotient singularity.}

{\footnotesize The arrows of various types are drawn with different 
shades of gray: light gray for type 1 (fields $X_{uv}$), dark for type 2
(fields $Y_{uv}$), medium for type 3 (fields $Z_{uv}$). 
In color rendering, these correspond to 3 different colors:
green for type 1, blue for type 2, red for type 3.}

\end{center}
\vskip 0.5 in
\fi

\subsection{The action of the surviving gauge group}
The gauge fields in the theory do not carry transverse space indices,
so the projection conditions on these are:
\be
\label{proj2}
R(u)UR^{-1}(u)=U~~.
\ee
If we index the entries of $U$ by the elements of $\Gamma$:
\be
U=(U_{v',v})_{v',v\in \Gamma}~~,
\ee
then (\ref{proj2}) becomes: 
\be
\chi_{v'-v}(u)U_{v',v}=U_{v',v}~~,
\ee
which shows that the surviving entries of $U$ are given by:
\be
U_v=U_{v,v} \in U(1)~~.
\ee
Therefore, the projected gauge group is given by diagonal unitary matrices 
and is isomorphic with $\Pi_{v \in \Gamma}U(1)\approx U(1)^\Gamma$. 
Its action on the surviving fields $x^i(u)$ is:
\be
\label{gauge}
x^i(u)\stackrel{U=(U_v)_{v \in \Gamma}\in U(1)^\Gamma}{\longrightarrow} 
U_{u-w_i}U_u^{-1}x^i(u)=e^{i(\phi_{u-w_i}-\phi_u)}x^i(u)
\ee
where we wrote $U_v=e^{i\phi_v}$ with $\phi_v\in \R$. Therefore, the 
charge of $x^i(u)$ with respect to the $v$-th $U(1)$ factor is:
\be
\label{charges}
q^i_v(u)=\delta_{v,u-w_i}-\delta_{v,u}~~.
\ee 
Note that the diagonal 
subgroup $U(1)_{diag}$ 
given by $U_v=e^{i\phi}$ with $\phi_v=\phi$ independent of 
$v$ acts trivially on all surviving fields, so the effective 
gauge group is given by $G=U(1)^\Gamma/U(1)_{diag}$.

The action (\ref{gauge}) can be translated into quiver language as follows. 
If $a \in {\cal Q}_1$ is an edge of the quiver, then the field $x(a)$ 
given by (\ref{edge_var}) transforms as:
\be
\label{edge_gauge}
x(a)\rightarrow e^{i(head(a)-tail(a))\phi}x(a)~~.
\ee
One can imagine the various $U(1)$ factors of the surviving gauge group to be 
sitting at the nodes of the quiver, having the natural action 
(\ref{edge_gauge}) on the variables $x(a)$ associated to its edges. 
Note that each edge is charged with respect to the $U(1)$ factors associated 
to its terminal points (its tail and its head). The action of $U(1)^\Gamma$ on 
$x(a)$ is encoded by the $|\Gamma|\times 3|\Gamma|$  matrix of charges 
$d=(d_{v,a})_{v\in \Gamma,a\in {\cal Q}_1}$ with entries:
\be
d_{v,a}=\delta_{v,head(a)}-\delta_{v,tail(a)}~~.
\ee 
(This is just another way of writing equation (\ref{charges})). 
In graph-theoretic language, $d$ is the 
{\em incidence matrix} 
of the quiver, where the incidence index of an arrow $a$ on a node $v$ is 
$-1$ if $v=tail(a)$, $+1$ if $v=head(a)$ and zero otherwise, while the 
incidence index of a loop with any node (including the node where the loop 
sits) is defined to be zero.

\subsection{The D-flatness constraints}

The moment map $\mu:\C^{{\cal Q}_1}\rightarrow \R^\Gamma$ for the action of 
$U(1)^\Gamma$ on the space $\C^{{\cal Q}_1}$ of all surviving fields is 
given by:
\be
\mu_v(x)=\sum_{\scriptsize \begin{array}{c}i=1..3\\u\in \Gamma\end{array}}
{q^i_v(u) |x^i(u)|^2}
\ee
which in quiver language becomes:
\be
\mu_v(x)=\sum_{a\in {\cal Q}_1}{d_{v,a}|x(a)|^2}=\sum_{a \in {\cal Q}_1, 
head(a)=v}{|x(a)|^2}-\sum_{a \in {\cal Q}_1,tail(a)=v}{|x(a)|^2}~~,
\ee
or, in matrix form:
\be
\mu=d~p~~,
\ee
where $p$ is the vector in $\R_+^{{\cal Q}_1}$ with components:
\be
p_a:=|x(a)|^2~~.
\ee

The D-flatness constraints in the presence of Fayet-Iliopoulos terms 
$(\xi_v)~(v \in \Gamma)$ read:
\be
\label{moment_eq0}
\mu_v(x)=\xi_v ~~\mbox{for~all~}v \in \Gamma~.
\ee
Note that the moment map satisfies the condition:
\be
\sum_{v \in \Gamma}{\mu_v(x)}=0~~,
\ee
which is a consequence of the trivial action of $U(1)_{diag}$. 
This is also reflected in the structure of the incidence 
matrix $d$, by the fact that the sum of all of its rows is zero. 
Consistency requires 
that the Fayet-Iliopoulos terms also satisfy:
\be
\sum_{v\in \Gamma}{\xi_v}=0~~.
\ee
Therefore, we can always express $\xi_0$ (where $0$ is the identity element of 
$\Gamma$) as :
\be
\xi_0=-\sum_{v \in \Gamma-\{0\}}{\xi_v}~~.
\ee
It follows that the space of allowed Fayet-Iliopoulos parameters is in fact 
only $(|\Gamma|-1)$-dimensional, and will be denoted by 
$\R^{|\Gamma|-1}(\xi)$, 
where the vector $\xi$ is given by $\xi=(\xi_v)_{v \in \Gamma-\{0\}}$. 
Similarly, the first equation of the system 
(\ref{moment_eq0}) (the one corresponding to $v=0$) 
is a consequence of the other $|\Gamma|-1$ equations. Eliminating it allows us 
to rewrite the moment map conditions in the form:
\be
\Delta p=\xi~~,
\ee
where $\Delta$ is the $(|\Gamma|-1)\times 3|\Gamma|$ matrix obtained by 
deleting the first row of $d$. 

\

{\bf Example}:$\Z_2\times \Z_2$ singularity

\

With the above enumeration of the group elements, the incidence matrix is:
\be
d=\left [\begin {array}{cccccccccccc} -1&0&0&1&-1&1&0&0&-1&0&1&0
\\\noalign{\medskip}0&-1&1&0&1&-1&0&0&0&-1&0&1\\\noalign{\medskip}0&1&
-1&0&0&0&-1&1&1&0&-1&0\\\noalign{\medskip}1&0&0&-1&0&0&1&-1&0&1&0&-1
\end {array}\right ]~~,                                 
\ee
while the matrix $\Delta$ is given by (the first row of $d$ 
corresponds to the neutral element of the group with our choice of 
enumeration):
\be
\Delta=\left [\begin {array}{cccccccccccc} 0&-1&1&0&1&-1&0&0&0&-1&0&1
\\\noalign{\medskip}0&1&-1&0&0&0&-1&1&1&0&-1&0\\\noalign{\medskip}1&0&0
&-1&0&0&1&-1&0&1&0&-1\end {array}\right ]~~.
\ee

\subsection{The F-flatness constraints}

The ${\cal N}=4$ theory includes a superpotential:
\be
\label{superpot}
{\cal W}=\epsilon_{ijk}Tr(X^iX^jX^k)~~.
\ee
This can be expressed in terms of the surviving fields as
\footnote{Note that the trace in (\ref{superpot}) closes on the projected 
fields precisely due to the special unitarity condition 
$w_1+w_2+w_3=0$.}:
\be
\label{superpot_projected}
{\cal W}=-\epsilon_{ijk}\sum_{v \in \Gamma}{x^k(v-w_i-w_j)x^j(v-w_i)x^i(v)}~~.
\ee

Supersymmetric vacua of the ${\cal N}=4$ theory satisfy the F-flatness
conditions: 
\be
[X^i,X^j]=0~~ ( 1\leq i < j \leq 3 ).
\ee
Taking (\ref{surviving}) into account these reduce to:
\be
\label{Fflat_projected}
x^j(v-w_i)x^i(v)-x^i(v-w_j)x^j(v) =0~~,\mbox{~for~all~} 1\leq i < j \leq 3~
\mbox{and}~v \in \Gamma~~,  
\ee
in agreement with the condition that (\ref{superpot_projected}) is
stationary. 

\iffigs
\vskip 1 in
\hskip 2in\resizebox{4cm}{!}{\includegraphics[0in,0in][2.5in,2in]{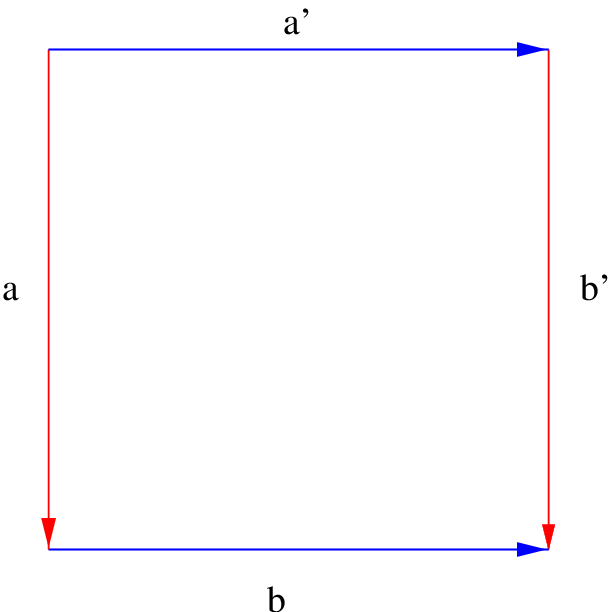}}
\vskip 0.1in
\begin{center}
Figure 4. {\footnotesize Pictorial description of the F-flatness constraints.

The arrows on opposite sides have the same types.}
\end{center}
\vskip 0.5 in
\fi

The constraints (\ref{Fflat_projected}) can be viewed as monomial 
relations:
\be
\label{mon_rels}
x^j(v-w_i)x^i(v)[x^i(v-w_j)]^{-1}[x^j(v)]^{-1}=1
\ee
among the surviving fields. The solution of these equations 
defines a complex subvariety ${\cal Z}$ of the space $\C^{{\cal Q}_1}$ of all 
variables $\{x^i(v)\}$, which we call 
{\em the variety of commuting matrices}. This algebraic variety can be 
described by the set of all its regular functions, which are obtained by 
restricting the regular functions defined on $\C^{{\cal Q}_1}$ (which are 
just the polynomials in the variables $\{x^i(v)\}$) to the subset ${\cal Z}$. 
The restriction of such a polynomial is achieved by simplifying each of its 
monomials with respect to the relations (\ref{mon_rels}). If 
$x=(x^i(v))_{i=1..3,v\in \Gamma}$ denotes the general point of 
$\C^{{\cal Q}_1}$, then a monomial of the polynomial ring $\C[x]$ has the 
form $\Pi_{i,v}{[x^i(v)]^{s^i(v)}}$, with $s^i(v)$ some nonnegative integer 
exponents. For simplicity of notation, we will write this as $x^s$, 
where $s=(s^i(v))_{i=1..3,v\in \Gamma}$ is a vector in the lattice 
$\Z^{{\cal Q}_1}$ whose components are nonnegative. The affine 
space $\C^{{\cal Q}_1}$ can be described by the collection of all such 
exponents, i.e. by the set of those 
points $m$ of the lattice $\Z^{{\cal Q}_1}$ which belong to the  
cone $C\subset \R^{{\cal Q}_1}$ 
defined by the inequalities $m^i(v)\geq 0$ for all $i,v$ 
($C$ is the first `octant' in $\R^{{\cal Q}_1}$). 

On the other hand, the relations (\ref{mon_rels}) are equivalent with:
\be
x^m=1~~\mbox{~ for~all~}~m \in R~~, 
\ee
where $R$ is the sublattice of $\Z^{Q_1}$ (which we will call 
{\em the lattice of relations}) spanned by the vectors:
\be
\label{rs}
r^{(ij)}:=e^j(v-w_i)+e^i(v)-e^i(v-w_j)-e^j(v)~~,
\ee
where $(e^i(v))_{i=1..3,v\in \Gamma}$ is the canonical basis of 
$\Z^{{\cal Q}_1}$. In general, the $3|\Gamma|$ vectors (\ref{rs}) 
are not linearly independent. In fact, it is not hard to see that they 
span a subspace of dimension $\rho=2(|\Gamma|-1)$, so that the lattice 
of relations has rank $\rho$. 

Picking such an integral basis  $\{r_1..r_\rho\}$ of $R$, 
we can form a $\rho$ by $3|\Gamma|$ 
matrix ${\hat R}$ whose rows are given by the components of $r_1..r_\rho$. 
The conditions (\ref{mon_rels}) show that two monomials $x^s,x^{s'}$ of the 
ambient space $\C^{{\cal Q}_1}$ will restrict to the same function on 
${\cal Z}$ if the (nonnegative) integral vectors $s,s' \in \Z^{{\cal Q}_1}$ 
are such that $s-s'$ belongs to the sublattice $R$. 
Therefore, one can describe ${\cal Z}$ by considering the collection 
of those points of the quotient lattice $\Z^{{\cal Q}_1}/R$ which lie in 
the image $\Sigma$ of the cone $C$ via the natural projection map 
$p:\Z^{{\cal Q}_1}\rightarrow \Z^{{\cal Q}_1}/R$. The lattice 
$\Z^{{\cal Q}_1}/R$ can be identified with the orthogonal complement 
$M$ of the lattice $R$ in $\Z^{{\cal Q}_1}$. Clearly $M$ coincides with  
the set of integral vectors which lie in the kernel of the matrix ${\hat R}$.
Therefore, a basis of the lattice $M$ can be obtained by computing a basis
of the integral kernel of ${\hat R}$. If $v_1..v_{|\Gamma|+2}$ is such a 
basis, then one can form a $3|\Gamma|$ by $|\Gamma|+2$ matrix $K$ 
whose columns are given by the the components of $v_1..v_{|\Gamma|+2}$. 
Then it is not hard to see that the projection map 
$p:\Z^{{\cal Q}_1}\rightarrow  \Z^{{\cal Q}_1}/R$ can be identified with the 
transpose $K^t$ of the matrix $K$, viewed as a linear map from 
$\Z^{{\cal Q}_1}$ to $M$. Therefore, once one has computed the matrix $K$, 
using its columns as a basis of $M$ identifies $M$ with $\Z^{|\Gamma|+2}$ 
and the projection $p(C)$ with the cone $\Sigma=K^t(C)$ generated by the 
columns of $K^t$, i.e. by the rows of $K$. In conclusion, the toric variety 
${\cal Z}$ is described by the cone of exponents in $\Z^{|\Gamma|+2}$ 
which is generated by the rows of $K$. 

\

{\bf Example}:$\Z_2\times \Z_2$ singularity

\

In this case, the number of independent monomial relations is 
$2(|\Gamma|-1)=6$, so the complex dimension of the variety of commuting 
matrixes is $dim{\cal Z}=|\Gamma|+2=6$.
The matrix of an integral basis of the lattice of monomial relations is:
\be
{\hat R}=\left [\begin {array}{cccccccccccc} 1&0&0&-1&0&0&-1&1&0&-1&2&-1
\\\noalign{\medskip}0&1&0&-1&0&0&0&0&0&-1&1&0\\\noalign{\medskip}0&0&1
&-1&0&0&-1&1&0&0&1&-1\\\noalign{\medskip}0&0&0&0&1&0&-1&0&0&0&1&-1
\\\noalign{\medskip}0&0&0&0&0&1&0&-1&0&0&-1&1\\\noalign{\medskip}0&0&0
&0&0&0&0&0&1&-1&1&-1\end {array}\right ]
\ee
and a basis of its integral kernel is given by the columns of:
\be
K=\left [\begin {array}{cccccc} 1&1&-1&1&-2&1\\\noalign{\medskip}1&0&0&1
&-1&0\\\noalign{\medskip}1&1&-1&0&-1&1\\\noalign{\medskip}1&0&0&0&0&0
\\\noalign{\medskip}0&1&0&0&-1&1\\\noalign{\medskip}0&0&1&0&1&-1
\\\noalign{\medskip}0&1&0&0&0&0\\\noalign{\medskip}0&0&1&0&0&0
\\\noalign{\medskip}0&0&0&1&-1&1\\\noalign{\medskip}0&0&0&1&0&0
\\\noalign{\medskip}0&0&0&0&1&0\\\noalign{\medskip}0&0&0&0&0&1
\end {array}\right ]~~.
\ee
The lattice $M$ is isomorphic with $\Z^6$ and the $12$ vectors given by the 
rows of $K$ generate the cone of exponents $\Sigma\subset \R^6$ of the 
toric variety ${\cal Z}$. 

\subsection{The action of the projected gauge group on the variety of 
commuting matrices}

Let $q:=|\Gamma|-1$ denote the rank of the effective gauge group. 
A monomial $x^m$ of the ambient space has a $U(1)^q$ charge 
given by the $q$-vector $\Delta m$. Since the exponent vectors 
are identified modulo the lattice $R$ when imposing the F-flatness constraints,
we can describe the restriction of the $U(1)^q$ action to the 
variety of commuting matrices by the descent of the map 
$\Delta :\Z^{{\cal Q}_1}\rightarrow \Z^q$ to a map 
$V:M\rightarrow \Z^q$. The fact that $d$ does indeed factor through the 
projection $p:\Z^{{\cal Q}_1}\rightarrow M$ follows from the fact that 
the relations (\ref{rs}) obviously lie in the kernel of $\Delta$. Since we 
identify $p$ with $K^t$, it follows that $V$ can be obtained as the unique 
matrix satisfying the condition:
\be
VK^t=\Delta ~~.
\ee
Note that $V$ has dimensions $q\times (|\Gamma|+2)$. 

\

{\bf Example}:$\Z_2\times \Z_2$ singularity

\

The integral matrix $V$ satisfying $VK^t=\Delta$ is given by:
\be
V=\left [\begin {array}{cccccc} 0&0&0&-1&0&1\\\noalign{\medskip}0&-1&1&0
&-1&0\\\noalign{\medskip}-1&1&-1&1&0&-1\end {array}\right ]
\ee

\subsection{The degenerate holomorphic quotient 
presentation of the moduli space}

        Once we have obtained the generators $K$ of the cone of exponents for 
${\cal Z}$ and the restriction $V$ of the $U(1)^q$ action to ${\cal Z}$, a 
symplectic quotient presentation of the moduli space can be obtained by the 
methods of \cite{branes3}. We will simply 
summarize the steps of that construction, referring the reader to 
\cite{branes3} for details. 

Since ${\cal Z}$ is defined by monomial relations inside of the affine 
space $\C^{{\cal Q}_1}$, it will be an affine toric variety 
(of complex dimension $|\Gamma|+2$) whose 
toric generators are the generators of the cone $\Sigma^{\rm v}$ dual to the 
cone of exponents $\Sigma $. If $c$ is the number of these vectors%
\footnote{$c$ coincides with the number of facets of $\Sigma$. This is the 
integer mentioned in subsection 1.2, which essentially controls the 
computational complexity of the problem.}  
and $T$ is the $(|\Gamma|+2)\times c$ 
matrix having them  as its columns, then one can construct a 
$(c-|\Gamma|-2)\times c$  matrix of  charges $Q$ whose 
rows form an integral basis for the kernel of $T$. 
Note that we do not 
have an analytic expression for the number $c$ of toric generators, 
since this number depends in a complicated way on the combinatorial 
properties of the cone $\Sigma$. 

At this stage, we have a toric variety ${\cal Z}$, presented as a holomorphic 
quotient $\C^c/(\C^*)^{c-|\Gamma|-2}$, which is further 
divided by the $(\C^*)^q=(\C^*)^{|\Gamma|-1}$ action given by the charges 
encoded by the rows of 
$V$. This double quotient can be reduced to the form 
$\C^c/(\C^*)^{c-3}$
by choosing a lift of the $(\C^*)^q$ action to the space $\C^c$. Such 
a lift can be given by first choosing a left inverse $U$ of the matrix $T^t$
(i.e $UT^t=id$, with $U$ a $(|\Gamma|+2)\times c$ matrix)
and then lifting the $(\C^*)^q$ action to $\C^c$ as the 
action specified by the charge matrix $VU$. It follows that the moduli space 
can be described by the holomorphic quotient $\C^c/(\C^*)^{c-3}$ 
associated to the $(c-3)\times c$ 
charge matrix $Q_{total}$ obtained by stacking $Q$ and 
$VU$. In what follows, we will always place the matrix $Q$ {\em above} 
the matrix $VU$ when constructing the matrix $Q_{total}$. 

\

{\bf Example}:$\Z_2\times \Z_2$ singularity

\

A choice for the matrices $T,U,Q,Q_{total}$ is:
\bea
\label{2x2}
T&=&\left [\begin {array}{ccccccccc} 0&1&0&1&1&0&0&0&0\\\noalign{\medskip}0
&0&1&0&0&1&0&1&0\\\noalign{\medskip}0&0&0&0&1&1&0&0&1
\\\noalign{\medskip}1&0&0&0&0&0&1&1&0\\\noalign{\medskip}0&0&0&1&0&0&1
&1&0\\\noalign{\medskip}0&0&0&1&0&0&1&0&1\end {array}\right ]\nn\\
Q&=&\left [\begin {array}{ccccccccc} 1&0&-1&1&-1&0&-2&1&1
\\\noalign{\medskip}0&1&-1&0&-1&0&-1&1&1\\\noalign{\medskip}0&0&0&0&0&
1&1&-1&-1\end {array}\right ]\nn\\
U&=&\left [\begin {array}{ccccccccc} 0&1&0&0&0&0&0&0&0\\\noalign{\medskip}0
&0&1&0&0&0&0&0&0\\\noalign{\medskip}0&-1&0&0&1&0&0&0&0
\\\noalign{\medskip}1&0&0&0&0&0&0&0&0\\\noalign{\medskip}-1&0&-1&0&0&0
&0&1&0\\\noalign{\medskip}1&-1&1&1&0&0&0&-1&0\end {array}\right ]\\
Q_{total}&=&\left [\begin {array}{ccccccccc} 1&0&-1&1&-1&0&-2&1&1
\\\noalign{\medskip}0&1&-1&0&-1&0&-1&1&1\\\noalign{\medskip}0&0&0&0&0&
1&1&-1&-1\\\noalign{\medskip}0&-1&1&1&0&0&0&-1&0\\\noalign{\medskip}1&
-1&0&0&1&0&0&-1&0\\\noalign{\medskip}0&1&0&-1&-1&0&0&1&0\end {array}
\right ]~~.
\eea
In particular, in this example we have $c=9$.

\subsection{The reduction of the holomorphic quotient}

To simplify notation in the sequel, we let 
$r=c-3$ denote the number of rows of $Q_{total}$. 
Taking the transpose of the kernel of $Q_{total}$ 
gives a $3\times c$ matrix $G_{total}$ such that $Q_{total}(G_{total})^t=
0_{r\times 3}$ (note that the columns of $G_{total}$, 
which play the role of toric generators,
are naturally associated with the columns of $Q_{total}$). 
The rows of  $Q_{total}$ form a basis of integral linear 
relations among these generators. 
In general, the holomorphic quotient description of the moduli space 
given by the charge matrix $Q_{total}$ is degenerate (not minimal) in the 
sense that the toric generators {\em are not all distinct}. 

In order to make the following discussion clear, note that one can always 
reorder the columns of $G_{total}$ such that identical generators appear 
consecutively 
(for example, one can sort $G_{total}$ in decreasing lexicographic
order on its columns, a convention which we will follow everywhere in this 
paper). While performing this rearrangement, 
one must also reorder the columns of $Q_{total}$ accordingly, 
since each generator is associated to such a column. 
Hence we let $G_t$ be the matrix obtained from $G_{total}$ by sorting its 
columns in decreasing lexicographic order, and $Q_t$ be the matrix obtained 
from $Q_{total}$ by performing the same permutation on its columns. 

The columns of $G_t$ will form a number $b$ of blocks 
$\Gamma^{(1)}...\Gamma^{(b)}$ (appearing in $G_t$ in this order), such 
that each block $\Gamma^{(k)}$ consists of  $m_k$ copies of the same column 
$\gamma^{(k)}$, and such that the 3-vectors $\gamma^{(1)}..\gamma^{(b)}$ are 
all distinct. 
Hence each block $\Gamma^{(k)}$ is a matrix of dimensions $3\times m_k$. 
In general, some of these will consist of one column only 
(then $m_k=1$), while other blocks will be multiple, i.e. will consist of 
$m_k\geq 2$ repetitions of $\gamma^{(k)}$. We let $n$ be the 
number of non-multiple blocks and $s$ the number of multiple blocks, 
so that $n+s=b$. Then the list of all toric generators consists of 
$m_1$ copies of $\gamma^{(1)}$, $m_2$ copies of $\gamma^{(2)}$...
$m_b$ copies of $\gamma^{(b)}$. 
It is convenient to index these as $\gamma^{(k)}_i$~
($i=1..m_k, k=1..b$) where 
$\gamma^{(k)}_1=...=\gamma^{(k)}_{m_k}=\gamma^{(k)}$ 
are the $m_k$ copies of $\gamma^{(k)}$ appearing in the block 
$\Gamma^{(k)}$. We also index the homogeneous coordinates of the holomorphic 
quotient by $z^{(k)}_i$~($k=1..b,i=1..m_k$) where $z^{(k)}_i$ corresponds to 
the column $\gamma^{(k)}_i$ of $Q_t$. If we also define 
$p^{(k)}_i=|z^{(k)}_i|^2\geq 0$, we can write the moment map equations for the 
symplectic description of our quotient as:
\be
\label{moment_t}
Q_tp={\tilde \xi}~~,
\ee
where:
\be
p=\left[\begin{array}{c}p^{(1)}_1\\...\\p^{(1)}_{m_1}\\...\\p^{(b)}_1\\...
\\p^{(b)}_{m_b}\end{array}\right]=
\left[\begin{array}{c}p^{(1)}\\...\\p^{(b)}\end{array}\right]~~,
\ee
with $p^{(k)}=
\left[\begin{array}{c}p^{(k)}_1\\...\\p^{(k)}_{m_k}\end{array}\right]$
and where ${\tilde \xi}$ is the $r$-vector whose first $r-q$ components 
are zero and whose last $q$ components are given by the vector of D-brane 
Fayet-Iliopoulos parameters $\xi$.

        The rows of the matrix $Q_t$ form a basis for the lattice $S$ of 
integral linear relations among the toric generators. In fact, this 
is the only piece of information needed to reconstruct the 
holomorphic quotient, and any basis of the lattice $S$ contains the same data.
In particular, one always has the freedom to perform invertible 
row operations on $Q_t$. Such a transformation (which 
corresponds to a change of integral basis of $S$) 
has the form:
\be
\label{charge_tf}
Q_t\rightarrow WQ_t~~,
\ee
with $W$ an $r\times r$ integral matrix which is invertible over the integers
\footnote{That is, the determinant of $W$ must be $\pm 1$.}. 
The presence of multiplicities in the list of toric generators implies that 
one can always find a transformation $W$ which brings the matrix $Q_t$ 
to a form which is particularly suited for discussing the reduction procedure. 
This `canonical' form can be found as follows. Since each {\em multiple} 
generator $\gamma^{(k)}$ appears $m_k\geq 2$ times, 
we can always find $m_k-1$ elements of the lattice $S$ which correspond to the 
following $m_k-1$ obvious relations:
\bea
\label{can_rels}
\gamma^{(k)}_1=\gamma^{(k)}_2\nn\\
\gamma^{(k)}_2=\gamma^{(k)}_3\nn\\
......\nn\\
\gamma^{(k)}_{m_k-1}=\gamma^{(k)}_{m_k}\nn\\
\eea
Certainly not all relations among $\gamma^{(k)}_i$ are of this form, but it 
is clear that all of the remaining relations must come from whatever linear 
relations exist among the {\em distinct} vectors 
$\gamma^{(1)}...\gamma^{(b)}$ obtained by eliminating all multiplicities in 
our list. A basis for the latter type of relations is given by the rows 
of the reduced charge matrix $Q_{reduced}$, which is defined as follows. 
Define the matrix $G_{reduced}$ to be obtained from $G_{total}$ by removing 
all column multiplicities (that is, $G_{reduced}$ consists of the distinct 
columns $\gamma^{(1)}..\gamma^{(b)}$ in this order (the inverse lexicographic 
order, in our conventions). Then $Q_{reduced}$ is any matrix whose rows 
give an integral 
basis for the kernel of $G_{reduced}$. Note that $Q_{reduced}$ has $b=n+s$ 
columns and let $p$ be the number of its rows. We clearly have $p=b-3$. 
We conclude that one can always 
find a basis of linear relations among $\gamma^{(k)}_i$ which consists of all 
of the `equality' relations (\ref{can_rels}) together with the relations given 
by the rows of $Q_{reduced}$. 
 
Remember that $q=|\Gamma|-1$ denotes the number of $U(1)$ factors of the
projected gauge group on the worldvolume 
(which coincides with the number of rows of the matrix $V$).
The arguments above show that $Q_t$ can always be brought to 
the form:
\be
Q_{can}=
\left[\begin{array}{cccccc}
V^{(1)}&V^{(2)}&V^{(3)}&V^{(4)}&...&V^{(b)}\\
c(m_1)&0_{(m_1-1)\times m_2}&0_{(m_1-1)\times m_3}&0_{(m_1-1)\times m_4}
&...&0_{(m_1-1)\times m_b}\\
0_{(m_2-1)\times m_1}&c(m_2)&0_{(m_2-1)\times m_3}&0_{(m_2-1)\times m_4}
&...&0_{(m_2-1)\times m_b}\\
0_{(m_3-1)\times m_1}&0_{(m_3-1)\times m_2}&c(m_3)&0_{(m_3-1)\times m_4}
&..&0_{(m_3-1)\times m_b}\\
0_{(m_4-1)\times m_1}&0_{(m_4-1)\times m_2}&0_{(m_4-1)\times m_3}&
c(m_4)&...&0_{(m_4-1)\times m_b}\\
...&...&...&...&...&...\\
0_{(m_b-1)\times m_1}&0_{(m_b-1)\times m_2}&
0_{(m_b-1)\times m_3}&0_{(m_b-1)\times m_4}&...&c(m_b)
\end{array}\right]
\ee
where we defined the {\em canonical $(m-1)\times m$ block} 
$c(m)$ to be given by:
\be
c(m):=\left[\begin{array}{ccccccc}
1&-1&0&0&..&0&0\\
0&1&-1&0&..&0&0\\
0&0&1&-1&..&0&0\\
..&..&..&..&..&..&..\\
0&0&0&0&..&1&-1
\end{array}\right]
\ee
if $m\geq 1$ and to be a `zero by $1$ block 
(i.e. a missing element in a column)
if $m=1$ (similarly, $0_{(m-1)\times m}$ is defined to be a missing element 
in a column if $m=1$). The $p\times m_k$ blocks $V^{(k)}$ appearing in 
$Q_{can}$ are defined by:
\be
V^{(k)}=
\left[\begin{array}{ccccc}
v^{(k)}_1&0&0&..&0\\
v_2^{(k)}&0&0&..&0\\    
..&..&..&..&..\\
v_p^{(k)}&0&0&..&0
\end{array}\right]~~
\ee
(no zero columns are present in the case $m_k=1$),
where the $p$-vector $v^{(k)}=\left[\begin{array}{c}v^{(k)}_1\\..\\v^{(k)}_p
\end{array}\right]$ is the $k$-th column of $Q_{reduced}$. 
Note that the rows of $Q_{can}$ are naturally divided into $s+1$ blocks, 
with the first block consisting of the first $p$ rows and the next $s$ blocks 
each consisting of $m_k-1$ rows for those $m_k$ which are different from $1$. 
In particular, we have 
$r=
p+\sum_{\scriptsize \begin{array}{c}k=1..b\\m_k\geq 2\end{array}}{(m_k-1)}$ 
and $c=\sum_{k=1..b}{m_k}$. If we let let $M=\sum_{\scriptsize
\begin{array}{c}k=1..b\\m_k\geq 2\end{array}}{(m_k-1)}$, 
then we can write the row and column dimensions of $Q_{can}$ 
as $r=p+M$ and $c=b+M$, and since $r=c-3$ we deduce that $p=b-3$.

The transformation (\ref{charge_tf}) performed in order to bring $Q_t$ 
to the form $Q_{can}$ brings the moment map equations 
(\ref{moment_t}) to the form:
\be
\label{moment_can}
Q_{can}p=\eta~~,
\ee 
with the vector $\eta\in \R^c$ given by:
\be
\eta=W_0\xi~~,
\ee
where $W_0$ is the $r\times q$ matrix obtained by keeping only the 
last $q$ columns of $W$. 

Since the rows of $Q_{can}$ are naturally divided into blocks, 
we divide the $r=p+M$-vector $\eta$ accordingly into a $p$-vector $\eta^{(0)}$ 
and $s$~~ $(m_k-1)$-vectors  $\eta^{(k)}$~(for those $k$ associated to 
multiple blocks, which we will call $k_1..k_s$) such that:
\be
\eta=\left[\begin{array}{c}\eta^{(0)}\\\eta^{(k_1)}\\...\\\eta^{(k_s)}
\end{array}\right]~~.
\ee
Then we can write:
\bea
\label{xi_eta}
\eta^{(0)}&=&w^{(0)}\xi~~\\
\eta^{(k)}&=&w^{(k)}\xi~
\mbox{~~for~all}~k~\mbox{associated~to~multiple~blocks},\nn
\eea
where $w^{(0)}$ is the $p\times q$ matrix formed by the first 
$p$ rows of $W_0$, and $w^{(k)}$ (for those $k$ corresponding to multiple 
blocks) are  the $(m_k-1)\times q$ matrices given 
by the rows of $W_0$ associated to the other row blocks of $Q_{can}$.

In order to reduce the holomorphic quotient, we must eliminate $m_k-1$ 
homogeneous variables out of the $m_k$ variables associated with each {\em 
multiple} block ($m_k \geq 2$). 
This is possible provided that the levels $\eta$ of the moment map 
are such that the $m_k-1$ variables to be eliminated in each multiple 
block are assured to be nonzero. 
Due to the structure of $Q_{can}$, 
the various canonical blocks are `decoupled' from each other, and we can 
discuss reduction within each canonical block separately. Indeed, the 
moment map equations (\ref{moment_can}) have the form:
\bea
\label{red_eq}
Q_{reduced}\left[\begin{array}{c}p^{(0)}_1\\p^{(1)}_1\\...\\p^{(b)}_1
\end{array}\right]=\eta^{(0)}\nn\\
c(m_{k})\left[\begin{array}{c}p^{(k)}_1\\p^{(k)}_2\\...\\p^{(k)}_{m_{k}}
\end{array}\right]=\eta^{(k)}\\
\eea
where $k$ runs over all {\em multiple} blocks. Let us consider 
the equations involving the homogeneous variables 
$z^{(k)}_1..z^{(k)}_{m_k}$ associated with the multiple block $k$.
The equations $c(m_k)p^{(k)}=\eta^{(k)}$ can be solved in terms of 
$p^{(k)}_1$ and $\eta^{(k)}$:
\bea
p^{(k)}_1=p^{(k)}_1\nn\\
p^{(k)}_2=p^{(k)}_1-\eta^{(k)}_1\nn\\
p^{(k)}_3=p^{(k)}_1-\eta^{(k)}_1-\eta^{(k)}_2\nn\\
.......\\
p^{(k)}_i=p^{(k)}_1-\eta^{(k)}_1 -...-\eta^{(k)}_{j-1}\nn\\
.......\nn\\
p^{(k)}_{m_k}=p^{(k)}_1-\eta^{(k)}_1-...-\eta^{(k)}_{m_k-1}~~.\nn
\eea
The values of $p^{(k)}_1$ are constrained by the conditions 
$p^{(k)}_1\geq0, p^{(k)}_2\geq 0,...,
p^{(k)}_{m_k}\geq 0$, which are equivalent to:
\be
\label{p1bound}
p^{(k)}_1\geq {\rm max}
(0,\eta^{(k)}_1, \eta^{(k)}_1+\eta^{(k)}_2,...,\eta^{(k)}_1+..+
\eta^{(k)}_{m-1})~~.
\ee
If the maximum in the right hand side is attained precisely at 
$\eta^{(k)}_1+...+\eta^{(k)}_{i-1}$ (and at no other point), i.e. if 
the strict inequalities:
\bea
\label{chambers}
\eta^{(k)}_1+..+\eta^{(k)}_{i-1}> 0,\nn \\
\eta^{(k)}_1+..+\eta^{(k)}_{i-1}>\eta^{(k)}_1 \nn \\ 
\eta^{(k)}_1+..+\eta^{(k)}_{i-1}>\eta^{(k)}_1+\eta^{(k)}_2 \nn \\
...\nn \\
\eta^{(k)}_1+..+\eta^{(k)}_{i-1}> \eta^{(k)}_1+...+\eta^{(k)}_{i-2} \\
\eta^{(k)}_1+..+\eta^{(k)}_{i-1}> \eta^{(k)}_1+..+\eta^{(k)}_{i}\nn \\
 ... \nn \\
\eta^{(k)}_1+..+\eta^{(k)}_{i-1}> \eta^{(k)}_1+..+\eta^{(k)}_{m_k-1} \nn 
\eea
hold, then equation (\ref{p1bound}) 
assures that $p^{(k)}_j=p^{(k)}_1-\sum_{l=1..j-1}{\eta^{(k)}_l} >0$ for all 
$j \neq i$, so that 
we can eliminate all homogeneous variables associated to the canonical 
block except for that associated to its $i$-th column.  
(If the maximum is attained precisely at $0$, then $p^{(k)}_j>0$ for all 
$j\neq 1$ and we can eliminate all variables except for that associated to 
the first column. This case corresponds to $i=1$). 
The inequalities (\ref{chambers}) 
(taken for all values of $i$ in turn) divide the space 
$\R^{m_k-1}(\eta^{(k)})$ of values of 
$\eta^{(k)}$ into $m_k$ distinct 
chambers $\sigma^{(k)}_i~(i=1..m_k)$, which are maximal-dimensional 
polyhedral cones in $\R^{m_k-1}(\eta^{(k)})$. 
If $\eta^{(k)}$ belongs to the interior of 
$\sigma^{(k)}_i$, then we can eliminate 
all of the homogeneous variables $z^{(k)}_j$ except for the 
variable $z^{(k)}_i$. 
In order to perform this reduction, we must first eliminate the variables 
$z^{(k)}_j (~j\neq i)$ from the first equations of (\ref{red_eq}). 
This can be done as follows. Since the the rows of 
$c(m_k)$ are linearly independent, there 
exists a unique $p\times (m_k-1)$ matrix $F_i(m_k)$ such that:
\be
F_i(m_k) c(m_k)=\left[
\begin{array}{cccccccc}
-v^{(k)}_1&0&..&0& v^{(k)}_1&0&..&0\\
-v^{(k)}_2&0&..&0& v^{(k)}_2&0&..&0\\   
..&..&..&..&..&..&..&..\\
-v^{(k)}_p&0&..&0&v^{(k)}_p&0&..&0
\end{array}\right]~~
\ee
(for $i=1$, the matrix on the right hand side is defined to be the null 
$p\times (m_k-1)$ matrix).
In fact, it is not hard to see that $F_i(m_k)$ is given by:
\be
F_i(m_k)=\left[\begin{array}{cccccccc}
-v^{(k)}_1&-v^{(k)}_1&..&-v^{(k)}_1&0&0&..&0\\
-v^{(k)}_2&-v^{(k)}_2&..&-v^{(k)}_2&0&0&..&0\\
..&..&..&..&..&..&..&0\\
-v^{(k)}_p&-v^{(k)}_p&..&-v^{(k)}_p&0&0&..&0\\
\end{array}\right]~~,
\ee
where the first $i-1$ columns are copies of the vector $v^{(k)}$ and the other 
columns are zero (for $i=1$, $F_1(m_k)$ is defined to be the null 
$p\times (m_k-1)$ matrix). 
Multiplying $c(m_k)$ to the left by $F_i(m_k)$ and adding 
the result to $V^{(k)}$ produces the matrix:
\be
\left[
\begin{array}{cccccccc}
0&0&..&0& v^{(k)}_i&0&..&0\\
0&0&..&0& v^{(k)}_i&0&..&0\\    
..&..&..&..&..&..&..&..\\
0&0&..&0&v^{(k)}_i&0&..&0\\
\end{array}\right]~~
\ee
Hence performing this invertible row operation allows us 
to bring $Q_{can}$ to a form in which all entries associated to the 
homogeneous variables $z^{(k)}_j~(j\neq i)$ are zero except for those 
appearing in the canonical block $c(m_k)$. Once $Q_{can}$ has been 
brought to this form, and once we know that $\eta^{(k)}$ lies in the 
interior of the cone $\sigma^{(k)}_i$ (so that $z^{(k)}_j\neq 0$ for all 
$j\neq i$), then we can 
eliminate these variables by using the $(\C^*)^{m_k-1}$ 
subgroup associated with the rows of $c(m_k)$ in order to set each of them 
equal to $1$\footnote{Here we are tacitly using the well-known 
equivalence between the symplectic and holomorphic quotient \cite{Audin}.}. 
Since these variables have been eliminated from the rest of the charge 
matrix, this will not have any effect on the remaining part of the 
holomorphic quotient. Note, however, that performing the row operations 
$V^{(k)}\rightarrow V^{(k)}+F_i(m_k)c(m_k)$ 
(which is needed in order eliminate 
our variables from the first $p$ rows of $Q_{can}$) will induce a 
redefinition of $\eta^{(0)}$ given by:
\be
\eta^{(0)}\rightarrow \eta^{(0)}+F_i(m_k)\eta^{(k)}~~.
\ee

Applying the above discussion to each of the $s$ {\em multiple} blocks 
leads to the following pattern of reduction. For each multiple block 
$k$~($m_k\geq 2$) we have a partition of  the space 
$\R^{m_k-1}(\eta^{(k)})$ into $m_k$ chambers 
$\sigma^{(k)}_i~(i=1..m_k)$ which adjoin along common faces. 
If, for each $k$ with $m_k\geq 2$, 
$\eta^{(k)}$ lies in the interior of one of these chambers
$\sigma^{(k)}_{i_k}$, 
then we can reduce each of the multiple blocks $k$ to its $i_k$-th column. 
This is achieved by performing the following 
invertible row operation on $Q_{can}$:
\be
Q_{can}\longrightarrow Q'_{can}=
\left[\begin{array}{cccc}
F_{i_1}(m_1)&0&..&0\\
0&F_{i_2}(m_2)&..&0\\
..&..&..&..\\
0&0&..&F_{i_s}(m_s)
\end{array}\right] Q_{can}~~,
\ee
which replaces the first $p$ components $\eta^{(0)}$ of $\eta$ with the 
$p$-vector given by:
\be
\label{zetas}
\zeta=\zeta_{i_1...i_s}=\eta^{(0)}+
\sum_{\scriptsize\begin{array}{c}k=1..b\\m_k\neq 1\end{array}}
{F_{i_k}(m_k)\eta^{(k)}}~~.
\ee
Performing the reduction of $z^{(k)}_j~(j\neq i_k)$ 
eliminates all except for the first 
$p$ rows of $Q'_{can}$ and all of its columns except for the non-multiple 
columns and those containing the $i_k$-th column of each multiple block. 
The result is a toric variety based on the reduced matrices of generators and 
charges $G_{reduced}$ and $Q_{reduced}$, in a phase given by the moment map 
level $\zeta_{i_1..i_k}$.

The crucial equations (\ref{zetas}) can be described with the help of a 
piecewise-linear function 
$\pi:\R^r\rightarrow \R^p$ which we define as follows. 
For each set of $s$ indices 
$i_1=1..m_{k_1}$..$i_s=1..m_{k_s}$ (remember that 
$k_1..k_s\in \{1..b\}$ index the {\em multiple} blocks), 
we let $\Sigma_{i_1..i_s}$ be the cone (or `wedge') in $\R^r$ 
defined by:
\be
\Sigma_{i_1..i_s}=\R^p\times 
\sigma^{(1)}_{i_1}\times...\times \sigma^{(s)}_{i_s}~~.
\ee
The collection of these cones (which has 
$\Pi_{\scriptsize \begin{array}{c}k=1..b\\m_k\geq 1\end{array}}
{m_k}=\Pi_{k=1..b}{m_k}$ elements) 
divides the space $\R^r(\eta)$ into chambers which adjoin along 
common walls. 
If $\eta$ belongs to the chamber $\Sigma_{i_1..i_s}$, then the value of 
$\pi$ at $\eta$ is given by the linear expression:
\be
\pi(\eta)=\eta^{(0)}+
\sum_{\scriptsize\begin{array}{c}k=1..b\\m_k\neq 1\end{array}}
{F_{i_k}(m_k)\eta^{(k)}}~~
\ee
which appears in the right hand side of (\ref{zetas}). 
It is obvious that these expressions agree on the walls, so that $\pi$ is 
a continuous piecewise linear function. In fact, one can give an analytic 
expression for $\pi$, if one notices that:
\be
F_{i_k}(m_k)\eta^{(k)}=-(\sum_{j=1..i_k-1}{\eta^{(k)}_j})v^{(k)}~~,
\ee
where the sum is defined to be zero if $i_k=1$. Since for 
$\eta \in \Sigma_{i_1..i_s}$ we have $\eta^{(k)}\in \sigma^{(k)}_{i_k}$, 
and since by the definition of the cones $\sigma^{(k)}_{i}$ this implies that:
\be
\sum_{j=1..i_k-1}{\eta^{(k)}_j}=
{\rm max}(0,\eta^{(k)}_1,\eta^{(k)}_1+\eta^{(k)}_2,...,
\eta^{(k)}_1+...+\eta^{(k)}_{m^{(k)}-1})~~,
\ee
it follows that for {\em any} $\eta \in \R^r$, the value of $\pi$ at $\eta$ is 
given by:
\be
\pi(\eta)=\eta^{(0)}-
\sum_{\scriptsize\begin{array}{c}k=1..b\\m_k\neq 1\end{array}}
{{\rm max}(0,\eta^{(k)}_1,\eta^{(k)}_1+
\eta^{(k)}_2,...,\eta^{(k)}_1+...+\eta^{(k)}_{m^{(k)}-1})v^{(k)}}~~.
\ee

The important information for us is the map 
$\phi:\R^q\rightarrow \R^p$ 
from the  D-brane Fayet-Iliopoulos 
parameters $\xi$ to the effective moment map levels 
$\zeta$ in the reduced toric presentation of the moduli space. 
This immediately follows from the above and from (\ref{xi_eta}):
\be
\phi(\xi)=w^{(0)}\xi -
\sum_{\scriptsize\begin{array}{c}k=1..b\\m_k\neq 1\end{array}}
{{\rm max}[0,
\sum_{j=1..q}{w^{(k)}_{1j}\xi_j},
\sum_{j=1..q}{(w^{(k)}_{1j}+w^{(k)}_{2j})\xi_j} 
,..., 
\sum_{j=1..q}{(w^{(k)}_{1j}+..+w^{(k)}_{m^{(k)}-1,j})\xi_j}
]v^{(k)}}~~.
\ee
In conclusion, the relation between the D-brane Fayet-Iliopoulos parameters
and the effective moment map levels is given by the piecewise-linear function 
$\phi$.  

Since $\phi$ is clearly continuous, its linear regions form a subdivision of 
$\R^q(\xi)$ into chambers which are polyhedral cones\footnote{More precisely, 
these regions form an integral polyhedral fan in $\R^q(\xi)$.}. 
These chambers can be found from (\ref{chambers}) as follows. 
For each $k$ associated to a multiple block, consider the vectors 
$e_1(m_k)...e_{m_k}(m_k)$ in $\R^{m_k-1}(\eta^{(k)})$ given by:
\bea
e_j(m_k)=\left[\begin{array}{c}
1\\1\\...\\1\\0\\0\\...\\0
\end{array}\right]~~,
\eea
where the first $j-1$ entries are equal to $1$ (for $j=1$, we define 
$e_1(m_k)$ to be the null $(m_k-1)$-vector). Then the equations 
(\ref{chambers}) (for a fixed $i\in \{1..m_k\}$) can be rewritten as follows:
\be
\label{chambers_dual}
\langle e_i(m_k)-e_j(m_k),\eta^{(k)}\rangle 
>0 \mbox{~~for~all~}j=1..m_k~,j\neq i~~.
\ee
Using $\eta^{(k)}=w^{(k)}\xi$ reduces these equations to:
\be
\label{chambers_phi}
\langle f^{(k)}_i-f^{(k)}_j,\xi\rangle  
>0 \mbox{~~for~all~}j=1..m_k~,j\neq i~~,
\ee
where the $q$-vectors $f^{(k)}_i$ are given by:
\be
f^{(k)}_i=[w^{(k)}]^te_i(m_k)~~.
\ee
It is now clear that $\phi(\xi)$ will belong to $\Sigma_{i_1..i_s}$ 
if and only if $\xi$ belongs to the cone 
$\Xi_{i_1..i_s}=\Xi^{(k_1)}_{i_1}\cap..\cap \Xi^{(k_s)}_{i_s}$, where 
$\Xi^{(k)}_i$ is the cone in $\R^q(\xi)$ defined by the $m_k-1$ inequalities 
(\ref{chambers_phi}). In general, the set of all cones $\Xi_{i_1..i_s}$ 
will be a refinement of the true linear chamber structure of $\phi$, since 
the action of the  matrices $[w^{(k)}]^t$ may 
`collapse' or identify some of the cones $\Xi^{(k)}_i$.

\subsection{Determining the preimage of an effective wall}

For the purposes of the present paper, an important question is the following. 
Given a convex subset $H$ of the space of effective moment map levels 
$\R^p(\zeta)$, what is its preimage via the map $\phi$ ? In particular, 
is this preimage nonzero ? That is, are the values of $\zeta$ associated 
to $H$ indeed realized by the D-brane theory ?. 

Once the map $\phi$ has been identified, this question can be answered as 
follows. For simplicity of notation, consider only the case when $H$ is 
determined by a set of linear equations\footnote{In general, $H$ is given by a 
set of linear equalities and inequalities; the exposition can be generalized 
immediately to this situation.}:
\be
\langle a(t),\zeta\rangle = 0 ~~(t=1..g).
\ee
Then $\phi(\xi)$ belongs to $H$ if and only if $\langle a(t),\phi(\xi)\rangle 
=0$ for all $t$.
In order to solve these conditions, one can simply look for the solution 
in each of the cones $\Xi_{i_1..i_s}$, where $\phi$ is given by a linear 
expression. Fixing such a cone, it is not hard to see that the set 
$\phi^{-1}(H)\cap \Xi_{i_1..i_s}$ consists of all values of $\xi$ which 
satisfy
the inequalities (\ref{chambers_phi}) together with the equalities:
\be
\label{cut}
\langle c(t)_{i_1..i_s},\xi\rangle =0~~,~(t=1..g)
\ee
where the components $\alpha=1..q$ of the 
$q$-vector $c(t)_{i_1..i_s}$ are given by:
\be
c(t)^\alpha_{i_1..i_s}=\sum_{\beta=1..p}{w^{(0)}_{\beta\alpha}a(t)^\beta}-
\sum_{\scriptsize\begin{array}{c}k=1..b\\m_k\neq 1\end{array}}
{\sum_{j=1..i_k-1}{w^{(k)}_{j\alpha}\langle v^{(k)},a(t)\rangle }}~~.
\ee
For each $i_1..i_k$, this system gives a subcone of the cone 
$\Xi_{i_1..i_s}$ (the cut of this cone with the subspace (\ref{cut})). 
Running over all such cuts allows for a complete solution of the problem, 
although the resulting presentation of the solution 
need not be the most economical one.

\

{\bf Example}:$\Z_2\times \Z_2$ singularity

\

In this case, a basis for the kernel of the matrix $Q_{total}$ of 
(\ref{2x2}) is given by the rows of:
\be
G_{total}=\left [\begin {array}{ccccccccc} 1&0&0&1&0&0&2&1&1\\\noalign{\medskip}0
&1&0&0&0&-1&-1&-1&-1\\\noalign{\medskip}0&0&1&0&1&2&0&1&1\end {array}
\right ]~~.
\ee
Sorting the columns of $G_{total}$ in decreasing lexicographic order gives 
the matrix:
\be
G_t=\left [\begin {array}{ccccccccc} 2&1&1&1&1&0&0&0&0\\\noalign{\medskip}
-1&0&0&-1&-1&1&0&0&-1\\\noalign{\medskip}0&0&0&1&1&0&1&1&2\end {array}
\right ]~~,
\ee
while doing the same permutation on the columns of $Q_{total}$ gives:
\be
Q_t=\left [\begin {array}{ccccccccc} -2&1&1&1&1&0&-1&-1&0
\\\noalign{\medskip}-1&0&0&1&1&1&-1&-1&0\\\noalign{\medskip}1&0&0&-1&-
1&0&0&0&1\\\noalign{\medskip}0&1&0&0&-1&-1&1&0&0\\\noalign{\medskip}0&0
&1&0&-1&-1&0&1&0\\\noalign{\medskip}0&-1&0&0&1&1&0&-1&0\end {array}
\right ]~~.
\ee
The matrix $G_t$ has only $6$ distinct columns, appearing with 
multiplicities:
\be
\begin{array}{ccccccc}
m_1 &=& 1~~m_2 &=& 2~~m_3 &=& 2 \\
m_4 &=& 1~~m_5 &=& 2~~m_6 &=& 1
\end{array}
\ee
Thus $G_t$ is formed of $4$ blocks of columns given as follows:
\be
\begin{array}{ccccccccc}
\Gamma^{(1)}&=&\left[\begin{array}{c} 2\\\noalign{\medskip}-1\\
\noalign{\medskip}0\end{array}\right] ~~&~~
\Gamma^{(2)}&=&\left[\begin{array}{cc} 1&1\\\noalign{\medskip}0&0
\\\noalign{\medskip}0&0\end{array}\right] ~~&~~
\Gamma^{(3)}&=&\left[\begin{array}{cc} 1&1\\\noalign{\medskip}-1&-1
\\\noalign{\medskip}1&1\end{array}\right]  \\
\Gamma^{(4)}&=&\left[\begin{array}{c} 0\\\noalign{\medskip}1\\\noalign{\medskip}0\end{array}\right] ~~&~~
\Gamma^{(5)}&=&\left[\begin{array}{cc} 0&0\\\noalign{\medskip}0&0
\\\noalign{\medskip}1&1\end{array}\right] ~~&~~
\Gamma^{(6)}&=&\left[\begin{array}{c} 0\\\noalign{\medskip}-1\\\noalign{\medskip}2\end{array}\right] 
\end{array}~~.
\ee
The multiple blocks are $\Gamma^{(2)}$,~$\Gamma^{(3)}$ and $\Gamma^{(5)}$. 
Keeping only one copy of each distinct column 
(without changing the decreasing lexicographic order) gives the matrix:
\be
G_{red}=\left [\begin {array}{cccccc} 2&1&1&0&0&0\\\noalign{\medskip}0&1&0&2&1
&0\\\noalign{\medskip}1&1&1&1&1&1\end {array}\right ]~~.
\ee
The columns of $G_{red}$ generate the cone over the following two-dimensional 
lattice polytope which lies in the hyperplane $z=1$ of $\R^3$:

\iffigs
$$\vbox{
\hskip 1.5 in \hbox{\epsfxsize=5cm\epsfbox{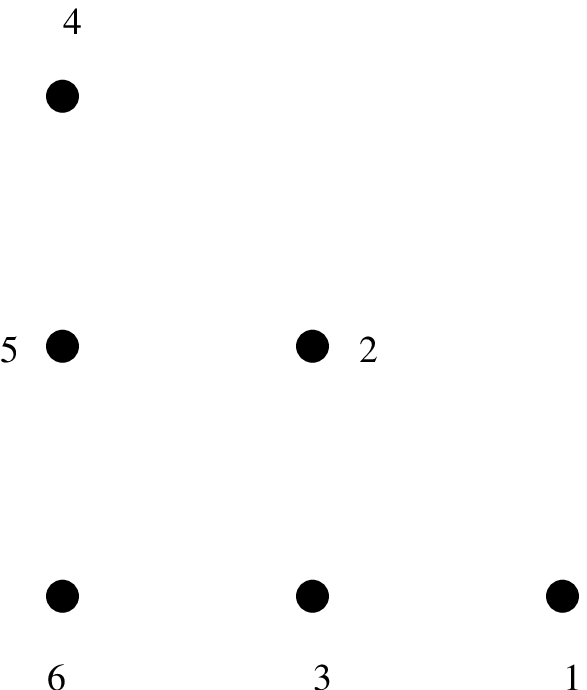}}
\vskip 0.3 in 

\hskip 0.4 in\hbox{Figure 5. {\footnotesize  The polytope associated to the 
$\Z_2\times \Z_2$ quotient singularity.}}}$$
\fi

A basis of linear relations between the columns of $G_{red}$ is given 
by the rows of the reduced charge matrix:
\be
Q_{red}=\left [\begin {array}{cccccc} 1&0&-2&0&0&1\\\noalign{\medskip}0&1&-1&0
&-1&1\\\noalign{\medskip}0&0&0&1&-2&1\end {array}\right ]~~,
\ee
whose number of rows is $p=3$.
The canonical form of $Q_{total}$ is:
\be
Q_{can}=\left [\begin {array}{ccccccccc} 1&0&0&-2&0&0&0&0&1
\\\noalign{\medskip}0&1&0&-1&0&0&-1&0&1\\\noalign{\medskip}0&0&0&0&0&1
&-2&0&1\\\noalign{\medskip}0&1&-1&0&0&0&0&0&0\\\noalign{\medskip}0&0&0
&1&-1&0&0&0&0\\\noalign{\medskip}0&0&0&0&0&0&1&-1&0\end {array}\right] 
\ee
and it involves three copies of  the canonical block:
\be
c(2)=\left [\begin {array}{cc} 1&-1\end {array}\right]~~.
\ee
The transition matrix $W$ satisfying $WQ_{total}=Q_{can}$ is given by:
\be
W=\left [\begin {array}{cccccc} 1&-2&1&-1&-1&0\\\noalign{\medskip}1&-1&1
&-1&-1&-1\\\noalign{\medskip}0&1&1&-1&0&-1\\\noalign{\medskip}0&0&0&0&
-1&-1\\\noalign{\medskip}-1&2&0&1&1&0\\\noalign{\medskip}0&0&0&1&0&1
\end {array}\right ]
\ee
and has determinant $+1$. Its last $q=|\Gamma|-1=3$ columns give the matrix:
\be
W_0=\left [\begin {array}{ccc} -1&-1&0\\\noalign{\medskip}-1&-1&-1
\\\noalign{\medskip}-1&0&-1\\\noalign{\medskip}0&-1&-1
\\\noalign{\medskip}1&1&0\\\noalign{\medskip}1&0&1\end {array}\right ]
\ee
which is formed of 4 row blocks: 
\be
w^{(0)}=\left [\begin {array}{ccc} -1&-1&0\\\noalign{\medskip}-1&-1&-1
\\\noalign{\medskip}-1&0&-1\end {array}\right ]
\ee
(associated to the first $p=3$ rows), 
\be
w^{(2)}=\left [\begin {array}{ccc} 0&-1&-1\end {array}\right ]
\ee
(associated to the next $m_2-1=1$ rows),
\be
w^{(3)}=\left [\begin {array}{ccc} 1&1&0\end {array}\right ]
\ee
(associated to the next $m_3-1=1$ rows),
\be
w^{(5)}=\left [\begin {array}{ccc} 1&0&1\end {array}\right ]
\ee
(associated to the next $m_5-1=1$ rows). 
The piecewise-linear function $\phi$ is given by:
\be
\phi(\xi)=\left [\begin {array}{c} -\xi_{{1}}-\xi_{{2}}\\
\noalign{\medskip}-\xi_{{1}}-\xi_{{2}}-\xi_{{3}}\\\noalign{\medskip}-
\xi_{{1}}-\xi_{{3}}\end {array}\right]
-{\rm max}[0,-\xi_{{2}}-\xi_{{3}}]
\left [\begin {array}{c} 0\\\noalign{\medskip}1\\\noalign{\medskip}0
\end {array}\right ]
-{\rm max}[0,\xi_{{1}}+\xi_{{2}}] 
\left [\begin {array}{c} -2\\\noalign{\medskip}-1\\\noalign{\medskip}0
\end {array}\right ]
-{\rm max}[0,\xi_{{1}}+\xi_{{3}}]
\left [\begin {array}{c} 0\\\noalign{\medskip}-1\\\noalign{\medskip}-2
\end {array}\right ]~~,
\ee 
where we used the vectors:
\be
\begin{array}{ccc}
v^{(2)}=\left[\begin{array}{c}0\\\noalign{\medskip}1\\\noalign{\medskip}0\end{array}\right]~, &
v^{(3)}=\left[\begin{array}{c}-2\\\noalign{\medskip}-1\\\noalign{\medskip}0\end{array}\right]~,& 
v^{(5)}=\left[\begin{array}{c}0\\\noalign{\medskip}-1\\\noalign{\medskip}-2\end{array}\right]
\end{array}~~,
\ee
given by the second, third and fifths columns of $Q_{reduced}$. 
This function has $8$ linear pieces, which are the octants in $\R^3(\xi)$
determined by the vectors (see Figure 6):
\be
\begin{array}{ccccccccc}
g_1&=&\left [\begin {array}{c} 1\\\noalign{\medskip}1\\\noalign{\medskip}-1
\end {array}\right ], &
g_2&=&\left [\begin {array}{c} 1\\\noalign{\medskip}-1\\\noalign{\medskip}1
\end {array}\right ], &
g_3&=&\left [\begin {array}{c} 1\\\noalign{\medskip}-1\\\noalign{\medskip}-1
\end {array}\right ]~~.
\end{array}
\ee

\hskip 0.8 in\scalebox{0.6}{\begin{picture}(0,0)%
\epsfbox{sourcecones.pstex}%
\end{picture}%
\setlength{\unitlength}{3947sp}%
\begingroup\makeatletter\ifx\SetFigFont\undefined%
\gdef\SetFigFont#1#2#3#4#5{%
  \reset@font\fontsize{#1}{#2pt}%
  \fontfamily{#3}\fontseries{#4}\fontshape{#5}%
  \selectfont}%
\fi\endgroup%
\begin{picture}(5499,7108)(1039,-6571)
\end{picture}
}

\hskip 1 in Figure 6. {\footnotesize The linear regions of 
$\phi$.}

\

The images of these vectors and their opposites under the map $\phi$ are:
\be
\begin{array}{ccccccccc}
f_1=\phi(\pm g_1)&=&\left [\begin {array}{c} 2\\\noalign{\medskip}1\\\noalign{\medskip}0\end {array}\right ]~, &
f_2=\phi(\pm g_2)&=&
\left [\begin {array}{c} 0\\\noalign{\medskip}1\\\noalign{\medskip}2
\end {array}\right ]~, &
f_3=\phi(\pm g_3)&=&\left [\begin {array}{c} 0\\\noalign{\medskip}-1\\\noalign{\medskip}0\end {array}\right ]~,
\end{array}~~
\ee
so that all of the 8 linear chambers of $\phi$ in $\R^3(\xi)$ are 
mapped onto the cone in $\R^3(\zeta)$ generated by $f_1,f_2$ and $f_3$
(these vectors are drawn in Figure 7).

\

\hskip 0.8 in\scalebox{0.6}{\begin{picture}(0,0)%
\epsfbox{image.pstex}%
\end{picture}%
\setlength{\unitlength}{3947sp}%
\begingroup\makeatletter\ifx\SetFigFont\undefined%
\gdef\SetFigFont#1#2#3#4#5{%
  \reset@font\fontsize{#1}{#2pt}%
  \fontfamily{#3}\fontseries{#4}\fontshape{#5}%
  \selectfont}%
\fi\endgroup%
\begin{picture}(5550,5085)(1201,-5011)
\put(1726,-4936){\makebox(0,0)[lb]{\smash{\SetFigFont{14}{16.8}{\familydefault}{\mddefault}{\updefault}
$\zeta_1$
}}}
\put(3376,-136){\makebox(0,0)[lb]{\smash{\SetFigFont{14}{16.8}{\familydefault}{\mddefault}{\updefault}
$\zeta_3$
}}}
\put(6751,-3436){\makebox(0,0)[lb]{\smash{\SetFigFont{14}{16.8}{\familydefault}{\mddefault}{\updefault}
$\zeta_2$
}}}
\put(3451,-5011){\makebox(0,0)[lb]{\smash{\SetFigFont{20}{24.0}{\familydefault}{\mddefault}{\updefault}
$f_1$
}}}
\put(1201,-3511){\makebox(0,0)[lb]{\smash{\SetFigFont{20}{24.0}{\familydefault}{\mddefault}{\updefault}
$f_3$
}}}
\put(5026,-661){\makebox(0,0)[lb]{\smash{\SetFigFont{20}{24.0}{\familydefault}{\mddefault}{\updefault}
$f_2$
}}}
\end{picture}
}

\

\hskip 1 in Figure 7. {\footnotesize The image of the linear regions of 
$\phi$.}

\

The conifold singularity can be realized in the region 
$\zeta_2=0,\zeta_1>0,\zeta_3>0$ of $\R^3(\zeta)$, 
which defines a two-dimensional cone in the space of effective moment map 
levels
(this particular realization corresponds to eliminating the points 1 and 4 in 
Figure 5). 
The preimage of this cone via the map $\phi$  
is the the union of the boundaries of the opposite cones 
$\Xi'_1=\langle h_1,h_2,h_3,h_4\rangle_{+}$ and 
$\Xi'_2=-\Xi'_1=\langle -h_1,-h_2,-h_3,-h_4\rangle_{+}$ in $\R^3(\xi)$, 
given by the generators:
\bea
\begin{array}{cccccccccccc}
h_1&=&\left [\begin {array}{c} -1\\\noalign{\medskip}0\\\noalign{\medskip}1
\end {array}\right ]~,&
h_2&=&\left [\begin {array}{c} -1\\\noalign{\medskip}1\\\noalign{\medskip}0
\end {array}\right ]~,&
h_3&=&\left [\begin {array}{c} 0\\\noalign{\medskip}1\\\noalign{\medskip}0
\end {array}\right ]~,&
h_4=&\left [\begin {array}{c} 0\\\noalign{\medskip}0\\\noalign{\medskip}1
\end {array}\right ]
\end{array}~~.
\eea
This region in $\R^3(\xi)$ is shown in Figure 8. The conifold transition 
is realized in the D-brane theory when we vary $\xi$ such as to cross this 
boundary. 

\iffigs
$$\vbox{
\hskip 1.4 in \hbox{\epsfxsize=7cm\epsfbox{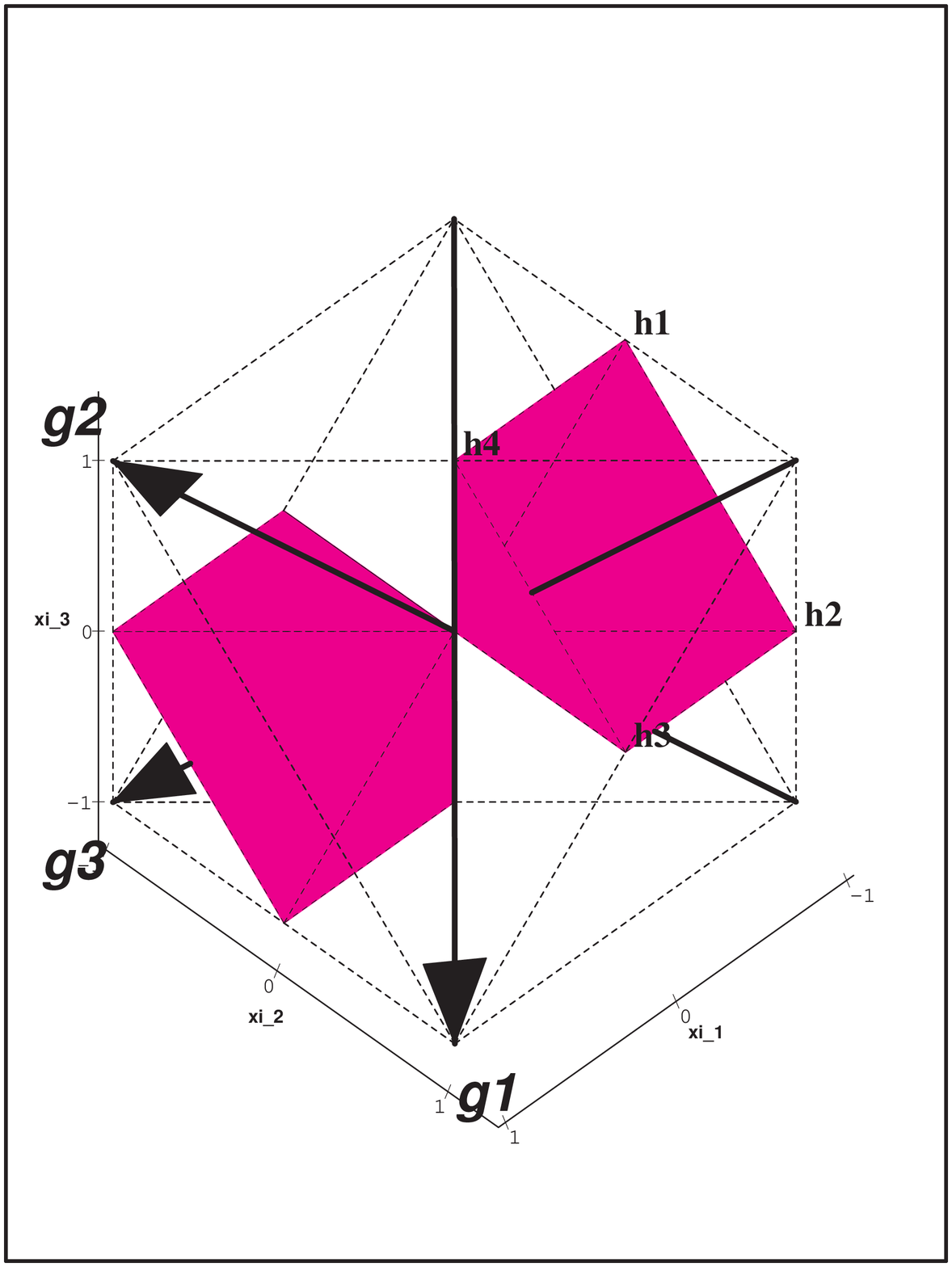}}

\vskip 0.3 in 

\hskip 1.25in Figure 8. {\footnotesize 
 The realization of the conifold region}

\hskip 1.25in {\footnotesize in the space of D-brane Fayet-Iliopoulos 
parameters.}}$$
\fi

\

\

\section{The case $\Gamma=\Z_3\times \Z_3$}

In this section, we apply the algorithm discussed above 
to the example of interest in this paper 
-- the worldvolume realization of partial resolutions of the 
$\C^3/\Z_3\times \Z_3$ quotient singularity.

The group $\Gamma=\Z_3\times \Z_3$ has $d=2$ torsion indices $t_1=3$, $t_2=3$. 
We consider its action on $\C^3$ given by the weights 
$w_1=(1,1)$, $w_2=(2,0)$, $w_3=(0,3)$ 
(which satisfy $w_1+w_2+w_3=(0,0)$ in our group). 

In this case, we have $|\Gamma|=9$ so the quiver will have $9$ nodes. 
The edges $v\rightarrow v-w_i$ can be obtained as above.
Indexing the group elements as follows:
\bea
(0,0)&\leftrightarrow 1&\nn\\
(0,1)&\leftrightarrow 4&\nn\\
(0,2)&\leftrightarrow 7&\nn\\
(1,0)&\leftrightarrow 2&\nn\\
(1,1)&\leftrightarrow 5&\nn\\
(1,2)&\leftrightarrow 8&\nn\\
(2,0)&\leftrightarrow 3&\nn\\
(2,1)&\leftrightarrow 6&\nn\\
(2,2)&\leftrightarrow 9&\nn,
\eea
we obtain the quiver drawn below:

\iffigs
\vskip 1 in
\hskip 0.3in\resizebox{4cm}{!}{\includegraphics[0in,4in][3in,8in]{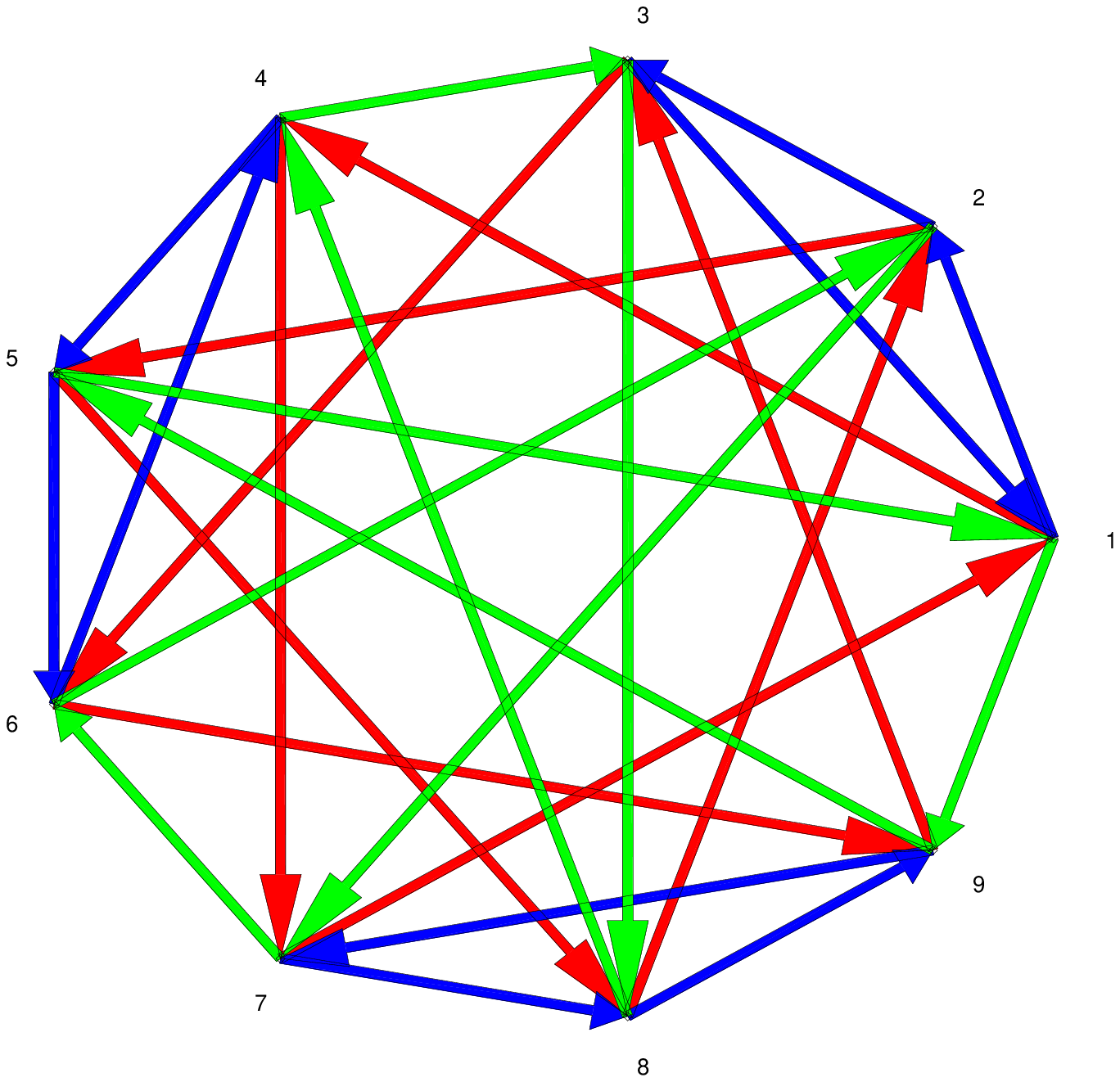}}\begin{center}
Figure 9. {\footnotesize The quiver describing the 
$\Z_3\times \Z_3$ orbifold theory.}

\end{center}
\vskip 0.5 in
\fi

With the above enumeration of the group elements, one obtains 
the $9 \times 27$ incidence matrix $d$ and the $8 \times 27$ 
matrix $\Delta$ given in the appendix (the first row of $d$ 
corresponds to the neutral element of the group with our choice of 
enumeration).

The number of independent monomial relations is 
$2(|\Gamma|-1)=16$, so the complex dimension of the variety of commuting 
matrixes is ${\rm dim}{\cal Z}=|\Gamma|+2=11$.
The matrix of an integral basis of the lattice of monomial relations is:
{\tiny \be
{\hat R}=\left [\begin {array}{ccccccccccccccccccccccccccc} 1&0&0
&0&0&0&0&0&-1&0&0&0&0&0&0&0&-1&1&0&0&0&0&-1&0&1&0&0
\\0&1&0&0&0&0&0&0&-1&0&0&0&0&0&0&1&-1&0&0&0&0&0&-1&0
&0&1&0\\0&0&1&0&0&0&0&0&-1&0&0&0&0&0&0&0&0&0&0&0&0&0
&-1&0&0&0&1\\0&0&0&1&0&0&0&0&-1&0&0&0&0&0&0&0&-1&1&0
&0&1&-1&-1&1&0&0&0\\0&0&0&0&1&0&0&0&-1&0&0&0&0&0&0&1
&-1&0&0&0&1&0&-2&1&-1&0&1\\0&0&0&0&0&1&0&0&-1&0&0&0&0
&0&0&0&0&0&0&0&1&0&-1&0&0&-1&1\\0&0&0&0&0&0&1&0&-1&0
&0&0&0&0&0&0&-1&1&0&0&0&0&-1&1&0&0&0\\0&0&0&0&0&0&0&
1&-1&0&0&0&0&0&0&1&-1&0&0&0&0&1&-1&0&0&0&0\\0&0&0&0&0
&0&0&0&0&1&0&0&0&0&0&-1&0&0&0&0&0&0&0&0&1&-1&0\\0&0&0
&0&0&0&0&0&0&0&1&0&0&0&0&0&-1&0&0&0&0&0&0&0&0&1&-1\\0
&0&0&0&0&0&0&0&0&0&0&1&0&0&0&0&0&-1&0&0&0&0&0&0&-1&0&1
\\0&0&0&0&0&0&0&0&0&0&0&0&1&0&0&-1&0&0&0&0&0&-1&1&0&0
&0&0\\0&0&0&0&0&0&0&0&0&0&0&0&0&1&0&0&-1&0&0&0&0&0&-
1&1&0&0&0\\0&0&0&0&0&0&0&0&0&0&0&0&0&0&1&0&0&-1&0&0&0
&1&0&-1&0&0&0\\0&0&0&0&0&0&0&0&0&0&0&0&0&0&0&0&0&0&1
&0&-1&1&0&-1&1&0&-1\\0&0&0&0&0&0&0&0&0&0&0&0&0&0&0&0
&0&0&0&1&-1&0&1&-1&0&1&-1\end {array}\right ]~~,
\ee}
(this has type $16\times 27$) and a basis of its integral kernel  is
given by the columns of the $27\times 11$ matrix: {\tiny \be K= \left
[\begin {array}{ccccccccccc} 1&0&1&-1&0&0&1&0&-1&0&0
\\1&-1&1&0&0&0&1&0&0&-1&0\\1&0&0&0
&0&0&1&0&0&0&-1\\1&0&1&-1&-1&1&1&-1&0&0&0
\\1&-1&1&0&-1&0&2&-1&1&0&-1\\1&0&0
&0&-1&0&1&0&0&1&-1\\1&0&1&-1&0&0&1&-1&0&0&0
\\1&-1&1&0&0&-1&1&0&0&0&0\\1&0&0&0
&0&0&0&0&0&0&0\\0&1&0&0&0&0&0&0&-1&1&0
\\0&0&1&0&0&0&0&0&0&-1&1\\0&0&0&1&0
&0&0&0&1&0&-1\\0&1&0&0&0&1&-1&0&0&0&0
\\0&0&1&0&0&0&1&-1&0&0&0\\0&0&0&1&0
&-1&0&1&0&0&0\\0&1&0&0&0&0&0&0&0&0&0
\\0&0&1&0&0&0&0&0&0&0&0\\0&0&0&1&0
&0&0&0&0&0&0\\0&0&0&0&1&-1&0&1&-1&0&1
\\0&0&0&0&1&0&-1&1&0&-1&1\\0&0&0&0
&1&0&0&0&0&0&0\\0&0&0&0&0&1&0&0&0&0&0
\\0&0&0&0&0&0&1&0&0&0&0\\0&0&0&0&0 &0&0&1&0&0&0\\0&0&0&0&0&0&0&0&1&0&0
\\0&0&0&0&0&0&0&0&0&1&0\\0&0&0&0&0 &0&0&0&0&0&1\end {array}\right ]~~.
\ee}
\noindent The lattice $M$ is isomorphic with $\Z^{11}$ and the $16$ vectors 
given by the rows of $K$ generate the cone of exponents 
$\Sigma\subset \R^{11}$ of the toric variety ${\cal Z}$.

The integral matrix $V$ satisfying $VK^t=\Delta$ is given by:
{\tiny\be
V=\left [\begin {array}{ccccccccccc} 0&0&0&0&0&0&0&0&0&1&0
\\0&0&0&0&-1&0&0&0&0&0&1\\0&0&0&0&0
&-1&0&0&0&0&0\\1&0&0&0&0&0&-1&0&0&0&0
\\0&0&0&0&1&0&0&-1&0&0&0\\0&-1&0&1
&0&1&0&0&-1&0&0\\0&1&-1&0&0&0&1&0&0&-1&0
\\-1&0&1&-1&0&0&0&1&0&0&-1\end {array}\right ]~~.
\ee}
A choice for the matrices $T,U, Q, Q_{total}$ results in the manner explained 
above. The types of these matrices are:
\be
\begin{array}{cccc}
T: 11\times 42 & U: 11 \times 42 &Q:31\times 42   &Q_{total}: 39\times 42~~.
\end{array}
\ee
\noindent 
In particular, in this example we have $c=42$.
These matrices, as well as  
the $3 \times 39$ matrix $G_t$ and the $39\times 42$ matrix 
$Q_t$, are listed in the appendix. 
 
The matrix $G_t$ has only $6$ distinct columns, appearing with 
multiplicities:
\be
\begin{array}{ccccccccccc}
m_1 &=& 1~~m_2 &=& 3~~m_3 &=& 3~~ m_4 &=& 3~~m_5 &=& 21\\
m_6 &=& 3 ~~m_7 &=& 1~~m_8 &=& 3~~m_9 &=& 3~~m_{10} &=& 1
\end{array}~~.
\ee
Thus $G_t$ has $7$ multiple blocks and $3$ non-multiple columns.
Keeping only one copy of each distinct column 
(without changing the decreasing lexicographic order) gives the matrix:
\be
G_{red}=\left [\begin {array}{cccccccccc} 
3&2&2&1&1&1&0&0&0&0\\
-1&0&-1&1&0&-1&2&1&0&-1\\
-1&-1&0&-1&0&1&-1&0&1&2\end {array}\right]~~.
\ee
This allows us to present the moduli space as a $3$-dimensional 
toric variety with $10$ toric generators and matrix of charges:
\be
Q_{red}=\left [\begin {array}{cccccccccc} 
1&0&0&0&0&-3&0&0&0&2\\
0&1&0&0&0&-2&0&0&-1&2\\
0&0&1&0&0&-2&0&0&0&1\\
0&0&0&1&0&-1&0&0&-2&2\\
0&0&0&0&1&-1&0&0&-1&1\\
0&0&0&0&0&0&1&0&-3&2\\
0&0&0&0&0&0&0&1&-2&1
\end {array}\right]~~,
\ee
which has $p=7$ rows.
The canonical form of $Q_{total}$, as well as the transition matrix $W$ 
(which has determinant $-1$) are listed in the appendix. 
The the $39\times 8$ matrix $W_0$ given by the last $8$ columns of $W$ 
can now be used to determine the piecewise-linear function $\phi$,which 
we display in the appendix.
Our procedure above gives $15309$ cones $\Xi$ in $\R^8(\xi)$, which form a 
refinement of the fan associated to $\phi$. By intersecting them with the 
preimages of the regions $H$ of $\R^7(\zeta)$ listed in Table 1, we find:

(a)$936$ maximal-dimensional cones in $\phi^{-1}(H_{F_0})$ leading to the 
complex cone over $F_0$

(b)$864$ maximal-dimensional cones in $\phi^{-1}(H_{dP_1})$ leading to the 
complex cone over $dP_1$

(c)$1152$ maximal-dimensional cones in $\phi^{-1}(H_{dP_2})$  leading to the 
complex cone over $dP_2$

(d)$1602$ maximal-dimensional cones in $\phi^{-1}(H_{dP_3})$ leading to the 
complex cone over $dP_3$

\noindent We consider each case in turn:

(a) A maximal-dimensional cone in $\phi^{-1}(H_{F_0})$ is generated by the 
columns of the matrix:
\be
\left[\begin {array}{ccccc} 
  1&1&1&0&0
\\0&-1&0&0&0
\\0&0&0&0&0
\\0&0&0&1&0
\\0&1&0&0&0
\\1&0&0&0&1
\\0&-1&-1&0&0
\\-1&0&0&-1&-1
\end {array}\right]
\ee
and maps to the cone in $\R^7(\zeta)$ generated by the columns of:
\be
\left[\begin {array}{ccccc} 
  2&0&0&2&1
\\0&0&0&2&0
\\1&0&0&1&0
\\0&0&0&2&0
\\0&0&0&1&0
\\0&2&1&2&0
\\0&1&0&1&0
\end {array}\right]
\ee~~.

(b)One maximal-dimensional cone in this class is generated by the columns 
of the matrix:
\be
\left[\begin {array}{ccccc} 
  1&1&0&0&0
\\0&0&0&0&-1
\\0&0&0&0&0
\\0&1&1&0&0
\\0&-1&0&0&0
\\-1&0&0&-1&0
\\-1&0&0&0&0
\\0&-1&-1&0&0
\end {array}\right]
\ee
and maps to the cone in $R^7(\zeta)$ generated by the columns of:
\be
\left[\begin {array}{ccccc} 
  0&4&2&0&1
\\0&3&2&0&0
\\0&2&1&0&0
\\1&2&2&0&0
\\0&1&1&0&0
\\2&1&2&1&0
\\0&0&1&0&0
\end {array}\right ]~~.
\ee

(c)One maximal-dimensional cone in this class is generated by the columns 
of the matrix:
\be
\left [\begin {array}{cccc} 
  1&0&1&1
\\0&0&-1&0
\\0&0&0&0
\\0&0&0&0
\\0&0&1&0
\\1&1&0&0
\\0&0&-1&-1
\\-1&-1&0&0
\end {array}\right]
\ee
and maps to the cone in $R^7(\zeta)$ generated by the columns of:
\be
\left[\begin {array}{cccc} 
  2&1&0&0
\\0&0&0&0
\\1&0&0&0
\\0&0&0&0
\\0&0&0&0
\\0&0&2&1
\\0&0&1&0
\end {array}\right ]
\ee

(d)One maximal-dimensional cone in this class is generated by the columns of 
the matrix:
\be
\left[\begin {array}{ccc} 
  0&0&1
\\0&0&0
\\0&0&0
\\1&0&0
\\0&0&0
\\0&1&0
\\0&0&-1
\\-1&-1&0
\end {array}\right]
\ee
and maps to the cone in $R^7(\zeta)$ generated by the columns of:
\be
\left[\begin {array}{ccc} 
  2&1&0
\\2&0&0
\\1&0&0
\\2&0&0
\\1&0&0
\\2&0&1
\\1&0&0
\end {array}\right]~~.
\ee

\section{Conclusions}

We considered the status of the AdS/CFT conjecture for 
nontrivial horizons built as $U(1)$ bundles over toric del Pezzo surfaces. 
By explicit computation, we discovered that all such geometries 
(including those cases for which the existence of an Einstein-Sasaki structure
is problematic)
can in fact be realized in the moduli space of D3-branes transverse to a 
Calabi-Yau quotient singularity. By investigating the classical moduli spaces
of the associated worldvolume theories, we discovered a highly intricate 
situation, whose complexity is markedly greater than in cases considered 
before. This required the development of a systematic approach to the problem,
thus improving on the methods presented in \cite{branes3}.

In this paper, we confined ourselves to geometric aspects and to classical 
properties of the moduli space of the associated field theories. The 
quantum-mechanical aspects of these theories and especially of their 
conformal limits are currently under investigation \cite{us}.

\appendix
\section{Various matrices relevant for the case $\C^3/\Z_3\times\Z_3$}
\rotatebox{270}{\tiny 
$\begin{array}{c}
d=\left [\begin {array}{ccccccccccccccccccccccccccc} 
\bar{1}&0&0&0&1&0&0&0&0&\bar{1}&0&1&0&0&0&0&0&0&\bar{1}&0&0&0&0&0&1&0&0\\0&\bar{1}&0&0&0&1
&0&0&0&1&\bar{1}&0&0&0&0&0&0&0&0&\bar{1}&0&0&0&0&0&1&0\\0&0&\bar{1}
&1&0&0&0&0&0&0&1&\bar{1}&0&0&0&0&0&0&0&0&\bar{1}&0&0&0&0&0&1\\0
&0&0&\bar{1}&0&0&0&1&0&0&0&0&\bar{1}&0&1&0&0&0&1&0&0&\bar{1}&0&0&0&0&0
\\0&0&0&0&\bar{1}&0&0&0&1&0&0&0&1&\bar{1}&0&0&0&0&0&1&0&0&\bar{1}&0
&0&0&0\\0&0&0&0&0&\bar{1}&1&0&0&0&0&0&0&1&\bar{1}&0&0&0&0&0&1&0
&0&\bar{1}&0&0&0\\0&1&0&0&0&0&\bar{1}&0&0&0&0&0&0&0&0&\bar{1}&0&1&0
&0&0&1&0&0&\bar{1}&0&0\\0&0&1&0&0&0&0&\bar{1}&0&0&0&0&0&0&0&1&
\bar{1}&0&0&0&0&0&1&0&0&\bar{1}&0\\1&0&0&0&0&0&0&0&\bar{1}&0&0&0&0&0
&0&0&1&\bar{1}&0&0&0&0&0&1&0&0&\bar{1}\end {array}\right ]~~\\
\  \\
\Delta=\left [\begin {array}{ccccccccccccccccccccccccccc} 0&\bar{1}&0&0&0&1
&0&0&0&1&\bar{1}&0&0&0&0&0&0&0&0&\bar{1}&0&0&0&0&0&1&0\\0&0&\bar{1}
&1&0&0&0&0&0&0&1&\bar{1}&0&0&0&0&0&0&0&0&\bar{1}&0&0&0&0&0&1\\0
&0&0&\bar{1}&0&0&0&1&0&0&0&0&\bar{1}&0&1&0&0&0&1&0&0&\bar{1}&0&0&0&0&0
\\0&0&0&0&\bar{1}&0&0&0&1&0&0&0&1&\bar{1}&0&0&0&0&0&1&0&0&\bar{1}&0
&0&0&0\\0&0&0&0&0&\bar{1}&1&0&0&0&0&0&0&1&\bar{1}&0&0&0&0&0&1&0
&0&\bar{1}&0&0&0\\0&1&0&0&0&0&\bar{1}&0&0&0&0&0&0&0&0&\bar{1}&0&1&0
&0&0&1&0&0&\bar{1}&0&0\\0&0&1&0&0&0&0&\bar{1}&0&0&0&0&0&0&0&1&
\bar{1}&0&0&0&0&0&1&0&0&\bar{1}&0\\1&0&0&0&0&0&0&0&\bar{1}&0&0&0&0&0
&0&0&1&\bar{1}&0&0&0&0&0&1&0&0&\bar{1}\end {array}\right ]~~\\
\  \\
T=\left [\begin {array}{cccccccccccccccccccccccccccccccccccccccccc} 
0&0 &0&0&0&0&0&0&0&1&1&0&1&0&0&1&1&0&1&1&0&0&0&1&1&1&1&0&0&0&0&1&1&1&0&0&0
&0&0&0&0&0\\0&0&1&0&0&0&0&1&0&0&1&1&0&0&0&0&1&1&1&0&
1&0&0&0&0&0&1&0&1&0&1&0&0&0&0&0&0&0&1&1&1&0\\1&1&0&0
&0&1&1&1&1&0&0&0&0&1&1&0&0&0&0&0&0&0&0&0&0&0&0&1&1&0&0&0&0&0&1&0&1&1&1
&0&0&0\\0&0&0&0&0&0&0&0&0&0&0&0&0&0&0&0&0&0&0&0&0&1&0
&1&1&1&1&1&1&1&0&1&1&1&0&0&0&1&0&1&0&1\\0&0&0&1&0&0&0
&0&0&0&0&1&1&1&1&1&1&1&0&0&0&0&0&0&0&0&0&0&0&1&0&1&0&0&0&1&1&1&0&0&0&1
\\0&0&0&1&1&1&1&0&0&0&0&0&0&0&0&0&0&0&0&0&0&1&1&1&1&0
&0&0&0&1&0&1&0&0&0&1&1&1&0&0&0&1\\0&0&1&1&1&0&0&0&0&0
&0&1&0&0&0&0&0&1&0&0&1&1&1&0&0&0&0&0&0&1&1&0&0&0&0&1&0&0&0&1&1&1
\\1&1&1&1&1&1&1&0&0&0&0&0&0&0&0&0&0&0&0&0&0&0&1&0&0&0
&0&0&0&0&1&0&0&0&1&1&1&0&1&0&1&0\\0&0&0&0&0&0&0&0&0&0
&0&0&0&0&1&1&1&1&1&1&1&0&1&0&0&0&0&0&0&0&1&0&0&0&1&1&1&0&1&0&1&0
\\0&1&0&0&0&1&0&0&0&0&0&0&0&1&1&1&0&0&0&1&0&0&1&1&0&0
&0&0&0&0&0&0&0&1&1&1&1&1&0&0&0&1\\0&0&0&0&0&0&0&0&0&0
&0&0&0&0&0&1&0&0&1&1&1&1&1&1&1&0&0&0&0&0&0&0&1&1&0&1&0&0&0&1&1&1
\end {array}\right ]~~\\
\  \\
U=\left [\begin {array}{cccccccccccccccccccccccccccccccccccccccccc} 0&0
&0&0&0&0&0&0&0&1&0&0&0&0&0&0&0&0&0&0&0&0&0&0&0&0&0&0&0&0&0&0&0&0&0&0&0
&0&0&0&0&0\\0&\bar{1}&1&0&\bar{1}&1&0&0&0&0&0&0&0&0&0&0&0&0&0&0
&0&0&0&0&0&0&0&0&0&0&0&0&0&0&0&0&0&0&0&0&0&0\\0&1&\bar{1}
&0&1&\bar{1}&0&1&0&0&0&0&0&0&0&0&0&0&0&0&0&0&0&0&0&0&0&0&0&0&0&0&0&0&0&0&0&0
&0&0&0&0\\1&\bar{2}&2&0&\bar{3}&2&0&\bar{2}&0&1&0&0&0&0&1&\bar{1}&0&0&0&0
&0&1&0&0&0&0&0&0&0&0&0&0&0&0&0&0&0&0&0&0&0&0\\0&0&0&
1&\bar{1}&0&0&0&0&0&0&0&0&0&0&0&0&0&0&0&0&0&0&0&0&0&0&0&0&0&0&0&0&0&0&0&0&0
&0&0&0&0\\0&\bar{1}&0&0&0&1&0&0&0&0&0&0&0&0&0&0&0&0&0&0&0
&0&0&0&0&0&0&0&0&0&0&0&0&0&0&0&0&0&0&0&0&0\\\bar{1}&2&\bar{1}&0
&2&\bar{2}&0&1&0&0&0&0&0&0&0&0&0&0&0&0&0&0&0&0&0&0&0&0&0&0&0&0&0&0&0&0&0&0&0
&0&0&0\\1&\bar{1}&1&0&\bar{1}&1&0&\bar{1}&0&0&0&0&0&0&0&0&0&0&0&0&0
&0&0&0&0&0&0&0&0&0&0&0&0&0&0&0&0&0&0&0&0&0\\1&\bar{2}&1&
\bar{1}&0&1&0&\bar{1}&0&0&0&0&0&0&1&0&0&0&0&0&0&0&0&0&0&0&0&0&0&0&0&0&0&0&0&0&0&0
&0&0&0&0\\\bar{1}&1&0&0&0&0&0&0&0&0&0&0&0&0&0&0&0&0&0&0&0
&0&0&0&0&0&0&0&0&0&0&0&0&0&0&0&0&0&0&0&0&0\\0&1&\bar{1}&0
&1&\bar{1}&0&1&0&\bar{1}&0&0&0&0&\bar{1}&1&0&0&0&0&0&0&0&0&0&0&0&0&0&0&0&0&0&0&0&0&0&0
&0&0&0&0\end {array}\right ]~~
\end{array}$}

\rotatebox{270}{\tiny 
$\begin{array}{c}
Q=\left [\begin {array}{cccccccccccccccccccccccccccccccccccccccccc} 1&0
&0&0&0&0&0&0&0&0&0&0&0&\bar{1}&0&0&0&0&0&0&1&0&0&0&\bar{1}&0&0&0&0&0&0&0&0&1&\bar{1}&0
&0&0&1&\bar{1}&\bar{1}&1\\0&1&0&0&0&0&0&0&0&0&0&0&0&\bar{1}&0&0&0&0
&0&0&1&0&0&0&\bar{1}&0&0&0&0&0&0&0&1&0&\bar{1}&0&0&0&1&\bar{1}&\bar{1}&1
\\0&0&1&0&0&0&0&0&0&0&0&0&0&\bar{1}&0&0&0&0&0&0&1&0&0&0&\bar{1}&0&0&0&0&0&0&\bar{1}&2&0&\bar{1}&0&0&0&2&\bar{2}&\bar{2}&2\\0&0&0&1&0&0
&0&0&0&0&0&0&0&\bar{1}&0&0&0&0&0&0&1&0&0&0&\bar{1}&0&0&0&0&0&0&\bar{1}&1&1&\bar{1}&0&0&0&2
&\bar{1}&\bar{2}&1\\0&0&0&0&1&0&0&0&0&0&0&0&0&\bar{1}&0&0&0&0&0&0&1
&0&0&0&\bar{2}&0&0&0&0&0&0&\bar{1}&2&1&\bar{2}&0&0&0&3&\bar{2}&\bar{2}&2\\0&0
&0&0&0&1&0&0&0&0&0&0&0&\bar{1}&0&0&0&0&0&0&1&0&0&0&\bar{2}&0&0&0&0&0&0&0&2&0&\bar{1}&0
&0&0&1&\bar{1}&\bar{1}&1\\0&0&0&0&0&0&1&0&0&0&0&0&0&\bar{1}&0&0&0&0
&0&0&1&0&0&0&\bar{2}&0&0&0&0&0&0&0&1&1&\bar{1}&0&0&0&1&\bar{1}&\bar{1}&1
\\0&0&0&0&0&0&0&1&0&0&0&0&0&\bar{1}&0&0&0&0&0&0&0&0&0&0&\bar{1}&0&0&0&0&0&0&0&1&0&0&0&0&0&0&\bar{1}&0&1\\0&0&0&0&0&0&0&0
&1&0&0&0&0&\bar{1}&0&0&0&0&0&0&0&0&0&0&\bar{1}&0&0&0&0&0&0&0&0&1&\bar{1}&0&0&0&1&\bar{1}&0
&1\\0&0&0&0&0&0&0&0&0&1&0&0&0&\bar{1}&0&0&0&0&0&0&0&0&0&0
&\bar{1}&0&0&0&0&0&0&\bar{1}&1&0&\bar{1}&0&0&0&2&\bar{1}&\bar{1}&2\\0&0&0&0&0
&0&0&0&0&0&1&0&0&\bar{1}&0&0&0&0&0&0&0&0&0&0&\bar{1}&0&0&0&0&0&0&\bar{1}&2&\bar{1}&0&0&0&0
&1&\bar{1}&\bar{1}&2\\0&0&0&0&0&0&0&0&0&0&0&1&0&\bar{1}&0&0&0&0&0&0
&0&0&0&0&0&0&0&0&0&0&0&\bar{1}&1&0&0&0&0&0&1&\bar{1}&\bar{1}&1\\0&0
&0&0&0&0&0&0&0&0&0&0&1&\bar{1}&0&0&0&0&0&0&0&0&0&0&0&0&0&0&0&0&0&\bar{1}&0&0&0&0
&0&0&1&0&\bar{1}&1\\0&0&0&0&0&0&0&0&0&0&0&0&0&0&1&0&0&0&0
&0&\bar{1}&0&0&0&1&0&0&0&0&0&0&0&\bar{1}&0&0&0&0&0&\bar{1}&1&1&\bar{1}\\0
&0&0&0&0&0&0&0&0&0&0&0&0&0&0&1&0&0&0&0&\bar{1}&0&0&0&1&0&0&0&0&0&0&0&\bar{1}&\bar{1}&
1&0&0&0&\bar{1}&2&0&\bar{1}\\0&0&0&0&0&0&0&0&0&0&0&0&0&0&0&0&1
&0&0&0&\bar{1}&0&0&0&1&0&0&0&0&0&0&\bar{1}&0&\bar{1}&1&0&0&0&\bar{1}&1&0&0
\\0&0&0&0&0&0&0&0&0&0&0&0&0&0&0&0&0&1&0&0&\bar{1}&0&0&0&1
&0&0&0&0&0&0&\bar{1}&0&0&0&0&0&0&0&0&0&0\\0&0&0&0&0&0&0&0
&0&0&0&0&0&0&0&0&0&0&1&0&\bar{1}&0&0&0&0&0&0&0&0&0&0&0&0&\bar{1}&1&0&0&0&\bar{1}&1&0&0
\\0&0&0&0&0&0&0&0&0&0&0&0&0&0&0&0&0&0&0&1&\bar{1}&0&0&0&0
&0&0&0&0&0&0&0&0&\bar{1}&0&0&0&0&0&1&0&0\\0&0&0&0&0&0&0&0
&0&0&0&0&0&0&0&0&0&0&0&0&0&1&0&0&\bar{1}&0&0&0&0&0&0&0&0&1&\bar{1}&0&0&0&1&\bar{1}&0&0
\\0&0&0&0&0&0&0&0&0&0&0&0&0&0&0&0&0&0&0&0&0&0&1&0&\bar{1}
&0&0&0&0&0&0&0&1&0&\bar{1}&0&0&0&1&0&\bar{1}&0\\0&0&0&0&0&0&0&0
&0&0&0&0&0&0&0&0&0&0&0&0&0&0&0&1&\bar{1}&0&0&0&0&0&0&0&1&\bar{1}&0&0&0&0&0&0&0&0
\\0&0&0&0&0&0&0&0&0&0&0&0&0&0&0&0&0&0&0&0&0&0&0&0&0&
1&0&0&0&0&0&\bar{1}&0&0&\bar{1}&0&0&0&1&\bar{1}&0&1\\0&0&0&0&0&0&0&0
&0&0&0&0&0&0&0&0&0&0&0&0&0&0&0&0&0&0&1&0&0&0&0&\bar{1}&1&\bar{1}&0&0&0&0&0&\bar{1}&0&
1\\0&0&0&0&0&0&0&0&0&0&0&0&0&0&0&0&0&0&0&0&0&0&0&0&0
&0&0&1&0&0&0&0&\bar{1}&1&\bar{1}&0&0&0&0&\bar{1}&1&0\\0&0&0&0&0&0&0
&0&0&0&0&0&0&0&0&0&0&0&0&0&0&0&0&0&0&0&0&0&1&0&0&0&0&0&0&0&0&0&\bar{1}&\bar{1}&1
&0\\0&0&0&0&0&0&0&0&0&0&0&0&0&0&0&0&0&0&0&0&0&0&0&0&0
&0&0&0&0&1&0&\bar{1}&0&1&\bar{1}&0&0&0&1&\bar{1}&0&0\\0&0&0&0&0&0&0
&0&0&0&0&0&0&0&0&0&0&0&0&0&0&0&0&0&0&0&0&0&0&0&1&\bar{1}&1&0&\bar{1}&0&0&0&1&\bar{1}&
\bar{1}&1\\0&0&0&0&0&0&0&0&0&0&0&0&0&0&0&0&0&0&0&0&0&0&0&0
&0&0&0&0&0&0&0&0&0&0&0&1&0&0&0&1&\bar{1}&\bar{1}\\0&0&0&0&0&0&0
&0&0&0&0&0&0&0&0&0&0&0&0&0&0&0&0&0&0&0&0&0&0&0&0&0&0&0&0&0&1&0&\bar{1}&1&0&
\bar{1}\\0&0&0&0&0&0&0&0&0&0&0&0&0&0&0&0&0&0&0&0&0&0&0&0&0
&0&0&0&0&0&0&0&0&0&0&0&0&1&\bar{1}&0&1&\bar{1}\end {array}\right ]~~\\
\ \\
Q_{total}=\left [\begin {array}
{cccccccccccccccccccccccccccccccccccccccccc} 1&0&0&0&0&0&0&0&0&0&0&0&0
&\bar{1}&0&0&0&0&0&0&1&0&0&0&\bar{1}&0&0&0&0&0&0&0&0&1&\bar{1}&0&0&0&1&\bar{1}&\bar{1}&1
\\0&1&0&0&0&0&0&0&0&0&0&0&0&\bar{1}&0&0&0&0&0&0&1&0&0&0&\bar{1}&0&0&0&0&0&0&0&1&0&\bar{1}&0&0&0&1&\bar{1}&\bar{1}&1\\0&0&1&0&0&0&0
&0&0&0&0&0&0&\bar{1}&0&0&0&0&0&0&1&0&0&0&\bar{1}&0&0&0&0&0&0&\bar{1}&2&0&\bar{1}&0&0&0&2&\bar{2}&\bar{2}&2\\0&0&0&1&0&0&0&0&0&0&0&0&0&\bar{1}&0&0&0&0&0&0&1&0
&0&0&\bar{1}&0&0&0&0&0&0&\bar{1}&1&1&\bar{1}&0&0&0&2&\bar{1}&\bar{2}&1\\0&0&0
&0&1&0&0&0&0&0&0&0&0&\bar{1}&0&0&0&0&0&0&1&0&0&0&\bar{2}&0&0&0&0&0&0&\bar{1}&2&1&\bar{2}&0
&0&0&3&\bar{2}&\bar{2}&2\\0&0&0&0&0&1&0&0&0&0&0&0&0&\bar{1}&0&0&0&0
&0&0&1&0&0&0&\bar{2}&0&0&0&0&0&0&0&2&0&\bar{1}&0&0&0&1&\bar{1}&\bar{1}&1
\\0&0&0&0&0&0&1&0&0&0&0&0&0&\bar{1}&0&0&0&0&0&0&1&0&0&0&\bar{2}&0&0&0&0&0&0&0&1&1&\bar{1}&0&0&0&1&\bar{1}&\bar{1}&1\\0&0&0&0&0&0&0
&1&0&0&0&0&0&\bar{1}&0&0&0&0&0&0&0&0&0&0&\bar{1}&0&0&0&0&0&0&0&1&0&0&0&0&0&0&\bar{1}&0
&1\\0&0&0&0&0&0&0&0&1&0&0&0&0&\bar{1}&0&0&0&0&0&0&0&0&0&0
&\bar{1}&0&0&0&0&0&0&0&0&1&\bar{1}&0&0&0&1&\bar{1}&0&1\\0&0&0&0&0&0
&0&0&0&1&0&0&0&\bar{1}&0&0&0&0&0&0&0&0&0&0&\bar{1}&0&0&0&0&0&0&\bar{1}&1&0&\bar{1}&0&0&0&2
&\bar{1}&\bar{1}&2\\0&0&0&0&0&0&0&0&0&0&1&0&0&\bar{1}&0&0&0&0&0&0&0
&0&0&0&\bar{1}&0&0&0&0&0&0&\bar{1}&2&\bar{1}&0&0&0&0&1&\bar{1}&\bar{1}&2\\0&0
&0&0&0&0&0&0&0&0&0&1&0&\bar{1}&0&0&0&0&0&0&0&0&0&0&0&0&0&0&0&0&0&\bar{1}&1&0&0&0
&0&0&1&\bar{1}&\bar{1}&1\\0&0&0&0&0&0&0&0&0&0&0&0&1&\bar{1}&0&0&0&0
&0&0&0&0&0&0&0&0&0&0&0&0&0&\bar{1}&0&0&0&0&0&0&1&0&\bar{1}&1\\0
&0&0&0&0&0&0&0&0&0&0&0&0&0&1&0&0&0&0&0&\bar{1}&0&0&0&1&0&0&0&0&0&0&0&\bar{1}&0&0
&0&0&0&\bar{1}&1&1&\bar{1}\\0&0&0&0&0&0&0&0&0&0&0&0&0&0&0&1&0&0
&0&0&\bar{1}&0&0&0&1&0&0&0&0&0&0&0&\bar{1}&\bar{1}&1&0&0&0&\bar{1}&2&0&\bar{1}
\\0&0&0&0&0&0&0&0&0&0&0&0&0&0&0&0&1&0&0&0&\bar{1}&0&0&0&1
&0&0&0&0&0&0&\bar{1}&0&\bar{1}&1&0&0&0&\bar{1}&1&0&0\\0&0&0&0&0&0&0
&0&0&0&0&0&0&0&0&0&0&1&0&0&\bar{1}&0&0&0&1&0&0&0&0&0&0&\bar{1}&0&0&0&0&0&0&0&0&0
&0\\0&0&0&0&0&0&0&0&0&0&0&0&0&0&0&0&0&0&1&0&\bar{1}&0&0&0
&0&0&0&0&0&0&0&0&0&\bar{1}&1&0&0&0&\bar{1}&1&0&0\\0&0&0&0&0&0&0
&0&0&0&0&0&0&0&0&0&0&0&0&1&\bar{1}&0&0&0&0&0&0&0&0&0&0&0&0&\bar{1}&0&0&0&0&0&1&0
&0\\0&0&0&0&0&0&0&0&0&0&0&0&0&0&0&0&0&0&0&0&0&1&0&0&
\bar{1}&0&0&0&0&0&0&0&0&1&\bar{1}&0&0&0&1&\bar{1}&0&0\\0&0&0&0&0&0&0
&0&0&0&0&0&0&0&0&0&0&0&0&0&0&0&1&0&\bar{1}&0&0&0&0&0&0&0&1&0&\bar{1}&0&0&0&1&0&\bar{1}&0\\0&0&0&0&0&0&0&0&0&0&0&0&0&0&0&0&0&0&0&0&0&0&0&1
&\bar{1}&0&0&0&0&0&0&0&1&\bar{1}&0&0&0&0&0&0&0&0\\0&0&0&0&0&0&0
&0&0&0&0&0&0&0&0&0&0&0&0&0&0&0&0&0&0&1&0&0&0&0&0&\bar{1}&0&0&\bar{1}&0&0&0&1&\bar{1}&0
&1\\0&0&0&0&0&0&0&0&0&0&0&0&0&0&0&0&0&0&0&0&0&0&0&0&0
&0&1&0&0&0&0&\bar{1}&1&\bar{1}&0&0&0&0&0&\bar{1}&0&1\\0&0&0&0&0&0&0
&0&0&0&0&0&0&0&0&0&0&0&0&0&0&0&0&0&0&0&0&1&0&0&0&0&\bar{1}&1&\bar{1}&0&0&0&0&\bar{1}&
1&0\\0&0&0&0&0&0&0&0&0&0&0&0&0&0&0&0&0&0&0&0&0&0&0&0
&0&0&0&0&1&0&0&0&0&0&0&0&0&0&\bar{1}&\bar{1}&1&0\\0&0&0&0&0&0&0
&0&0&0&0&0&0&0&0&0&0&0&0&0&0&0&0&0&0&0&0&0&0&1&0&\bar{1}&0&1&\bar{1}&0&0&0&1&\bar{1}&0
&0\\0&0&0&0&0&0&0&0&0&0&0&0&0&0&0&0&0&0&0&0&0&0&0&0&0
&0&0&0&0&0&1&\bar{1}&1&0&\bar{1}&0&0&0&1&\bar{1}&\bar{1}&1\\0&0&0&0&0&0&0
&0&0&0&0&0&0&0&0&0&0&0&0&0&0&0&0&0&0&0&0&0&0&0&0&0&0&0&0&1&0&0&0&1&\bar{1}&
\bar{1}\\0&0&0&0&0&0&0&0&0&0&0&0&0&0&0&0&0&0&0&0&0&0&0&0&0
&0&0&0&0&0&0&0&0&0&0&0&1&0&\bar{1}&1&0&\bar{1}\\0&0&0&0&0&0&0&0
&0&0&0&0&0&0&0&0&0&0&0&0&0&0&0&0&0&0&0&0&0&0&0&0&0&0&0&0&0&1&\bar{1}&0&1&\bar{1}
\\\bar{1}&1&0&0&0&0&0&0&0&0&0&0&0&0&0&0&0&0&0&0&0&0&0&0&0
&0&0&0&0&0&0&0&0&0&0&0&0&0&0&0&0&0\\0&1&\bar{1}&\bar{1}&2&\bar{1}&0
&1&0&\bar{1}&0&0&0&0&\bar{1}&1&0&0&0&0&0&0&0&0&0&0&0&0&0&0&0&0&0&0&0&0&0&0&0&0&0
&0\\0&1&0&0&0&\bar{1}&0&0&0&0&0&0&0&0&0&0&0&0&0&0&0&0&0&0
&0&0&0&0&0&0&0&0&0&0&0&0&0&0&0&0&0&0\\1&\bar{2}&1&0&\bar{2}&2&0
&\bar{1}&0&1&0&0&0&0&0&0&0&0&0&0&0&0&0&0&0&0&0&0&0&0&0&0&0&0&0&0&0&0&0&0&0&0
\\\bar{1}&1&\bar{1}&1&0&\bar{1}&0&1&0&0&0&0&0&0&0&0&0&0&0&0&0&0&0&0
&0&0&0&0&0&0&0&0&0&0&0&0&0&0&0&0&0&0\\0&0&0&1&\bar{2}&1&0
&\bar{1}&0&1&0&0&0&0&0&\bar{1}&0&0&0&0&0&1&0&0&0&0&0&0&0&0&0&0&0&0&0&0&0&0&0&0&0
&0\\0&\bar{1}&1&0&0&0&0&0&0&0&0&0&0&0&0&0&0&0&0&0&0&0&0&0
&0&0&0&0&0&0&0&0&0&0&0&0&0&0&0&0&0&0\\0&1&\bar{1}&0&2&\bar{1}&0
&1&0&\bar{1}&0&0&0&0&0&0&0&0&0&0&0&\bar{1}&0&0&0&0&0&0&0&0&0&0&0&0&0&0&0&0&0&0&0
&0\end {array}\right ]~~
\end{array}$}

\rotatebox{270}{\tiny
$\begin{array}{ccc}
G\_t=\left [\begin {array}{cccccccccccccccccccccccccccccccccccccccccc} 
3&2&2&2&2&2&2&1&1&1&1&1&1
&1&1&1&1&1&1&1&1&1&1&1&1&1&1&1&1&1&1&1&1&1&0&0&0&0&0&0&0&0\\
\bar{1}&0&0&0&\bar{1}&\bar{1}&\bar{1}&1&1&1&0&0&0&0&0&0&0&0&0&0&0&0&0&0
&0&0&0&0&0&0&0&\bar{1}&\bar{1}&\bar{1}&2&1&1&1&0&0&0&\bar{1}\\\bar{1}&\bar{1}&\bar{1}
&\bar{1}1&0&0&0&\bar{1}&\bar{1}&\bar{1}&0&0&0&0&0&0&0&0&0&0&0&0&0&0&0&0&0&0&0&0&0&1&1&1&\bar{1}&0&0
&0&1&1&1&2\end {array}\right ]~~\\
\ \\
Q\_t=\left [\begin {array}{cccccccccccccccccccccccccccccccccccccccccc}
0&1&0&\bar{1}&0&0&0&1&0&\bar{1}&1&1&1&0&0&0&0&0&0&0&0&0&0&0&0&0&0&0&\bar{1}&\bar{1}&\bar{1}
&0&0&0&0&0&0&0&0&0&0&0\\0&1&0&\bar{1}&0&0&0&1&0&\bar{1}&1&0&0&1&1&0&0&0&0&0&0&0&0&0&0
&0&0&0&\bar{1}&\bar{1}&\bar{1}&0&0&0&0&0&0&0&0&0&0&0\\0&2&0&\bar{2}&0&0&0
&2&0&\bar{2}&1&0&0&2&0&1&0&0&0&0&0&0&0&0&0&0&0&\bar{1}&\bar{1}&\bar{1}&\bar{1}&0&0&0&0&0&0&0&0&0
&0&0\\0&2&0&\bar{1}&0&0&0&1&0&\bar{2}&1&1&0&1&0&0&0&0&0&0&0&0&0
&0&0&0&0&\bar{1}&\bar{1}&\bar{1}&\bar{1}&0&0&0&0&1&0&0&0&0&0&0\\0&3&0&\bar{2}
&0&0&0&2&0&\bar{2}&1&1&0&2&0&0&0&0&0&0&0&0&0&0&0&0&0&\bar{1}&\bar{2}&\bar{2}&\bar{1}&0&0&0&0&0&0
&0&1&0&0&0\\0&1&0&\bar{1}&0&0&0&1&0&\bar{1}&1&0&0&2&0&0&1&0&0&0
&0&0&0&0&0&0&0&0&\bar{2}&\bar{1}&\bar{1}&0&0&0&0&0&0&0&0&0&0&0\\0&1
&0&\bar{1}&0&0&0&1&0&\bar{1}&1&1&0&1&0&0&0&1&0&0&0&0&0&0&0&0&0&0&\bar{2}&\bar{1}&\bar{1}&0&0&0&0
&0&0&0&0&0&0&0\\0&0&0&\bar{1}&1&0&0&1&0&0&0&0&0&1&0&0&0&0
&0&0&0&0&0&0&0&0&0&0&\bar{1}&0&\bar{1}&0&0&0&0&0&0&0&0&0&0&0\\0
&1&0&\bar{1}&0&0&0&1&0&0&0&1&0&0&0&0&0&0&0&0&0&0&0&0&0&0&0&0&\bar{1}&\bar{1}&\bar{1}&1&0&0
&0&0&0&0&0&0&0&0\\0&2&0&\bar{1}&0&0&0&2&0&\bar{1}&0&0&0&1&0&0&0
&0&0&0&0&0&0&0&0&0&0&\bar{1}&\bar{1}&\bar{1}&\bar{1}&0&0&0&0&0&0&0&0&0&0&1
\\0&1&0&\bar{1}&0&0&0&2&0&\bar{1}&0&\bar{1}&0&2&0&0&0&0&0&0&0&0&0&0
&0&0&0&\bar{1}&\bar{1}&0&\bar{1}&0&1&0&0&0&0&0&0&0&0&0\\0&1&0&\bar{1}&0&0
&0&1&0&\bar{1}&0&0&0&1&0&0&0&0&1&0&0&0&0&0&0&0&0&\bar{1}&0&0&\bar{1}&0&0&0&0&0&0&0&0&0
&0&0\\0&1&0&0&0&0&0&1&0&\bar{1}&0&0&0&0&0&0&0&0&0&0&0&0&0
&0&0&0&0&\bar{1}&0&0&\bar{1}&0&0&0&0&0&0&0&0&1&0&0\\0&\bar{1}&0&1&0
&0&0&\bar{1}&0&1&\bar{1}&0&0&\bar{1}&0&0&0&0&0&1&0&0&0&0&0&0&0&0&1&0&0&0&0&0&0&0&0&0&0
&0&0&0\\0&\bar{1}&0&2&0&0&0&\bar{1}&0&0&\bar{1}&\bar{1}&0&\bar{1}&0&0&0&0&0&0
&0&0&0&0&0&0&0&0&1&1&0&0&0&0&0&0&1&0&0&0&0&0\\0&\bar{1}&0
&1&0&0&0&0&0&0&\bar{1}&\bar{1}&0&0&0&0&0&0&0&0&1&0&0&0&0&0&0&\bar{1}&1&1&0&0&0&0&0&0&0
&0&0&0&0&0\\0&0&0&0&0&0&0&0&0&0&\bar{1}&0&0&0&0&0&0&0&0&0
&0&1&0&0&0&0&0&\bar{1}&1&0&0&0&0&0&0&0&0&0&0&0&0&0\\0&\bar{1}&0
&1&0&0&0&0&0&0&\bar{1}&\bar{1}&0&0&0&0&0&0&0&0&0&0&1&0&0&0&0&0&0&1&0&0&0&0&0&0&0
&0&0&0&0&0\\0&0&0&1&0&0&0&0&0&0&\bar{1}&\bar{1}&0&0&0&0&0&0&0&0
&0&0&0&0&0&0&0&0&0&0&0&0&0&0&0&0&0&0&0&0&1&0\\0&1&0&
\bar{1}&0&0&0&0&0&0&0&1&0&0&0&0&0&0&0&0&0&0&0&1&0&0&0&0&\bar{1}&\bar{1}&0&0&0&0&0&0&0
&0&0&0&0&0\\0&1&0&0&0&0&0&0&0&\bar{1}&0&0&0&1&0&0&0&0&0&0
&0&0&0&0&0&0&0&0&\bar{1}&\bar{1}&0&0&0&0&0&0&0&1&0&0&0&0\\0&0&0
&0&0&0&0&0&0&0&0&\bar{1}&0&1&0&0&0&0&0&0&0&0&0&0&1&0&0&0&\bar{1}&0&0&0&0&0&0&0&0
&0&0&0&0&0\\0&1&0&\bar{1}&0&0&0&1&0&0&0&0&0&0&0&0&0&0&0&0
&0&0&0&0&0&0&0&\bar{1}&0&\bar{1}&0&0&0&1&0&0&0&0&0&0&0&0\\0&0&0
&\bar{1}&0&1&0&1&0&0&0&\bar{1}&0&1&0&0&0&0&0&0&0&0&0&0&0&0&0&\bar{1}&0&0&0&0&0&0&0&0&0
&0&0&0&0&0\\0&0&0&\bar{1}&0&0&1&0&0&1&0&1&0&\bar{1}&0&0&0&0&0&0
&0&0&0&0&0&0&0&0&0&\bar{1}&0&0&0&0&0&0&0&0&0&0&0&0\\1&\bar{1}&0
&\bar{1}&0&0&0&0&0&1&0&0&0&0&0&0&0&0&0&0&0&0&0&0&0&0&0&0&0&0&0&0&0&0&0&0&0&0
&0&0&0&0\\0&1&0&\bar{1}&0&0&0&0&0&0&0&1&0&0&0&0&0&0&0&0&0
&0&0&0&0&1&0&\bar{1}&0&\bar{1}&0&0&0&0&0&0&0&0&0&0&0&0\\0&1&0&
\bar{1}&0&0&0&1&0&\bar{1}&0&0&0&1&0&0&0&0&0&0&0&0&0&0&0&0&1&\bar{1}&0&\bar{1}&0&0&0&0&0&0&0
&0&0&0&0&0\\0&0&0&1&0&0&0&\bar{1}&0&\bar{1}&0&0&0&0&0&0&0&0&0&0
&0&0&0&0&0&0&0&0&0&0&0&0&0&0&1&0&0&0&0&0&0&0\\0&\bar{1}&0
&1&0&0&0&\bar{1}&1&0&0&0&0&0&0&0&0&0&0&0&0&0&0&0&0&0&0&0&0&0&0&0&0&0&0&0&0&0
&0&0&0&0\\0&\bar{1}&1&0&0&0&0&\bar{1}&0&1&0&0&0&0&0&0&0&0&0&0&0
&0&0&0&0&0&0&0&0&0&0&0&0&0&0&0&0&0&0&0&0&0\\0&0&0&0&0
&0&0&0&0&0&0&0&\bar{1}&0&1&0&0&0&0&0&0&0&0&0&0&0&0&0&0&0&0&0&0&0&0&0&0&0&0&0
&0&0\\0&0&0&0&1&0&0&0&0&0&0&0&0&0&1&\bar{1}&\bar{1}&0&0&\bar{1}&0&0
&0&0&0&0&0&0&0&0&0&0&0&0&0&\bar{1}&1&0&2&0&0&\bar{1}\\0&0&0&0&0
&0&0&0&0&0&0&0&0&0&1&0&\bar{1}&0&0&0&0&0&0&0&0&0&0&0&0&0&0&0&0&0&0&0&0&0&0&0
&0&0\\0&0&0&0&\bar{1}&0&0&0&0&0&0&0&1&0&\bar{2}&1&2&0&0&0&0&0&0
&0&0&0&0&0&0&0&0&0&0&0&0&0&0&0&\bar{2}&0&0&1\\0&0&0&0&1&0
&0&0&0&0&0&0&\bar{1}&0&1&\bar{1}&\bar{1}&0&0&0&0&0&0&0&0&0&0&0&0&0&0&0&0&0&0&1&0&0&0&0
&0&0\\0&0&0&0&\bar{1}&0&0&0&0&0&0&0&0&0&0&0&1&0&0&0&0&0&0
&1&0&0&0&0&0&0&0&0&0&0&0&1&\bar{1}&0&\bar{2}&0&0&1\\0&0&0&0&0&0
&0&0&0&0&0&0&0&0&\bar{1}&1&0&0&0&0&0&0&0&0&0&0&0&0&0&0&0&0&0&0&0&0&0&0&0&0&0
&0\\0&0&0&0&1&0&0&0&0&0&0&0&0&0&1&\bar{1}&\bar{1}&0&0&0&0&0&0&
\bar{1}&0&0&0&0&0&0&0&0&0&0&0&0&0&0&2&0&0&\bar{1}\end {array}\right ]
\end{array}$}

\rotatebox{270}{\tiny 
$W=\left [\begin {array}{ccccccccccccccccccccccccccccccccccccccc} \bar{1}&5&
\bar{5}&0&6&\bar{5}&0&4&\bar{3}&\bar{1}&0&0&0&\bar{1}&1&0&0&0&0&\bar{1}&0&0&0&0&0&1&0&0&0&0&0&1&\bar{1}&0
&1&\bar{1}&0&1&\bar{1}\\\bar{2}&5&\bar{4}&1&3&\bar{4}&0&3&\bar{2}&0&0&0&0&\bar{1}&0&0&0
&0&0&0&0&0&0&0&0&0&0&0&0&0&0&0&\bar{1}&0&1&\bar{1}&\bar{1}&0&\bar{1}\\0&
2&\bar{2}&0&2&\bar{2}&0&2&\bar{2}&0&0&0&0&0&0&0&0&0&0&0&0&0&0&0&0&0&0&0&0&0&0&1&0&0&1
&0&0&1&0\\1&0&\bar{1}&\bar{1}&2&\bar{1}&0&1&\bar{1}&0&0&0&0&0&0&0&0&0&0&
\bar{1}&0&0&0&0&0&0&0&0&0&0&0&1&0&1&1&1&0&0&\bar{1}\\0&1&\bar{1}&1&
\bar{1}&0&0&0&\bar{1}&1&0&0&0&0&\bar{1}&0&0&0&0&0&0&0&0&0&0&0&0&0&0&0&0&0&0&0&0&0&\bar{1}&0
&\bar{1}\\0&0&0&0&\bar{1}&0&0&0&0&1&0&0&0&0&\bar{1}&0&0&0&0&0&0&0&0
&0&0&0&0&0&1&0&0&0&0&1&1&1&\bar{1}&\bar{1}&\bar{1}\\0&0&0&0&0&0&0&0
&0&0&0&0&0&0&0&0&0&0&0&0&0&0&0&0&0&0&0&0&0&0&0&0&0&1&1&1&0&0&0
\\\bar{2}&2&\bar{1}&2&\bar{2}&0&0&0&0&1&0&0&0&0&\bar{1}&0&0&0&0&1&0&0&0&0
&0&0&0&0&0&0&\bar{1}&\bar{1}&0&0&0&\bar{1}&\bar{1}&0&0\\0&1&0&\bar{1}&2&\bar{2}&0&
1&0&\bar{1}&0&0&0&\bar{1}&1&0&0&0&0&0&0&0&0&0&0&0&0&0&0&0&1&0&\bar{1}&\bar{1}&0&0&0&\bar{1}&0
\\1&\bar{3}&3&0&\bar{4}&3&0&\bar{2}&0&2&0&0&0&1&\bar{1}&0&0&0&0&1&0&0&0&
\bar{1}&0&0&0&0&0&0&0&0&1&0&0&1&0&0&1\\0&2&\bar{2}&0&2&\bar{2}&0&1&0
&\bar{1}&0&0&0&0&0&0&0&0&0&0&0&0&0&1&\bar{1}&0&0&0&0&0&0&1&0&0&1&0&0&1&0
\\2&\bar{3}&1&\bar{1}&0&2&0&\bar{1}&0&0&0&0&0&1&0&0&0&0&0&\bar{1}&0&0&0&0
&0&0&0&0&0&\bar{1}&0&1&1&1&0&1&1&1&0\\0&1&0&\bar{1}&2&\bar{2}&0&1&0
&\bar{1}&0&0&0&\bar{1}&1&0&0&0&0&0&0&0&0&0&0&0&0&0&0&1&0&0&\bar{1}&\bar{1}&0&0&0&\bar{1}&0
\\0&0&1&1&\bar{4}&2&0&\bar{2}&0&2&0&0&0&0&\bar{1}&0&0&0&0&1&0&0&0&0
&0&0&0&0&0&0&0&\bar{1}&0&\bar{1}&\bar{1}&0&\bar{1}&0&0\\0&0&\bar{1}&0&2&\bar{1}&0&
1&0&\bar{1}&0&0&0&0&0&0&0&0&0&\bar{1}&0&0&0&0&0&0&0&0&0&0&0&1&0&0&0&0&0&0&\bar{1}
\\1&\bar{1}&1&0&\bar{2}&1&0&\bar{1}&0&1&0&0&0&0&0&0&0&0&0&1&0&0&0&0
&0&0&0&0&0&0&0&0&0&0&0&0&0&0&1\\\bar{1}&1&\bar{1}&0&2&\bar{1}&0&1&0
&\bar{1}&0&0&0&0&0&0&0&0&0&\bar{1}&0&0&0&0&0&0&0&0&0&0&0&\bar{1}&0&0&0&0&0&0&\bar{1}
\\0&0&0&0&0&0&0&0&0&0&0&0&0&0&0&0&0&0&0&0&0&0&0&0&0&0
&0&0&0&0&0&0&0&0&0&0&0&\bar{1}&0\\0&0&0&0&0&0&0&0&0&0&0&0
&0&0&0&0&0&0&0&0&0&0&0&0&0&0&0&0&0&0&0&0&0&1&0&0&0&1&0
\\1&\bar{1}&0&0&0&1&\bar{1}&0&0&0&0&0&0&0&0&0&0&0&0&0&0&0&0&0&0
&0&0&0&0&0&0&1&0&0&0&0&0&0&0\\\bar{2}&2&0&1&0&\bar{2}&1&1&0&0&0
&\bar{1}&0&0&0&0&0&0&0&0&0&0&0&0&0&0&0&0&0&0&0&\bar{1}&0&\bar{1}&0&\bar{1}&0&\bar{1}&0
\\1&\bar{1}&0&\bar{1}&0&1&0&\bar{1}&0&0&0&1&0&0&0&0&0&0&0&0&0&0&0&0
&0&0&0&0&0&0&0&1&1&1&1&1&1&2&1\\1&\bar{3}&2&0&\bar{2}&2&0&\bar{1}&0
&1&0&0&0&1&0&\bar{1}&0&0&0&0&0&0&0&0&0&0&0&0&0&0&0&0&0&0&\bar{1}&0&0&\bar{1}&0
\\\bar{1}&2&\bar{1}&0&2&\bar{2}&0&1&0&\bar{1}&0&0&0&0&0&1&\bar{1}&0&0&0&0&0&0
&0&0&0&0&0&0&0&0&0&0&0&1&0&0&0&0\\1&\bar{1}&0&\bar{1}&0&1&0&0&0
&0&0&0&0&\bar{1}&1&0&1&\bar{1}&0&0&0&0&0&0&0&0&0&0&0&0&0&0&\bar{1}&0&\bar{1}&0&0&0&0
\\\bar{1}&2&\bar{1}&1&0&\bar{1}&0&0&0&0&0&0&0&0&\bar{1}&0&0&1&0&0&0&0&0&0
&0&0&0&0&0&0&0&0&0&0&1&0&\bar{1}&0&0\\0&\bar{1}&1&0&\bar{2}&2&0&\bar{1}&0
&1&0&0&0&0&0&0&0&0&0&1&0&\bar{1}&0&0&0&0&0&0&0&0&0&\bar{1}&0&0&\bar{1}&0&0&0&0
\\0&0&0&1&0&\bar{1}&0&0&0&0&0&0&0&1&\bar{1}&0&0&0&0&0&0&1&0&0&0
&0&\bar{1}&0&0&0&0&1&1&0&1&0&0&0&0\\1&\bar{1}&1&\bar{2}&2&0&0&0&0&
\bar{1}&0&0&0&0&1&0&0&0&0&\bar{1}&0&0&0&0&0&0&1&\bar{1}&0&0&0&0&0&0&0&1&1&0&0
\\0&\bar{1}&0&2&\bar{4}&2&0&\bar{1}&0&2&0&0&0&0&\bar{1}&0&0&0&0&1&0&0&0&0
&0&0&0&1&0&0&0&0&0&0&\bar{1}&\bar{1}&\bar{1}&0&0\\0&1&\bar{1}&\bar{1}&2&\bar{1}&0&
1&0&\bar{1}&0&0&0&\bar{1}&1&0&0&0&0&0&0&0&0&0&0&0&0&0&0&0&0&0&\bar{1}&0&0&0&0&0&0
\\0&1&0&\bar{1}&2&\bar{1}&0&0&0&\bar{1}&0&0&0&1&0&0&0&0&0&\bar{1}&0&0&0&0
&0&0&0&0&0&0&0&0&1&0&1&1&1&1&0\\\bar{2}&3&\bar{2}&2&0&\bar{2}&0&2&0
&0&0&0&0&\bar{1}&0&0&0&0&0&0&0&0&0&0&0&0&0&0&0&0&0&\bar{1}&\bar{1}&\bar{1}&\bar{1}&\bar{2}&\bar{1}&\bar{1}&\bar{1}
\\0&\bar{2}&2&0&\bar{2}&2&0&\bar{1}&1&1&\bar{1}&0&0&0&0&0&0&0&0&0&0&0&0&0
&0&0&0&0&0&0&0&\bar{1}&0&0&\bar{1}&0&0&\bar{1}&0\\1&\bar{1}&1&0&\bar{2}&1&0&
\bar{2}&0&1&1&0&0&1&\bar{1}&0&0&0&0&1&0&0&\bar{1}&0&0&0&0&0&0&0&0&1&1&0&1&1&0&1&1
\\0&0&0&0&0&0&0&0&0&0&0&0&0&0&0&0&0&0&0&0&0&0&0&0&0&0
&0&0&0&0&0&0&0&0&0&0&1&1&1\\0&0&0&\bar{1}&2&0&0&0&0&\bar{1}&0&0
&0&0&1&0&0&0&0&0&\bar{1}&0&0&0&0&0&0&0&0&0&0&0&0&1&1&1&0&0&0
\\0&0&0&1&\bar{1}&0&0&0&0&1&0&0&\bar{1}&0&0&0&0&0&0&0&0&0&0&0&0
&0&0&0&0&0&0&0&0&\bar{1}&\bar{1}&\bar{1}&0&0&0\\0&0&0&\bar{2}&2&0&0&0&0&
\bar{1}&0&0&1&0&1&0&0&0&\bar{1}&0&0&0&0&0&0&0&0&0&0&0&0&0&0&1&1&1&1&1&1
\end {array}\right ]$}

\

{\tiny $\phi(\xi)= \left [\begin {array}{c} \xi_{{2}}-\xi_{{3}}+\xi_{{5}}-\xi_{{6}}+\xi_{
{8}}-\xi_{{9}}+3\,M_1\\\noalign{\medskip}-\xi_{{3}}+\xi_{{5}}-\xi_{{6}}-\xi
_{{7}}-\xi_{{9}}-M_2+2\,M_3+M_4\\\noalign{\medskip}\xi
_{{2}}+\xi_{{5}}+\xi_{{8}}-M_5+2\,M_6
\\\noalign{\medskip}\xi_{{2}}+\xi_{{4}}+\xi_{{5}}+\xi_{{6}}-\xi_{{9}}-
M_7+M_8+2\,M_9\\\noalign{\medskip}-\xi_{{7}}-\xi_{{9}}
-M_{10}+M_{11}+M_{12}\\\noalign{\medskip}\xi_{{4}}+\xi_{{5}}+
\xi_{{6}}-\xi_{{7}}-\xi_{{8}}-\xi_{{9}}+3\,M_{13}\\\noalign{\medskip}\xi_{{4}}
+\xi_{{5}}+\xi_{{6}}-M_{14}+2\,M_{15}\end {array}\right ]$}~~,

\noindent where 

\noindent {\tiny $
M_1=\max(0,\xi_{{3}}+\xi_{{6}}+\xi_{{9}},-\xi_{{2}}-\xi_
{{5}}-\xi_{{8}})\\
M_2=\max(-\xi_{{2}}-\xi_{{6}}-\xi_{{7}}-\xi_{{3}}-\xi_{{4
}}-\xi_{{8}},0,-\xi_{{2}}-\xi_{{6}}-\xi_{{7}})\\
M_3=\max(0,\xi_{{3}}+\xi
_{{6}}+\xi_{{9}},-\xi_{{2}}-\xi_{{5}}-\xi_{{8}})\\
M_4=\max(0,\xi_{{7}}+\xi_
{{8}}+\xi_{{9}},-\xi_{{4}}-\xi_{{5}}-\xi_{{6}})\\
M_5=\max(0,\xi_{{3}}+\xi_{{6}}+\xi_{{9}},\xi_{{
2}}+\xi_{{3}}+\xi_{{5}}+\xi_{{6}}+\xi_{{8}}+\xi_{{9}})\\
M_6=\max(0,\xi_{
{3}}+\xi_{{6}}+\xi_{{9}},-\xi_{{2}}-\xi_{{5}}-\xi_{{8}})\\
M_7=\max(0,\xi_{{2}}+\xi_{{6}}+\xi_{{7}},\xi_{{2}}+\xi_{{3}}+\xi_{{4}}+\xi
_{{6}}+\xi_{{7}}+\xi_{{8}})\\
M_8=\max(0,\xi_{{3}}+\xi_{{6}}+\xi_{{9}},-\xi_
{{2}}-\xi_{{5}}-\xi_{{8}})\\
M_9=\max(0,\xi_{{7}}+\xi_{{8}}+\xi_{{9}},-
\xi_{{4}}-\xi_{{5}}-\xi_{{6}})\\
M_{10}=\max(\xi_{{3}},0,-\xi_{{2}}-\xi_{{4}}-\xi_{{5}}-\xi_{{7}}-\xi_{{9}},-
\xi_{{4}}-\xi_{{5}}-\xi_{{7}}-\xi_{{9}},-\xi_{{4}}-\xi_{{5}}-\xi_{{7}}
,-\xi_{{2}}-\xi_{{4}}-\xi_{{5}}-\xi_{{7}},-\xi_{{2}}-\xi_{{4}}-\xi_{{5
}}-\xi_{{7}}-\xi_{{9}}-\xi_{{8}},-\xi_{{2}}-\xi_{{5}}-\xi_{{7}}-\xi_{{
9}},-\xi_{{5}}-\xi_{{7}}-\xi_{{9}},-\xi_{{2}}-\xi_{{4}}-\xi_{{5}}-\xi_
{{7}}-\xi_{{9}}-\xi_{{8}}-\xi_{{6}},\xi_{{3}}+\xi_{{8}},\xi_{{3}}-\xi_
{{5}},-\xi_{{5}},-\xi_{{7}},-\xi_{{2}}-\xi_{{5}}-\xi_{{7}},\xi_{{3}}-
\xi_{{7}},\xi_{{3}}+\xi_{{6}},\xi_{{3}}-\xi_{{5}}-\xi_{{7}},-\xi_{{5}}
-\xi_{{7}},\xi_{{3}}+\xi_{{6}}+\xi_{{8}},-\xi_{{2}}-\xi_{{4}}-\xi_{{5}
}-\xi_{{6}}-\xi_{{7}}-\xi_{{9}})\\
M_{11}=\max(0,\xi_{{3}}+\xi_{{6}}+\xi_{{9}},-\xi_{{2}}-\xi_{{5}}-\xi_{{8}})\\
M_{12}=\max(0,\xi_{{7}}+\xi_{{8}}+\xi_{{9}},-\xi_{{4}}-\xi_{{5}}-\xi_{{6}})\\
M_{13}=\max(0,\xi_{{7}}+\xi_{{8}}+
\xi_{{9}},-\xi_{{4}}-\xi_{{5}}-\xi_{{6}})\\
M_{14}=\max(0,\xi_{{7}}+\xi_{{8}}+\xi_{{9}},\xi_{{4}}+
\xi_{{5}}+\xi_{{6}}+\xi_{{7}}+\xi_{{8}}+\xi_{{9}})\\
M_{15}=\max(0,\xi_{{7}}
+\xi_{{8}}+\xi_{{9}},-\xi_{{4}}-\xi_{{5}}-\xi_{{6}})\\
$}

\pagebreak

\pagebreak


\begin{thebibliography}{99}


\bibitem{branes3}{M.~R.~Douglas, B.~R.~Greene, D.~R.~Morrison,  
{\em Orbifold Resolution by 
D-Branes}, Nucl. Phys. {\bf B506} (1997) 84--106.}

\bibitem{branes3'}{B.~R.~Greene, {\em D-Brane Topology Changing Transitions}, 
hep-th/9711124,
T.~Muto, {\em D-branes on Orbifolds and Topology Change}, hep-th/9711090.}



\bibitem{Maldacena}{
J. Maldacena, {\em The large N limit of superconformal field theories 
and supergravity}, Adv. Theor. Math. Phys. {\bf 2} (1998) 231--252, 
hep-th/9711200.}


\bibitem{us}{Chris~Beasley, B.~Greene,C.~I.~Lazaroiu,R.~Plesser, 
{\em in preparation}.}

\bibitem{Figueroa}{
B.~S.~Acharya, J.~M.~Figueroa-O'Farrill, C.~M.~Hull, B.~Spence,
{\em Branes at conical singularities and holography}, 
Adv. Theor. Math. Phys. {\bf 2} (1998) 1249-1286, hep-th/9808014.}



\bibitem{MP}{
David R. Morrison and M. Ronen Plesser, {\em Non-Spherical Horizons, I,} 
hep-th/9810201.}


\bibitem{Tian_Yau}{G.~Tian, S.~T.~Yau, {\em K\"ahler-Einstein metrics on 
compact surfaces with $c_1>0$}, Commun Math Phys, {\bf 112}, 175-203 (1987).}

\bibitem{Tian}{G.~Tian,{\em K\"ahler-Einstein metrics with positive scalar 
curvature}, Invent. Math. {\bf 137}, 1--37(1997). }


\bibitem{Tian_moduli}{G.~Tian, {\em On Calabi's conjecture for complex 
surfaces with positive first Chern class},Invent. Math. {\bf 101},101--172
(1990). }


\bibitem{Altmann}{
K. Altmann, {\em Deformations of affine torus varieties}, 
 Beitrage Algebra Geom. 34 (1993), no. 1, 119--150; 
{\em The versal deformation of an isolated toric Gorenstein 
singularity}, Invent. Math. {\bf 128} (1997) 443--479.}


\bibitem{Klebanov}{
I.~R. Klebanov and E. Witten, {\em Superconformal field theory on 
threebranes at a Calabi-Yau singularity}, Nucl.Phys. {\bf B536} (1998) 
199-218, hep-th/9807080.}


\bibitem{Oda}{T.~Oda, {\em Convex bodies and algebraic geometry: 
an introduction to the
theory of toric varieties}, Ergebnisse der Mathematik und ihrer Grenzgebiete; 
3.Folge, {\bf 15}, Springer, 1988; }



\bibitem{top_change}{
P.~S. Aspinwall, B.~R. Greene, and D.~R. Morrison, {\em Calabi-Yau moduli 
space, mirror manifolds, and spacetime topology change in string theory}, 
Nucl. Phys. {\bf B416} (1994) 414--480, hep-th/9309097, 
E.~Witten,{\em Phases of $N=2$ Theories In Two Dimensions},Nucl.Phys. B403 (1993) 159-222.}


\bibitem{Fulton}{
W. Fulton, {\em Introduction to Toric Varieties }, 
Princeton University Press, 1993.}


\bibitem{Cox}{D.~A.~Cox, {\em The Homogeneous Coordinate Ring of a Toric 
Variety}, alg-geom/9210008.}

\bibitem{douglas-moore}{M.~R.~Douglas, G.~Moore, {\em D-branes,
Quivers, and ALE  Instantons}, hep-th/9603167.}

\bibitem{Infirri}{A.~V.~Sardo~Infirri, {\em Partial resolutions of orbifold 
singularities via moduli spaces of HYM-type bundles}, alg-geom/9610004; 
{\em Resolutions of orbifold singularities and the transportation problem on 
the McKay quiver}, alg-geom/9610005.}



\bibitem{nonabelian}{Brian R. Greene, C. I. Lazaroiu, Mark Raugas
{\em D-branes on Nonabelian Threefold Quotient Singularities}, hep-th/9811201.}



\bibitem{Audin}{M.~Audin, {\em The topology of torus actions on symplectic 
manifolds}, Progress in Mathematics {\bf 93}, Birkhauser, 1991.}





\end{thebibliography}
\end{document}